\documentclass[]{aa} 

\usepackage{epsfig}

\renewcommand{\d}{{\rm d}}
\newcommand{\pl}{\partial}
 
\newcommand{\beq}{\begin{equation}} 
\newcommand{\eeq}{\end{equation}} 
\newcommand{\beqa}{\begin{eqnarray}} 
\newcommand{\eeqa}{\end{eqnarray}} 
\newcommand{\bea}{\begin{array}} 
\newcommand{\ea}{\end{array}} 
\newcommand{\cH}{{\cal H}} 
 
\newcommand{\lag}{\langle} 
\newcommand{\rag}{\rangle}
\newcommand{\bx}{{\bf x}}
\newcommand{\bv}{{\bf v}}
\newcommand{\bk}{{\bf k}}
\newcommand{\Om}{\Omega_{\rm m}}
\newcommand{\OL}{\Omega_{\Lambda}}
\newcommand{\cO}{{\cal O}}
\newcommand{\tcO}{{\tilde{\cal O}}}
\newcommand{\tKs}{{\tilde{K}_s}}
\newcommand{\Tr}{{\rm Tr}}

\newcommand{\bq}{{\bf q}}
\newcommand{\br}{{\bf r}}
\newcommand{\cP}{{\cal P}}

\begin{document} 
 
\title{Large-$N$ expansions applied to gravitational clustering}    
\author{P. Valageas}   
\institute{Service de Physique Th\'eorique, CEA Saclay, 91191 Gif-sur-Yvette, 
France}  
\date{Received / Accepted } 
 
\abstract{

We develop a path-integral formalism to study the formation of large-scale 
structures in the universe. Starting from the equations of motion of 
hydrodynamics (single-stream approximation) we derive the action which
describes the statistical properties of the density and velocity fields
for Gaussian initial conditions. Then, we present large-$N$ expansions
(associated with a generalization to $N$ fields or with a semi-classical 
expansion) of the path-integral defined by this action. This provides a 
systematic expansion for two-point functions such as the response function
and the usual two-point correlation. We present the results of two such
expansions (and related variants) at one-loop order for a SCDM and 
a $\Lambda$CDM cosmology.
We find that the response function exhibits fast oscillations in the non-linear
regime with an amplitude which either follows the linear prediction (for the
direct steepest-descent scheme) or decays (for the 2PI effective action 
scheme). On the other hand, the correlation function agrees with the
standard one-loop result in the quasi-linear regime and remains well-behaved
in the highly non-linear regime. This suggests that these large-$N$ expansions
could provide a good framework to study the dynamics of gravitational
clustering in the non-linear regime. Moreover, the use of various expansion
schemes allows one to estimate their range of validity without the need of
$N-$body simulations and could provide a better accuracy in the weakly 
non-linear regime.

} 
 
\maketitle

\section{Introduction} 
\label{Introduction}

The large-scale structures we observe in the present universe (such as 
galaxies and clusters of galaxies) have formed thanks to gravitational 
instability (Pebbles 1980) which amplified the small density perturbations 
created in the early universe (e.g. through quantum fluctuations generated
during an inflationary phase, Liddle \& Lyth 1993). Moreover, the power
increases at small scales as in the CDM model (Peebles 1982) which leads to
a hierarchical scenario where small scales become non-linear first. Then,
small structures merge in the course of time to build increasingly large
objects. Therefore, at large scales or at early times one can use
a perturbative approach to describe the density field 
(e.g. Bernardeau et al. 2002 for a review). This is of great
practical interest for observations such as weak-lensing surveys (e.g.
Bernardeau et al. 1997) and CMB studies (e.g. Seljak \& Zaldarriaga 1996).
At smaller scales one uses N-body numerical simulations to obtain fits to
the density power-spectrum (e.g. Smith et al. 2003) which can be compared
with observations.

In the weakly non-linear regime one may apply the standard perturbative
expansion over powers of the initial density fluctuations (e.g., Fry 1984;
Goroff et al. 1986). This procedure is used within the hydrodynamical
framework where the system is described by density and velocity fields
rather than by the phase-space distribution $f(\bx,\bv,t)$, that is one
starts from the Euler equation of motion rather than the Vlasov equation.
This is quite successful at tree-order where one can 
obtain the leading-order contribution to all $p-$point correlations and 
reconstruct the density probability distribution (Bernardeau 1994), assuming
Gaussian initial fluctuations as in usual inflationary scenarios 
(Liddle \& Lyth 1993). However, several problems show up when one tries to 
compute higher-order corrections. First, ultraviolet divergences appear 
for linear power-spectra $P_L(k) \propto k^n$ with $n\geq -1$ 
(Scoccimarro \& Frieman 1996b) which would break the self-similarity seen 
in numerical simulations. This may also be interpreted as a consequence of
the breakdown of the hydrodynamical description beyond shell-crossing
(Valageas 2002). However, for $n<-1$ the one-loop correction yields a good
agreement with numerical simulation (Scoccimarro \& Frieman 1996b). Secondly,
being an expansion over powers of $P_L(k)$ this standard perturbative approach
yields a series of terms which grow increasingly fast into the non-linear 
regime at higher orders and cannot relax at small scales. This makes the 
series badly behaved and it seems difficult to go deeper into the non-linear
regime in this manner. 

On the other hand, a good description of the weakly 
non-linear regime becomes of great practical interest as cosmological probes
now aim to constrain cosmological parameters with an accuracy of the order
of $1\%$ to keep pace with CMB observations. In particular, weak-lensing surveys 
mainly probe this transition regime to non-linearity and a good accuracy
for the matter power-spectrum is required to make the most of observations and
constrain the dark energy equation of state (Munshi et al. 2007).
Another probe of the expansion history of the Universe is provided by the
measure of the baryon acoustic oscillations which also requires a good 
description of weakly non-linear effects 
(Seo \& Eisenstein 2007; Koehler et al. 2007).

As a consequence, it is worth investigating other approaches which may
partly cure some of these problems and provide a better description 
of the non-linear regime. For instance, McDonald (2006) proposed
a renormalization group method to improve the standard perturbative
approach. On the other hand, Crocce \& Scoccimarro (2006a,b)
presented a diagrammatic technique to organize the various terms which
arise in the standard perturbative expansion and to perform some infinite 
partial resummations. Besides, using the asymptotic behavior of the vertices 
of the theory they managed to complete the resummation in the small-scale
limit for the response function $R$ (also called the propagator).

In this article, we apply to the hydrodynamical framework the method 
presented in Valageas (2004) for the Vlasov equation. This approach
introduces a path-integral formalism to derive the statistical properties
of the system from its action $S$. Then, one applies a ``large-$N$ expansion''
(which can be seen as similar to the semi-classical expansion over powers
of $\hbar$ or can be derived from a generalization to $N$ fields) to compute
the quantities of interest such as the two-point correlation. 
This offers a different perspective to address the non-linear regime of 
gravitational clustering which complements other approaches and may also
serve as a basis for other approximation schemes than the ones described in this
article.
For instance, previous approaches can be recovered by expanding the path-integral
obtained in this paper in different manners (e.g. the usual perturbative expansion
corresponds to the expansion over the non-Gaussian part of the action $S$). The two
large-$N$ expansions discussed in this article are two other means of performing
systematic expansions to compute the required correlation functions. This also 
amounts to reorganize the standard perturbative expansion by performing some partial
infinite resummations. In this fashion, one can hope to obtain new expansion schemes
which are better suited to describe the non-linear regime of gravitational 
clustering. 

First, we recall in sect.~\ref{Equations-of-motion} the equations of motion
obtained within the hydrodynamical framework. Next, we derive the path-integral
formulation of the problem in sect.~\ref{Path-integral} and we describe
two possible large-$N$ expansions in sect.~\ref{Large-N-expansions}.
Finally, we present our numerical results in 
sects.~\ref{Direct-steepest-descent}-\ref{Simplified-response-approximations},
focusing on a critical-density universe. Since our formalism applies equally
well to any cosmology we describe our results for a $\Lambda$CDM universe
in sect.~\ref{LCDM}. Finally, we conclude in sect.~\ref{Conclusion}.

\section{Equations of motion}
\label{Equations-of-motion}

\subsection{Fluid approach}
\label{Fluid}

At scales much larger than the Jeans length both the cold dark matter and the 
baryons can be described as a pressureless dust. Moreover, before orbit 
crossing one can use a hydrodynamical approach (in the continuum limit where
the mass $m$ of particles goes to zero). Then, at scales much smaller than the
horizon where the Newtonian approximation is valid, the equations of motion
read (Peebles 1980):
\beq
\frac{\pl\delta}{\pl\tau} + \nabla.[(1+\delta) \bv] = 0,
\label{continuity1}
\eeq
\beq
\frac{\pl\bv}{\pl\tau} + \cH \bv + (\bv .\nabla) \bv = - \nabla \phi,
\label{Euler1}
\eeq
\beq
\Delta \phi = \frac{3}{2} \Om \cH^2 \delta , 
\label{Poisson1}
\eeq
where $\tau=\int \d t/a$ is the conformal time (and $a$ the scale factor), 
$\cH=\d\ln a/\d\tau$ the
conformal expansion rate and $\Om$ the matter density cosmological parameter.
Here $\delta$ is the matter density contrast and $\bv$ the peculiar velocity.
Since the vorticity field decays within linear theory (Peebles 1980) we take
the velocity to be a potential field so that $\bv$ is fully specified by its 
divergence $\theta=\nabla.\bv$. Then, in Fourier space 
eqs.(\ref{continuity1})-(\ref{Poisson1}) read (Goroff et al. 1986):
\beqa
\frac{\pl\delta(\bk,\tau)}{\pl\tau} + \theta(\bk,\tau) & = & 
- \int\d\bk_1\d\bk_2 \; \delta_D(\bk_1+\bk_2-\bk) \nonumber \\
&& \times \alpha(\bk_1,\bk_2) \theta(\bk_1,\tau) \delta(\bk_2,\tau) ,
\label{continuity2}
\eeqa
\beqa
\lefteqn{ \frac{\pl\theta(\bk,\tau)}{\pl\tau} + \cH \theta(\bk,\tau) 
+ \frac{3}{2} \Om \cH^2 \delta(\bk,\tau) = } \nonumber \\
&& \!\!\!\! - \!\! \int\d\bk_1\d\bk_2 \; \delta_D(\bk_1+\bk_2-\bk) 
\beta(\bk_1,\bk_2) \theta(\bk_1,\tau) \theta(\bk_2,\tau) ,
\label{EulerPoisson2}
\eeqa
where $\delta_D$ is the Dirac distribution, the coupling functions $\alpha$ 
and $\beta$ are given by:
\beq
\alpha(\bk_1,\bk_2)= \frac{(\bk_1+\bk_2).\bk_1}{k_1^2} ,
\beta(\bk_1,\bk_2)= \frac{|\bk_1+\bk_2|^2 (\bk_1.\bk_2)}{2 k_1^2 k_2^2} ,
\label{alphabeta}
\eeq
and we defined the Fourier transforms as:
\beq
\delta(\bk) = \int\frac{\d\bx}{(2\pi)^3} e^{-i\bk.\bx} \delta(\bx) .
\label{deltak}
\eeq
As in Crocce \& Scoccimarro (2006a,b) let us define the two-component vector
$\psi$ as:
\beq
\psi(\bk,\eta) = \left(\bea{c} \psi_1(\bk,\eta) \\ \psi_2(\bk,\eta)\ea \right)
= \left( \bea{c} \delta(\bk,\eta) \\ -\theta(\bk,\eta)/\cH f 
\ea \right) ,
\label{psidef}
\eeq
where we introduced the time coordinate $\eta$ defined from the linear 
growing rate $D_+$ of the density contrast:
\beq
\eta= \ln \frac{D_+(\tau)}{D_{+0}} , \;\;\; f=\frac{\d\ln D_+}{\d\ln a}
= \frac{1}{\cH}\frac{\d\ln D_+}{\d\tau} ,
\label{eta}
\eeq
with $D_{+0}$ is the value of $D_+$ today at redshift $z=0$ and:
\beq
\frac{\d^2 D_+}{\d\tau^2}+\cH\frac{\d D_+}{\d\tau} = \frac{3}{2}\Om\cH^2 D_+ .
\label{D+}
\eeq
Then, the equations of motion (\ref{continuity2})-(\ref{EulerPoisson2}) can be
written as:
\beq
\cO(x,x').\psi(x') = K_s(x;x_1,x_2) . \psi(x_1) \psi(x_2) ,
\label{OKsdef}
\eeq
where we introduced the coordinate $x=(\bk,\eta,i)$ where $i=1,2$ is the 
discrete index of the two-component vectors. In eq.(\ref{OKsdef}) and in the 
following we use the convention that repeated coordinates are integrated over
as:
\beq
\cO(x,x').\psi(x') = \!\! \int\!\d\bk'\d\eta'\sum_{i'=1}^2 
\cO_{i,i'}(\bk,\eta;\bk',\eta') \psi_{i'}(\bk',\eta') .
\label{defproduct}
\eeq
The matrix $\cO$ reads:
\beq
\! \cO(x,x') \! = \! \left( \bea{cc} \frac{\pl}{\pl\eta} & -1 \\ -\frac{3\Om}{2f^2} \;
& \; \frac{\pl}{\pl\eta} + \frac{3\Om}{2f^2} -1 \ea \right) \! \delta_D(\bk-\bk')
\delta_D(\eta-\eta')
\label{Odef}
\eeq
whereas the symmetric vertex $K_s(x;x_1,x_2)=K_s(x;x_2,x_1)$ writes:
\beqa
K_s(x;x_1,x_2) & = & \delta_D(\bk_1+\bk_2-\bk) \delta_D(\eta_1-\eta) 
\delta_D(\eta_2-\eta) \nonumber \\ && \times \gamma^s_{i;i_1,i_2}(\bk_1,\bk_2)
\label{Ksdef}
\eeqa
with:
\beq
\gamma^s_{1;1,2}(\bk_1,\bk_2)= \frac{\alpha(\bk_2,\bk_1)}{2} , \;\;
\gamma^s_{1;2,1}(\bk_1,\bk_2)= \frac{\alpha(\bk_1,\bk_2)}{2} ,
\label{gamma1}
\eeq
\beq
\gamma^s_{2;2,2}(\bk_1,\bk_2)= \beta(\bk_1,\bk_2) ,
\label{gamma2}
\eeq
and zero otherwise (Crocce \& Scoccimarro 2006a).

\subsection{Linear regime}
\label{Linear}

At large scales or at early times where the density and velocity fluctuations
are small one can linearize the equation of motion (\ref{OKsdef}) which yields
$\cO.\psi_L=0$. Then, the linear growing mode is merely:
\beq
\psi_L(x) = e^{\eta} \delta_{L0}(\bk) \left(\bea{c} 1 \\ 1 \ea\right) ,
\label{psiL}
\eeq
where $\delta_{L0}(\bk)$ is the linear density contrast today at redshift
$z=0$. We shall define the initial conditions by the linear growing mode
(\ref{psiL}) (which is consistent with the fact that we only kept the curl-free
part of the peculiar velocity in 
eqs.(\ref{continuity2})-(\ref{EulerPoisson2})). Moreover, we assume Gaussian
homogeneous and isotropic initial conditions defined by the linear 
power-spectrum $P_{L0}(k)$:
\beq
\lag \delta_{L0}(\bk_1) \delta_{L0}(\bk_2) \rag =  P_{L0}(k_1) 
\delta_D(\bk_1+\bk_2) .
\label{PL0}
\eeq
It is convenient to define the power per logarithmic wavenumber $\Delta^2(k)$
by:
\beq
\Delta^2(k) = 4\pi k^3 P(k) ,
\label{Delta2def}
\eeq
whence:
\beq
\lag\delta(\bx_1)\delta(\bx_2)\rag = \int_0^{\infty} \frac{\d k}{k} 
\Delta^2(k) \frac{\sin k|\bx_1-\bx_2|}{k|\bx_1-\bx_2|} .
\label{xir}
\eeq
Thus, for a CDM cosmology the linear power $\Delta^2_{L0}(k)$
grows as $k^4$ at small $k$ and as $\ln k$ at high $k$.
Then, the linear two-point correlation function 
$G_L(x_1,x_2)$ reads:
\beqa
G_L(x_1,x_2) & = & \lag \psi_L(x_1) \psi_L(x_2) \rag \nonumber \\
& = & e^{\eta_1+\eta_2} P_{L0}(k_1) \delta_D(\bk_1+\bk_2) 
\left(\bea{cc} 1 & 1 \\ 1 & 1 \ea\right) .
\label{GL}
\eeqa
As in Valageas (2004) it is convenient to introduce the response function
$R(x_1,x_2)$ (which is also called the propagator in Crocce \& Scoccimarro 
(2006a,b)) defined by the functional derivative:
\beq
R(x_1,x_2) = \lag \frac{\delta \psi(x_1)}{\delta\zeta(x_2)} \rag_{\zeta=0} ,
\label{Rdef}
\eeq
where $\zeta(x)$ is a ``noise'' added to the r.h.s. in eq.(\ref{OKsdef}).
Thus, $R(x_1,x_2)$ measures the linear response of the system to an external 
source of noise. Because of causality it contains an Heaviside factor
$\theta(\eta_1-\eta_2)$ since the field $\psi(x_1)$ can only depend on the
values of the ``noise'' at earlier times $\eta_2\leq\eta_1$. Moreover,
it satisfies the initial condition:
\beq
\eta_1\rightarrow\eta_2^+ : \; R(x_1,x_2) \rightarrow \delta_D(\bk_1-\bk_2) 
\delta_{i_1,i_2} .
\label{Requaltimes}
\eeq
In the linear regime, the response $R_L$ may be obtained by first calculating
the linear solution $\psi_L(x|\zeta)$ of the linearized equations of motion 
$\cO.\psi_L(x|\zeta) = \zeta$ and next taking the functional derivative 
(\ref{Rdef}). Alternatively, one can directly obtain the matrix $R_L$ from 
the initial condition (\ref{Requaltimes}) and the linear dynamics 
$\cO.R_L=0$ (as implied by the definition (\ref{Rdef}) and $\cO.\psi_L =0$). 
For the Einstein-de-Sitter cosmology ($\Om=1,\OL=0$) where the factors 
$\Om/f^2$ are constant and equal to unity, this can be easily solved
by looking for eigenmodes or performing a Laplace transform as in 
Crocce \& Scoccimarro (2006a). This yields:
\beqa
\lefteqn{R_L(x_1,x_2) = \theta(\eta_1-\eta_2) \delta_D(\bk_1-\bk_2)} 
\nonumber \\
&& \times \left\{ \frac{e^{\eta_1-\eta_2}}{5} \left(\bea{cc} 3 & 2 \\ 3 & 2 \ea\right)
+ \frac{e^{-3(\eta_1-\eta_2)/2}}{5} \left(\bea{cc} 2 & -2 \\ -3 & 3 \ea\right)
\right\} .
\label{RL}
\eeqa
For other cosmological parameters one must solve numerically the linearized
equation of motion for $R_L$. Note that in all cases it exhibits the factor
$\theta(\eta_1-\eta_2) \delta_D(\bk_1-\bk_2)$ times a term which is
independent of wavenumbers $\bk_1,\bk_2$.

\section{Path-integral formalism}
\label{Path-integral}

As in Valageas (2004) we can apply a path-integral approach to the 
hydrodynamical system (\ref{OKsdef}) since we are only interested in the
statistical properties of the density and velocity fields (we do not look for
peculiar solutions of the equations of motion). Let us briefly recall how this
can be done (also Martin et al. 1973; Phythian 1977) . 
In order to include explicitly the initial conditions we rewrite
eq.(\ref{OKsdef}) as:
\beq
\cO . \psi = K_s . \psi \psi + \mu_i
\label{OKsmu}
\eeq
with $\psi=0$ for $\eta<\eta_i$ and:
\beq
\mu_i(x) = \delta_D(\eta-\eta_i) e^{\eta_i} \delta_{L0}(\bk) \left(\bea{c} 1 \\ 1 \ea\right) .
\label{mui}
\eeq
Thus, the source $\mu_i$ merely provides the initial conditions at time 
$\eta_i$, obtained from the linear growing mode (\ref{psiL}). We shall 
eventually take the limit $\eta_i\rightarrow -\infty$. Next, we define the
generating functional $Z[j]$ by:
\beq
Z[j] = \lag e^{j.\psi} \rag = \int [\d\mu_i] \;
e^{j.\psi[\mu_i]-\frac{1}{2} \mu_i . \Delta_i^{-1} . \mu_i}
\label{Zjmui}
\eeq
where we took the average over the Gaussian initial conditions:
\beq
\lag\mu_i\rag=0, \;\;\; \lag\mu_i(x_1)\mu_i(x_2)\rag= \Delta_i(x_1,x_2) .
\label{Deltai}
\eeq
All statistical properties of the field $\psi$ may be obtained from $Z[j]$.
It is convenient to write eq.(\ref{Zjmui}) as:
\beqa
Z[j] & = &\int [\d\mu_i] [\d\psi] \; |\det M| \;
\delta_D(\mu_i-\cO.\psi+K_s.\psi\psi) \nonumber \\
&& \times e^{j.\psi-\frac{1}{2} \mu_i . \Delta_i^{-1} . \mu_i} ,
\label{Zjmuipsi}
\eeqa
where the Jacobian $|\det M|$ is defined by the functional derivative 
$M=\delta\mu_i/\delta\psi$. We can compute the latter by factorizing the
operator $\pl/\pl\eta$:
\beq
\! \det M \! = \det(\cO-2 K_s.\psi) = \det(\pl/\pl\eta) \det(1+\tcO-2 \tKs.\psi)
\label{detM}
\eeq
with:
\beq
\! \tcO(x,x')  \! = \! \delta_D(\bk-\bk') \theta(\eta-\eta') \! 
\left( \bea{cc} 0 & -1 \\ -\frac{3\Om}{2f^2}(\eta') 
&  \frac{3\Om}{2f^2}(\eta') -1 \ea \right) 
\label{tOdef}
\eeq
(note that we separated the identity part out of $\tcO$) and:
\beqa
\tKs(x;x_1,x_2) & = & \delta_D(\bk_1+\bk_2-\bk) \theta(\eta-\eta_1) 
\delta_D(\eta_2-\eta_1) \nonumber \\ && \times 
\gamma^s_{i;i_1,i_2}(\bk_1,\bk_2) .
\label{tKsdef}
\eeqa
Then, we have:
\beq
\det M = \det(\pl/\pl\eta) \; e^{\Tr\ln(1+\tcO-2 \tKs.\psi)}
\label{detMexp}
\eeq
and:
\beqa
\Tr\ln(1+\tcO-2 \tKs.\psi) & = & \sum_{q=1}^{\infty} \frac{(-1)^{q-1}}{q} 
\Tr(\tcO-2 \tKs.\psi)^q \nonumber \\ & = & \Tr(\tcO-2 \tKs.\psi) ,
\label{Trln}
\eeqa
where we used the fact that the trace of powers $q\geq 2$ vanishes because of
the Heaviside factors $\theta(\eta-\eta')$ (Zinn-Justin 1989, Valageas 2004). 
Next, the trace $\Tr\tcO$ is an
irrelevant constant whereas $\Tr(2 \tKs.\psi)=0$ (we regularize as 
usual the gravitational interaction at large distances to remove ambiguities,
e.g. perform angular integrals first, see also Valageas 2004). Therefore, 
the Jacobian in the 
functional integral (\ref{Zjmuipsi}) is a constant which can be absorbed into
the normalization. Then, introducing an imaginary ghost field $\lambda$ 
to express the Dirac as an exponential and performing the Gaussian 
integration over $\mu_i$ we obtain:
\beqa
Z[j] & = & \int [\d\mu_i] [\d\psi] [\d\lambda] \; e^{j.\psi 
+ \lambda.(\mu_i-\cO.\psi+K_s.\psi\psi) - \frac{1}{2}\mu_i.\Delta_i^{-1}.\mu_i}
\nonumber \\
&= & \int [\d\psi] [\d\lambda] \; e^{j.\psi 
+ \lambda.(-\cO.\psi+K_s.\psi\psi) + \frac{1}{2}\lambda.\Delta_i.\lambda} .
\label{Zjpsilambda}
\eeqa
Thus, the statistical properties of the system (\ref{OKsdef}) are described
by the action $S[\psi,\lambda]$ defined by:
\beq
S[\psi,\lambda] = \lambda.(\cO.\psi-K_s.\psi\psi) 
- \frac{1}{2}\lambda.\Delta_i.\lambda.
\label{Spsilambda}
\eeq
Moreover, we can note that adding a ``noise'' $\zeta$ to the r.h.s. of 
eq.(\ref{OKsdef}) amounts to change $\mu_i\rightarrow\mu_i+\zeta$ which
translates into $S\rightarrow S-\lambda.\zeta$. Therefore, functional 
derivatives with respect to $\zeta$ are equivalent to insertions of the ghost
field $\lambda$. In particular, we have:
\beq
R(x_1,x_2) = \lag \psi(x_1) \lambda(x_2) \rag, \;\;\; \lag\lambda\rag=0 , 
\;\;\; \lag\lambda\lambda\rag=0 .
\label{Rpsilambda}
\eeq

\section{Large-$N$ expansions}
\label{Large-N-expansions}

The path-integral (\ref{Zjpsilambda}) can be computed by expanding
over powers of its non-Gaussian part (i.e. over powers of $K_s$). This
actually yields the usual perturbative expansion over powers of the
linear power-spectrum $P_L$ (see also Valageas (2001, 2004) for the
case of the Vlasov equation of motion). 
On the other hand, the path-integral (\ref{Zjpsilambda}) may also be studied
through a large-$N$ expansion as in Valageas (2004). Thus, one considers 
the generating functional $Z_N[j,h]$ defined by:
\beq
Z_N[j,h] = \int [\d\psi] [\d\lambda] \; e^{N[j.\psi+h.\lambda-S[\psi,\lambda]]}
\label{ZN}
\eeq
and one looks for an expansion over powers of $1/N$, taking eventually $N=1$
into the results.
As discussed in Valageas (2004) the large-$N$ expansions may also be derived
from a generalization of the dynamics to $N$ fields $\psi^{(\alpha)}$. 
This yields the same results once we take care of the long-wavelength 
divergences which constrain which subsets of diagrams need to be gathered.

\subsection{Steepest-descent method}
\label{Steepest-descent}

A first approach to handle the large-$N$ limit of 
eq.(\ref{ZN}) is to use a steepest-descent method (also called a 
semi-classical or loopwise expansion in the case of usual Quantum field 
theory with $\hbar=1/N$). This yields for auxiliary correlation and response
functions $G_0$ and $R_0$ the equations (Valageas 2004):
\beqa
\cO(x,z).G_0(z,y) &=& 0 \label{G0eq} \\
\cO(x,z).R_0(z,y) &=& \delta_D(x-y) \label{R0forward} \\
R_0(x,z).\cO(z,y) &=& \delta_D(x-y) \label{R0eq} 
\eeqa
whereas the actual correlation and response functions obey:
\beqa
\cO(x,z).G(z,y) &=& \Sigma(x,z).G(z,y) + \Pi(x,z).R^T(z,y) \label{Geq} \\
\cO(x,z).R(z,y) &=& \delta_D(x-y) + \Sigma(x,z).R(z,y) \label{Rforward} \\
R(x,z).\cO(z,y) &=& \delta_D(x-y) + R(x,z).\Sigma(z,y) \label{Req}  
\eeqa
where the self-energy terms $\Sigma$ and $\Pi$ are given at one-loop order by:
\beqa
\!\!\!\Sigma(x,y) \!\!\! & = & \!\!\! 4 K_s(x;x_1,x_2) K_s(z;y,z_2) R_0(x_1,z) G_0(x_2,z_2) \label{Seq} \\
\!\!\!\Pi(x,y) \!\!\! & = & \!\!\! 2 K_s(x;x_1,x_2) K_s(y;y_1,y_2) G_0(x_1,y_1) G_0(x_2,y_2) \label{Peq} 
\eeqa
Note that eqs.(\ref{G0eq})-(\ref{Req}) are exact and that the expansion over 
powers of $1/N$ only enters the expression of the self-energy 
(\ref{Seq})-(\ref{Peq}). Here we only kept the lowest-order terms 
(see Valageas (2004) for the next-order terms). We also took the limit 
$\eta_i\rightarrow -\infty$ so that terms involving $\Delta_i$ vanish.
From sect.~\ref{Linear} and eqs.(\ref{G0eq})-(\ref{R0eq}) we find that the 
auxiliary matrices $G_0$ and $R_0$ are actually equal to their linear 
counterparts:
\beq
G_0= G_L, \;\;\; R_0 = R_L .
\label{G0GLR0RL}
\eeq

\subsection{2PI effective action method}
\label{2PI-effective-action}

As described in Valageas (2004) a second approach is to first introduce the
double Legendre transform $\Gamma[\psi,G]$ of the functional $W=\ln Z$ (with
respect to both the field $\psi$ and its two-point correlation $G$) and next 
to apply the $1/N$ expansion to $\Gamma$. 
This ``2PI effective action''method yields the
same equations (\ref{Geq})-(\ref{Req}) and the self-energy shows the same
structure as (\ref{Seq})-(\ref{Peq}) where $G_0$ and $R_0$ are replaced by 
$G$ and $R$. Thus, the direct steepest-descent method yields a series of
linear equations which can be solved directly whereas the 2PI effective 
action method gives a system of non-linear equations (through the dependence
on $G$ and $R$ of $\Sigma$ and $\Pi$) which must usually be solved numerically 
by an iterative scheme. However, thanks to the Heaviside factors appearing
in the response $R$ and the self-energy $\Sigma$ these equations can be solved
directly by integrating forward over time $\eta_1$.

\subsection{General properties}
\label{General-properties}

As discussed in Valageas (2004) 
both the steepest-descent and 2PI effective action methods agree with the
standard perturbative analysis over powers of $P_{L0}(k)$ up to the order
used for the self-energy (e.g. up to order $P_{L0}^2$ if we only consider
the one-loop terms (\ref{Seq})-(\ref{Peq})). As compared with the standard 
perturbative approach, the two schemes described above also include
two different infinite partial resummations, as can be seen from 
eqs.(\ref{Geq})-(\ref{Req}) which clearly generate terms at all orders over
$P_{L0}$ for $G$ and $R$ even if $\Sigma$ and $\Pi$ are only linear or
quadratic over $P_{L0}$. 

We can note that the equations we obtain for the hydrodynamical system 
(\ref{continuity1})-(\ref{Poisson1}) are simpler than for the collisionless
system studied in Valageas (2004) which is described by the Vlasov-Poisson 
system. Indeed, here the mean $\lag\psi\rag$ vanishes at all orders. 
This can be explicitly
checked from eqs.(\ref{continuity1})-(\ref{Poisson1}) and in the derivation
of eqs.(\ref{G0eq})-(\ref{Peq}).
This is not the case for the Vlasov dynamics where eq.(\ref{G0GLR0RL})
does not hold.

Using a diagrammatic technique Crocce \& Scoccimarro (2006a,b) derived 
equations (\ref{Geq})-(\ref{Peq}) in an integral form (i.e. without the 
differential operator $\cO$ in the l.h.s.) by first integrating the equation
of motion (\ref{OKsdef}). Of course, our path-integral approach can also
be applied to the integral form of eq.(\ref{OKsdef}). This amounts to
write eq.(\ref{OKsdef}) as $\psi=\psi_L+{\tilde K}_s \psi\psi$ where
$\psi_L$ is the linear growing mode and ${\tilde K}_s$ the integral form
of the vertex $K_s$. Then, the procedure presented in sect.~\ref{Path-integral}
can be applied to this integral form of the equation of motion, as described
for instance in Valageas (2001). In this article we preferred to keep
eq.(\ref{OKsdef}) in its differential form so that the r.h.s. does not contain
an integral over time. Then the self-energy terms $\Sigma(\eta_1,\eta_2)$
and $\Pi(\eta_1,\eta_2)$ only depend on the response and correlation at the
same times $(\eta_1,\eta_2)$, see eqs.(\ref{Seq})-(\ref{Peq}). By contrast,
the integrated vertex ${\tilde K}_s$ would lead to self-energies which depend
on the values of the response and correlation at all past times which would
entail less efficient numerical computations.

\subsection{Integral expression for the two-point correlation}
\label{Integral-expression}

Thanks to statistical homogeneity and isotropy the matrices $G_0,G$ and $\Pi$
are symmetric and of the form:
\beq
G(x_1,x_2) = \delta_D(\bk_1+\bk_2) G_{i_1,i_2}(k_1;\eta_1,\eta_2) ,
\label{Gk}
\eeq
with:
\beq
G_{i_1,i_2}(k;\eta_1,\eta_2) = G_{i_2,i_1}(k;\eta_2,\eta_1) ,
\label{Gsym}
\eeq
whereas the matrices $R_0,R$ and $\Sigma$ are of the form:
\beq
R(x_1,x_2) = \delta_D(\bk_1-\bk_2) \theta(\eta_1-\eta_2) 
R_{i_1,i_2}(k_1;\eta_1,\eta_2) .
\label{Rk}
\eeq
Besides, from eqs.(\ref{Geq})-(\ref{Req}) we see that the correlation $G$
can also be written in terms of the response $R$ as:
\beq
G(x_1,x_2) = R \times G_0(\eta_i) \times R^T + R.\Pi.R^T 
\label{GPi}
\eeq
where the second product is defined as in (\ref{defproduct}) whereas the
first product does not contain any integration over time:
\beqa
\lefteqn{ R \times G_0(\eta_i) \times R^T = \delta_D(\bk_1+\bk_2) 
\sum_{i_1',i_2'} R_{i_1,i_1'}(k_1;\eta_1,\eta_i) } \nonumber \\ 
&& \times G_{0;i_1',i_2'}(k_1;\eta_i,\eta_i) R_{i_2,i_2'}(k_1;\eta_2,\eta_i) ,
\label{GRPiR}
\eeqa
and we let $\eta_i\rightarrow -\infty$. 
The physical meaning of eq.(\ref{GPi}) is clear. The first term in the r.h.s. 
means that the fluctuations at the initial time $\eta_i$ are merely 
transported forward in time through the matrix $R$. This is the only non-zero 
term in the linear regime (with $R=R_L$ and $G_0=G_L$). The effect of the 
non-linear dynamics is to modify the transport matrix $R$ and to add a second 
term to the r.h.s. of eq.(\ref{GPi}). 
The latter has the meaning of a source term which produces fluctuations with
two-point correlation $\Pi(\eta_1',\eta_2')$ at the times $(\eta_1',\eta_2')$ 
which are next transported 
forward to later times $(\eta_1,\eta_2)$ by the matrices $R(\eta_1,\eta_1')$ 
and $R^T(\eta_2',\eta_2)$. Thus, the ``dot product'' means
that we need to integrate over all fluctuations produced at all earlier times
$\eta_1'<\eta_1$ and $\eta_2'<\eta_2$. 
This interpretation considers $\Pi$ as an external source for an ``effective''
linear dynamics. For the system which we study here the ``source'' 
$\Pi$ is not external but corresponds to the non-linear interaction of the 
field $\psi$, as seen from the explicit expression (\ref{Peq}). Moreover, 
the linear operator which would define the transport with time also depends on 
$G$ and $R$ through $\Sigma$, see eqs.(\ref{Seq}) and (\ref{Req}).

Alternatively, Crocce \& Scoccimarro (2006a) pointed out that the expression
(\ref{GPi}) is somewhat similar to the result obtained within a 
phenomenological halo model (Seljak 2000). Indeed, the latter model splits
the non-linear power-spectrum into two terms, one that dominates in the linear
regime ($2-$halo term) and the other that dominates in the highly non-linear
regime ($1-$halo term). They could be identified with the first and second
terms of eq.(\ref{GPi}). However, it is not clear whether the analogy can be
pursued much further. In particular, eq.(\ref{GPi}) is explicitly dynamical
(i.e. it explicitly involves integrations over past events) whereas the
halo model gives a static expression (it writes the two-point correlation
as integrals over the current halo distribution, which is the basic 
time-dependent quantity).

Note that within the 2PI effective action method, even if $R$ would be 
treated independently of $G$ the expression (\ref{GPi}) does not give a 
quadratic dependence of $G$ on $R$ since $\Pi$ depends quadratically on $G$. 
Besides, solving eq.(\ref{GPi}) perturbatively over $G_0$ at fixed $R$ 
actually generates terms at all orders $q$ over powers $G_0^q$.
By contrast, in the steepest-descent approach where $\Pi$ only depends on $G_0$
the expression (\ref{GPi}) is fully explicit and quadratic over both 
$G_0(=G_L)$ and $R$ if the latter are treated independently. This is a 
signature of the additional resummations involved in the 2PI effective action 
method as compared with the direct steepest-descent approach.

An advantage of eq.(\ref{GPi}) is that once
$\Pi$ and $R$ are known, we obtain an explicit expression for $G$. Hence
we do not need to solve eq.(\ref{Geq}) (by solving 
for $R$ we have already performed the ``inversion'' of $\cO-\Sigma$).
A second advantage of eq.(\ref{GPi}) is that it provides an expression for
$G$ which is clearly symmetric. Moreover, since we start from a two-point
correlation $G_0(=G_L)$ which is positive, as shown by eq.(\ref{GL}),
we see from eq.(\ref{Peq}) and eq.(\ref{GPi}) that both $\Pi$ and $G$ are
positive. This also holds for the 2PI effective action approach which may be 
obtained by iterating the system (\ref{Geq})-(\ref{Peq}) (substituting $G$ and
$R$ into the self-energy).

\subsection{Integration over wavenumber for the self-energy}
\label{Integration-over-wavenumber}

Using the symmetry (\ref{Gsym}) for $\Pi$ and the form (\ref{Rk}) for $\Sigma$
we see that $\Pi(x_1,x_2)$ and $\Sigma(x_1,x_2)$ only need to be computed at 
times $\eta_1\geq\eta_2$. Besides, thanks to the various Dirac factors
the expressions (\ref{Seq})-(\ref{Peq}) only involve an integral over one
wavenumber $\bk'$. Thus, using the expression (\ref{Ksdef}) of the kernel 
$K_s$ we can write from eqs.(\ref{Seq})-(\ref{Peq}) the self-energy as:
\beqa
\lefteqn{ \Sigma_{i_1 i_2}(k;\eta_1,\eta_2) = 4 
\gamma^s_{i_1;j_1,j_2}(\bk',\bk-\bk') \gamma^s_{l_1;i_2,l_2}(\bk,\bk'-\bk)} 
\nonumber \\
&& \times R_{j_1 l_1}(k';\eta_1,\eta_2) G_{j_2 l_2}(|\bk-\bk'|;\eta_1,\eta_2) 
\label{Sgg}
\eeqa
and:
\beqa
\lefteqn{\Pi_{i_1 i_2}(k;\eta_1,\eta_2) = 2 
\gamma^s_{i_1;j_1,j_2}(\bk',\bk-\bk')\gamma^s_{i_2;l_1,l_2}(-\bk',-\bk+\bk')} 
\nonumber \\
&& \times G_{j_1 l_1}(k';\eta_1,\eta_2) G_{j_2 l_2}(|\bk-\bk'|;\eta_1,\eta_2) ,
\label{Pgg}
\eeqa
where we integrate over $\bk'$ and we sum over $j_1,j_2,l_1,l_2$.
In order to keep the symmetry $\bk'\leftrightarrow\bk-\bk'$ in the numerical
integration we use symmetric elliptic coordinates:
\beq
\bk' = \frac{\bk}{2} + \bq \;\;\; \mbox{with} \;\;\; \bq= \frac{k}{2} \left|
\bea{l} \sinh\zeta\sin\mu\cos\phi \\ \sinh\zeta\sin\mu\sin\phi \\ 
\cosh\zeta\cos\mu \ea \right.
\label{kp}
\eeq
where $\bk$ is taken along the third axis and 
$\zeta\geq0, 0\leq\mu\leq\pi, 0\leq\phi<2\pi$.

At equal times $\eta_1=\eta_2$ we can obtain from eq.(\ref{Sgg}):
\beq
\Sigma(k;\eta,\eta) = - \frac{4\pi}{3} k^2 \int_0^{\infty} \d k' 
G_{22}(k';\eta,\eta) \; \times \; \left(\bea{cc} 1 & 0 \\ 0 & 1 \ea\right) .
\label{Sequaltimes}
\eeq
The explicit expression (\ref{Sequaltimes}) shows that within the direct
steepest-descent method we need a linear power-spectrum $P_{L0}(k)$ with a 
logarithmic slope $n=\d\ln P_{L0}/\d\ln k$ such that $n>-1$ at large scales 
and $n<-1$ at small scales to obtain finite results. These constraints are 
the same as those which appear in the standard perturbative approach. 
The divergence at long wavelengths
eventually cancels out if we only consider the equal-time correlations of
the density field, as seen in Jain \& Bertschinger (1996) (also 
Vishniac 1983). As pointed out by these authors, this can be understood from 
the fact that contributions to the velocity field from long-wave modes 
correspond to an almost uniform translation of the fluid and therefore should 
not affect the growth of density structures on small scales. However, these 
infrared contributions can no longer be ignored when one studies 
different-times correlations or the velocity field itself. The same problem
appears when one considers the collisionless dynamics as described by the
Vlasov equation (Valageas 2004). The small-scale divergences are more
harmful as they do not cancel in such a fashion. In fact, they may be
understood as the signature of anomalous terms (which do not scale as integer 
powers of $P_{L0}$) which cannot be recovered by
the standard perturbative scheme as discussed in Valageas (2002). It is 
probably necessary to go beyond the hydrodynamical approximation 
(\ref{continuity1})-(\ref{Poisson1}) to cure this small-scale problem (e.g.
use the Vlasov equation which does not break down at shell-crossing).
On the other hand, note that within the 2PI effective action scheme these
constraints apply to the non-linear power $G_{22}(k)$ which may have different
a different asymptotic behavior than the linear power-spectrum at small 
scales. Therefore, the range of initial conditions where the integral in 
eq.(\ref{Sequaltimes}) converges at high $k$ may be different within this 
scheme.

\section{Direct steepest-descent method}
\label{Direct-steepest-descent}

We present in the following sections our numerical results which we compare 
to fits obtained from direct numerical simulations and to the standard 
perturbative analysis. 
We consider an Einstein-de-Sitter cosmology with
$\Om=1, \OL=0$, a shape parameter $\Gamma=0.5$ and a normalization
$\sigma_8=0.51$ for the linear power-spectrum $P_{L0}(k)$ and a Hubble constant
$H_0=50$ km.s$^{-1}$.Mpc$^{-1}$ (i.e. the reduced Hubble parameter is $h=0.5$)
as in Smith et al.(2003) so as to compare our results with N-body simulations.
Since we mainly wish to investigate in this 
article the properties of the large-$N$ expansions described in 
sect.~\ref{Large-N-expansions} we first focus on the Einstein-de-Sitter 
cosmology where both $G_L$ and $R_L$ have simple analytical forms 
(\ref{GL})-(\ref{RL}). We shall discuss the case of a $\Lambda$CDM cosmology
in sect.~\ref{LCDM} below.

We first study in this section the results obtained from the direct 
steepest-descent method of sect.~\ref{Steepest-descent} which involves the 
auxiliary two-point functions $G_0$ and $R_0$. 
Since the latter are equal to their linear counterparts
(\ref{G0GLR0RL}) the correlation $G_0$ is given by eq.(\ref{GL}) whereas
$R_0$ is obtained by solving eq.(\ref{R0forward}) forward in time, in agreement
with the discussion below eq.(\ref{Requaltimes}). For the Einstein-de-Sitter 
cosmology which we consider here $R_0$ is actually given by the explicit 
expression (\ref{RL}).

\subsection{Self-energy}
\label{Self-energy0}

\begin{figure}[htb]
\begin{center}
\epsfxsize=8 cm \epsfysize=7 cm {\epsfbox{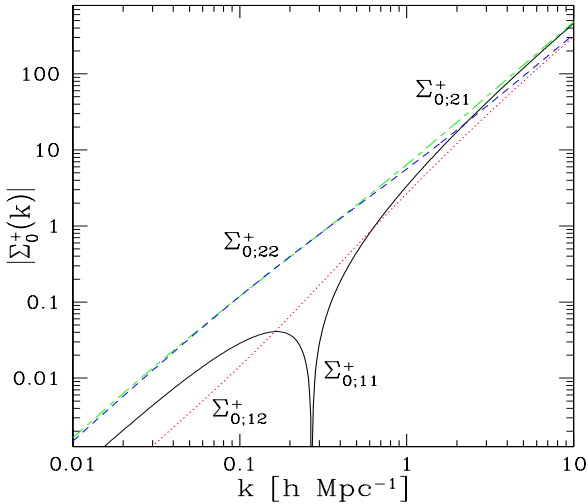}}
\end{center}
\caption{The self-energy terms $\Sigma_{0;ij}^+(k)$ of eq.(\ref{S0pS0m}) 
as a function of wavenumber $k$. We display the four components 
$\Sigma_{0;11}^+$ (solid line), $\Sigma_{0;21}^+$ (dot-dashed line), 
$\Sigma_{0;12}^+$ (dotted line) and $\Sigma_{0;22}^+$ (dashed line).
They are all negative except for $\Sigma_{0;11}^+$ which is positive at 
$k < 0.2 h$ Mpc$^{-1}$ and negative at higher $k$. All terms are of the same 
magnitude and grow roughly as $k^2$ in agreement with eq.(\ref{S01122}).}
\label{figS0pk}
\end{figure}

From $G_0$ and $R_0$ we need to compute the self-energy from 
eqs.(\ref{Seq})-(\ref{Peq}).
In fact, for the Einstein-de-Sitter cosmology the time-dependence of the 
self-energy can be factorized as follows using eqs.(\ref{GL}), (\ref{RL}).
First, the auxiliary response $R_0$ can be written from eq.(\ref{RL}) as:
\beqa
R_0(x_1,x_2) & = & \theta(\eta_1-\eta_2) \delta_D(\bk_1-\bk_2) \nonumber \\
&& \times \left[ e^{\eta_1-\eta_2} R_0^+ + e^{-3(\eta_1-\eta_2)/2} 
R_0^- \right],
\label{R0R0pR0m}
\eeqa
with:
\beq
R_0^+= \left(\bea{cc} 3/5 & 2/5 \\ 3/5 & 2/5 \ea\right) \;\; \mbox{and} \;\; 
R_0^-= \left(\bea{cc} 2/5 & -2/5 \\ -3/5 & 3/5 \ea\right) ,
\label{R0pR0m}
\eeq
where we separated the contributions from the growing mode $R_0^+$ and the 
decaying mode $R_0^-$. Then, from eq.(\ref{Sgg}) the self-energy term 
$\Sigma_0$ which is linear over $R_0$ also splits into two terms as:
\beqa
\Sigma_0(x_1,x_2) & = & \theta(\eta_1-\eta_2) \delta_D(\bk_1-\bk_2) 
\nonumber \\
&& \times \left[ e^{2\eta_1} \Sigma_0^+(k_1) + e^{-\eta_1/2+5\eta_2/2} 
\Sigma_0^-(k_1) \right],
\label{S0S0pS0m}
\eeqa
with:
\beqa
\Sigma_{0;i_1 i_2}^{\pm}(k) & = & 4 \sum_{j_1 j_2 l_1 l_2}\int\d\bq \; 
\gamma^s_{i_1;j_1,j_2}(\bk-\bq,\bq) \nonumber \\
&& \times \gamma^s_{l_1;i_2,l_2}(\bk,-\bq) \; P_{L0}(q) R_{0;j_1 l_1}^{\pm} 
\label{S0pS0m}
\eeqa
where we used eq.(\ref{GL}). Besides, from the expressions 
(\ref{gamma1})-(\ref{gamma2}) of the vertices $\gamma^s$ we obtain:
\beq
\Sigma_0^-(k) = \left(\bea{cc} \Sigma_{0;22}^+ & -\Sigma_{0;12}^+  
\\ -\Sigma_{0;21}^+ & \Sigma_{0;11}^+ \ea\right)
\label{S0mS0p}
\eeq
which is consistent with eqs.(\ref{Sequaltimes}), (\ref{S0S0pS0m}).
In particular, we have:
\beq
\Sigma_{0;11}^+ + \Sigma_{0;22}^+ = - \frac{4\pi}{3} k^2 
\int_0^{\infty} \d q P_{L0}(q) .
\label{S01122}
\eeq
Moreover, at high $k$ we have the asymptotic behaviors:
\beq
k\rightarrow\infty: \;\;\; \Sigma_{0;11}^+ \simeq \Sigma_{0;21}^+ 
\simeq -\frac{3}{5} \frac{4\pi}{3} k^2 \int_0^{\infty} \d q P_{L0}(q) 
\label{S0k1121}
\eeq
\beq
k\rightarrow\infty: \;\;\; \Sigma_{0;22}^+ \simeq \Sigma_{0;12}^+ 
\simeq -\frac{2}{5} \frac{4\pi}{3} k^2 \int_0^{\infty} \d q P_{L0}(q) 
\label{S0k2212}
\eeq
Of course, we can check that this is consistent with eq.(\ref{S01122}).

We display in Fig.~\ref{figS0pk} the self-energy term $\Sigma_0^+(k)$
as a function of wavenumber $k$, from which $\Sigma_0$ can be obtained using
eqs.(\ref{S0S0pS0m}), (\ref{S0mS0p}). All components $\Sigma_{0;ij}^+$ are of
the same magnitude and negative except for $\Sigma_{0;11}^+$ which is positive
at large scales $k\la 0.2 h$ Mpc$^{-1}$. All terms grow as $k^2$ at high
wavenumbers and we recover the asymptotic behaviors 
(\ref{S0k1121})-(\ref{S0k2212}). Note that the self-energy 
decreases at large scales so that the two-point functions obtained from 
eqs.(\ref{Geq})-(\ref{Req}) will converge to the linear asymptotics

\begin{figure}[htb]
\begin{center}
\epsfxsize=8 cm \epsfysize=7 cm {\epsfbox{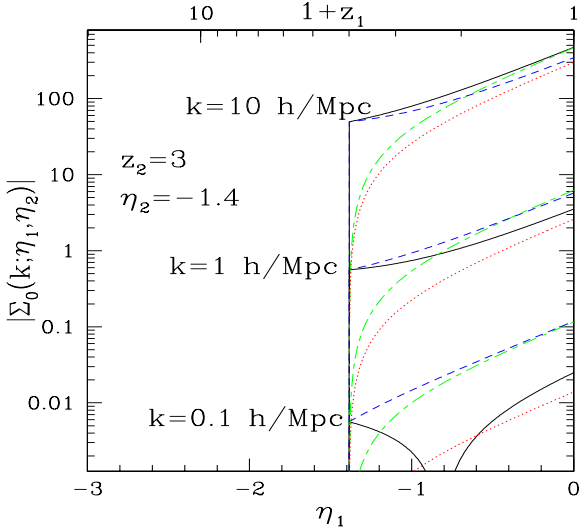}}
\end{center}
\caption{The self-energy $\Sigma_0(k;\eta_1,\eta_2)$ as a function of 
forward time $\eta_1$, for $\eta_2=-1.4$ (i.e. $z_2=3$) and wavenumbers 
$k=0.1,1$ and $10 \times h$ Mpc$^{-1}$. 
The line styles are as in Fig.~\ref{figS0pk}. 
The diagonal components vanish for $\eta_1<\eta_2$ whereas the off-diagonal 
components vanish for $\eta_1\leq\eta_2$. All terms are negative except for 
$\Sigma_{0;11}$ which becomes positive at $\eta_1>-0.8$ for $k=0.1 h$ 
Mpc$^{-1}$.}
\label{figS0t1}
\end{figure}

We show in Fig.~\ref{figS0t1} the evolution forward over time $\eta_1$ of the
self-energy $\Sigma_0(k;\eta_1,\eta_2)$ for time $\eta_2=-1.4$ 
(i.e. $z_2=3$) and wavenumbers $k=0.1,1$ and $10 \times h$ Mpc$^{-1}$ 
(from bottom to top).
The absolute value of the self-energy is larger for higher $k$ in agreement 
with Fig.~\ref{figS0pk}.
All components are negative (except for $\Sigma_{0;11}$ at $\eta_1>-0.8$ 
in the case $k=0.1 h$ Mpc$^{-1}$) and exhibit a smooth dependence with time 
following eq.(\ref{S0S0pS0m}) which goes to $e^{2\eta_1}$ at large $\eta_1$.

\begin{figure}[htb]
\begin{center}
\epsfxsize=8 cm \epsfysize=7 cm {\epsfbox{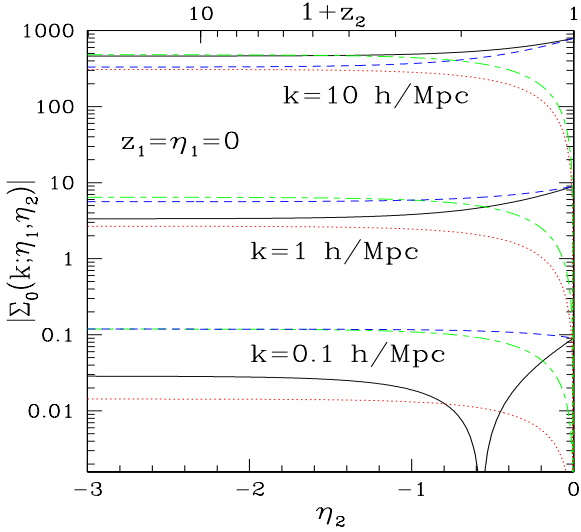}}
\end{center}
\caption{The self-energy $\Sigma_0(k;\eta_1,\eta_2)$ as a function of 
backward time $\eta_2$, for $\eta_1=0$ (i.e. $z_1=0$) and wavenumbers 
$k=0.1,1$ and $10\times h$ Mpc$^{-1}$. 
The line styles are as in Fig.~\ref{figS0pk}. 
All terms are negative except for $\Sigma_{0;11}$ which becomes positive at 
$\eta_2<-0.6$ for $k=0.1 h$ Mpc$^{-1}$.}
\label{figS0t2}
\end{figure}

We show in Fig.~\ref{figS0t2} the evolution backward over time $\eta_2$ of the
self-energy $\Sigma_0(k;\eta_1,\eta_2)$ for time $\eta_1=0$ 
(i.e. $z_1=0$) and wavenumbers $k=0.1,1$ and $10\times h$ Mpc$^{-1}$ 
(from bottom to top). The behavior agrees with 
Figs.~\ref{figS0pk},\ref{figS0t1}. 
Following eq.(\ref{S0S0pS0m}) $\Sigma_0$ converges to a constant for
$\eta_2\rightarrow-\infty$.

\begin{figure}[htb]
\begin{center}
\epsfxsize=8 cm \epsfysize=7 cm {\epsfbox{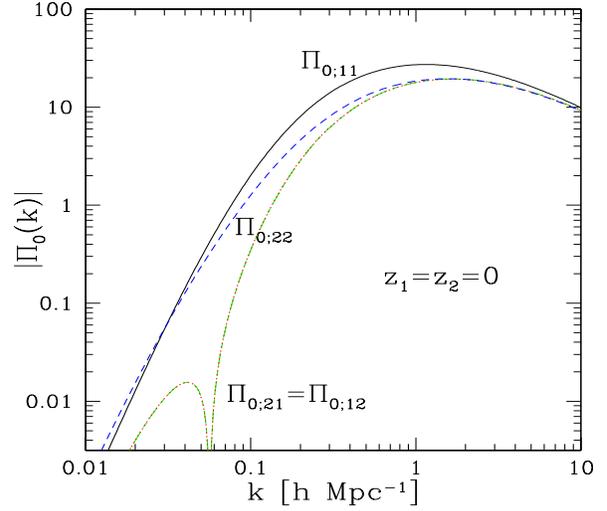}}
\end{center}
\caption{The self-energy $\Pi_0(k)$ as a function of wavenumber $k$.
At high $k$ all components are very close and positive. 
At $k\leq 0.06 h$ Mpc$^{-1}$ the off-diagonal components become negative.}
\label{figP0k}
\end{figure}

Finally, using the factorized form (\ref{GL}) of the auxiliary two-point
correlation the self-energy $\Pi_0$ can be written as:
\beq
\Pi_0(x_1,x_2) = \delta_D(\bk_1+\bk_2) e^{2\eta_1+2\eta_2} \Pi_0(k_1) ,
\label{Pi0}
\eeq
with:
\beqa
\lefteqn{\Pi_{0;i_1 i_2}(k) = 2 \sum_{j_1 j_2 l_1 l_2} \int\d\bq \; 
\gamma^s_{i_1;j_1,j_2}(\bq,\bk-\bq)} \nonumber \\
&& \times \gamma^s_{i_2;l_1,l_2}(-\bq,-\bk+\bq) P_{L0}(q) P_{L0}(|\bk-\bq|) .
\label{Pi0k}
\eeqa
In agreement with the symmetry
(\ref{Gsym}) we can check that $\Pi_{0;21}(k)=\Pi_{0;12}(k)$.

Since the time-dependence (\ref{Pi0}) is quite simple we only display 
the self-energy term $\Pi_0(k)$ of eq.(\ref{Pi0k}) as a function of 
wavenumber $k$ in Fig.~\ref{figP0k}.
At high $k$ all components are very close and positive. 
The dependence on wavenumber $k$ follows the smooth growth at small scales
of the logarithmic power $\Delta^2_{L0}(k)=k^3 P_{L0}(k)$.
As for $\Sigma_0$ the self-energy term $\Pi_0$ decreases very fast at low 
$k$ so that the two-point functions obtained from eqs.(\ref{Geq})-(\ref{Req}) 
will converge to the linear asymptotics at large scales.

\subsection{Response function}
\label{Response-function0}

From the self-energy obtained in sect.~\ref{Self-energy0} we compute the
response function from eq.(\ref{Rforward}). Thanks to the Heaviside
factor within the term $\Sigma_0(x,z)$ the r.h.s. only involves earlier times
hence eq.(\ref{Rforward}) can be integrated forward over time $\eta_1$ at
fixed $\eta_2$ to give $R(k;\eta_1,\eta_2)$. Note that each wavenumber $k$ 
evolves independently thanks to the Dirac factors $\delta_D(\bk_1-\bk_2)$ 
within both $\Sigma_0$ and $R$ as in eq.(\ref{Rk}). Thus all processes 
associated with mode-coupling between different wavenumbers $\bk$ are 
contained in the calculation of the self-energy $\Sigma_0$ (and $\Pi_0$), 
as in eqs.(\ref{S0pS0m}), (\ref{Pi0k}).

Eq.(\ref{Rforward}) also writes for $\eta_1>\eta_2$:
\beqa
\frac{\pl R_1}{\pl\eta_1} - R_2 = \Sigma_{0;11} . R_1 + \Sigma_{0;12} . R_2 
\label{R01} \\
\frac{\pl R_2}{\pl\eta_1} - \frac{3}{2} R_1 + \frac{1}{2} R_2 = 
\Sigma_{0;21} . R_1 + \Sigma_{0;22} . R_2 
\label{R02}
\eeqa
where $(R_1,R_2)=(R_{11},R_{21})$ or $(R_{12},R_{22})$, indeed in this case
the two columns of the matrix $R$ are not coupled. Using 
eqs.(\ref{S0S0pS0m}) and (\ref{S0mS0p}) the first eq.(\ref{R01}) also reads:
\beqa
\lefteqn{ \frac{\pl R_1}{\pl\eta_1}(\eta_1,\eta_2) - R_2(\eta_1,\eta_2) = } 
\nonumber \\ 
&& \int_{\eta_2}^{\eta_1} \d\eta \; \biggl\lbrace \Bigl( 
e^{2\eta_1}\Sigma_{0;11}^+ + e^{-\eta_1/2+5\eta/2} \Sigma_{0;22}^+ \Bigr) 
R_1(\eta,\eta_2) \nonumber \\ 
&& \hspace{0.5cm} + \Bigl( e^{2\eta_1} \Sigma_{0;12}^+ 
- e^{-\eta_1/2+5\eta/2} \Sigma_{0;12}^+ \Bigr) R_2(\eta,\eta_2) \biggr\rbrace
\label{R01int}
\eeqa
Taking advantage of the simple dependence on time $\eta_1$ of the r.h.s. it is 
possible to eliminate the integrals over $\eta$ by taking two derivatives of 
eq.(\ref{R01int}) with respect to $\eta_1$. This yields:
\beqa
\lefteqn{\frac{\pl^3 R_1}{\pl\eta_1^3} - \frac{3}{2} \frac{\pl^2 R_1}{\pl\eta_1^2} -\frac{\pl^2 R_2}{\pl\eta_1^2} - \frac{\pl R_1}{\pl\eta_1} + \frac{3}{2} \frac{\pl R_2}{\pl\eta_1} + R_2 = } \nonumber \\
&& e^{2\eta_1} \left[ (\Sigma_{0;11}^+ + \Sigma_{0;22}^+) \frac{\pl R_1}{\pl\eta_1} + \frac{5}{2} \Sigma_{0;11}^+ R_1 + \frac{5}{2} \Sigma_{0;12}^+ R_2 \right] .
\label{R01diff}
\eeqa
In a similar fashion, one obtains from eq.(\ref{R02}):
\beqa
\lefteqn{\!\!\!\!\frac{\pl^3 R_2}{\pl\eta_1^3} - \frac{3}{2} \frac{\pl^2 R_1}{\pl\eta_1^2} -\frac{\pl^2 R_2}{\pl\eta_1^2} + \frac{9}{4} \frac{\pl R_1}{\pl\eta_1} - \frac{7}{4} \frac{\pl R_2}{\pl\eta_1} + \frac{3}{2} R_1 - \frac{1}{2} R_2 = } \nonumber \\
&& e^{2\eta_1} \left[ (\Sigma_{0;11}^+ + \Sigma_{0;22}^+) \frac{\pl R_2}{\pl\eta_1} + \frac{5}{2} \Sigma_{0;21}^+ R_1 + \frac{5}{2} \Sigma_{0;22}^+ R_2 \right] .
\label{R02diff}
\eeqa
These equations apply to $\eta_1>\eta_2$ and need to be supplemented by 
initial conditions obtained from eqs.(\ref{R01})-(\ref{R02}) and 
eq.(\ref{Requaltimes}) which yield at point $\eta_1=\eta_2^+$:
\beqa
X^T & = & \left( \bea{cccccc} R_1 & R_2 & \frac{\pl R_1}{\pl\eta_1} & \frac{\pl R_2}{\pl\eta_1} & \frac{\pl^2 R_1}{\pl\eta_1^2} & \frac{\pl^2 R_2}{\pl\eta_1^2} \ea \right)
\label{Xdef} \\
& = & \left( \bea{cccccc} 1 & 0 & 0 & \frac{3}{2} & \frac{3}{2}+e^{2\eta_1}(\Sigma_{0;11}^+ + \Sigma_{0;22}^+) & - \frac{3}{4} \ea \right)
\label{X1}
\eeqa
for the pair $(R_{11},R_{21})$ and:
\beq
X^T = \left( \bea{cccccc} 0 & 1 & 1 & -\frac{1}{2} & -\frac{1}{2} & \frac{7}{4}+e^{2\eta_1}(\Sigma_{0;11}^+ + \Sigma_{0;22}^+) \ea \right)
\label{X2}
\eeq
for the pair $(R_{12},R_{22})$. One can easily check that the linear response 
$R_L$ of eq.(\ref{RL}) is a solution of eqs.(\ref{R01diff})-(\ref{R02diff}) 
with $\Sigma_0=0$. The advantage of these differential equations over the 
integro-differential eqs.(\ref{R01})-(\ref{R02}) is that they no longer 
require the computation of integrals over past values of the response $R$. 
Therefore, one does not need to save the values of $R$ on a grid
with a very fine mesh and one can advance over large time steps $\eta_1$.
Moreover, each time-step runs faster since one does not need to compute 
lengthy integrals as in eqs.(\ref{R01})-(\ref{R02}). We use an adaptative 
Runge-Kutta algorithm to solve numerically between two grid 
points the eqs.(\ref{R01diff})-(\ref{R02diff}), written as a system of coupled
first-order linear equations over the components of the vector $X$ of 
eq.(\ref{Xdef}).

\begin{figure}[htb]
\begin{center}
\epsfxsize=9 cm \epsfysize=8 cm {\epsfbox{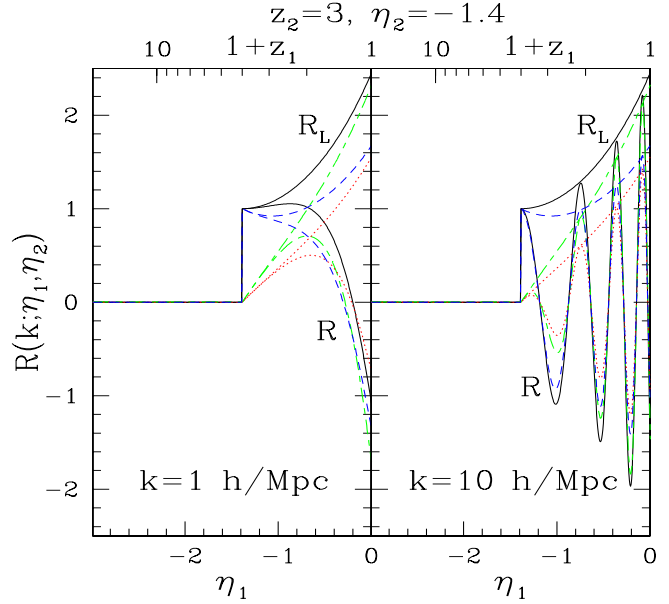}}
\end{center}
\caption{The non-linear response $R(k;\eta_1,\eta_2)$ as a function of forward
time $\eta_1$, for $\eta_2=-1.4$ (i.e. $z_2=3$) and wavenumbers $k=1$ 
(left panel) and $10\times h$ Mpc$^{-1}$ (right panel). 
We also plot the linear response $R_L$ which shows a simple exponential 
growth.}
\label{figR0t1}
\end{figure}

\begin{figure}[htb]
\begin{center}
\epsfxsize=9 cm \epsfysize=8 cm {\epsfbox{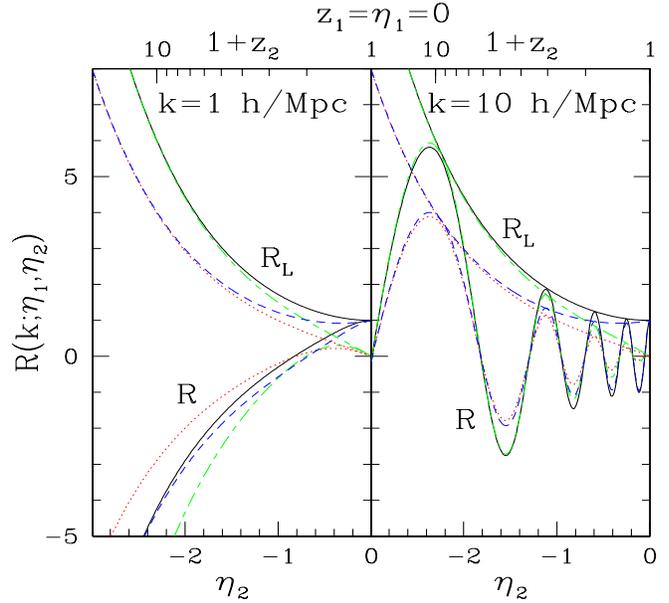}}
\end{center}
\caption{The non-linear response $R(k;\eta_1,\eta_2)$ as a function of backward
time $\eta_2$, for $\eta_1=0$ (i.e. $z_1=0$) and wavenumbers $k=1$ 
(left panel) and $10\times h$ Mpc$^{-1}$ (right panel). 
We also plot the linear response $R_L$ which shows a simple exponential 
growth.}
\label{figR0t2}
\end{figure}

We can note that eqs.(\ref{R01diff})-(\ref{R02diff}) are somewhat similar to
Bessel functions in terms of the scale factor $a_1=e^{\eta_1}$ but are of 
degree three rather than two. In the small-scale limit 
$|\Sigma_0^+|\rightarrow\infty$ we can look for an asymptotic solution of 
the form:
\beq
R_{1,2}(\eta_1) = \rho_{1,2}(\eta_1) e^{i \omega e^{\eta_1}} 
\label{Rasymp}
\eeq
where the functions $\rho_i(\eta_1)$ may be expanded as:
\beq
\rho_i(\eta_1)= \rho_i^{(0)} + \frac{1}{\omega} \rho_i^{(1)}
+ \frac{1}{\omega^2} \rho_i^{(2)} + ...
\label{rhoj}
\eeq
and $\omega^2 \sim |\Sigma_0^+|$. We obtain at order $\omega^3$ after 
substituting into eqs.(\ref{R01diff})-(\ref{R02diff}):
\beq
\omega= \pm \sqrt{-\Sigma_{0;11}^+ - \Sigma_{0;22}^+} 
= \pm k \left(\frac{4\pi}{3} \int \d q P_{L0}(q)\right)^{1/2} 
\label{omega}
\eeq
where we used eq.(\ref{S01122}). At order $\omega^2$ we obtain:
\beqa
\frac{\pl\rho_1^{(0)}}{\pl\eta_1} - \rho_2^{(0)} = 0
\label{rho1} \\
\frac{\pl\rho_2^{(0)}}{\pl\eta_1} - \frac{3}{2} \rho_1^{(0)} 
+ \frac{1}{2} \rho_2^{(0)} = 0
\label{rho2}
\eeqa
where we used the asymptotic behavior (\ref{S0k1121})-(\ref{S0k2212})
to simplify factors of the form $\Sigma_{0;ij}^+/\omega^2$.
By comparison with eqs.(\ref{R01})-(\ref{R02}), we see that these equations
are precisely the equations obeyed by the linear response $R_L=R_0$.
Moreover, the initial conditions at $\eta_1=\eta_2$ are precisely the same
if we redefine at order 0:
\beqa
R_i^{(0)}(\eta_1) & = & \rho_i^{(0)}(\eta_1)
\cos[\omega(e^{\eta_1}-e^{\eta_2})] \nonumber \\
& = & \rho_i^{(0)}(a_1) \cos[\omega(a_1-a_2)] .
\label{rho0}
\eeqa
Therefore, at high $k$ the ``envelope'' $\rho_i^{(0)}(\eta_1)$ of the 
non-linear response $R_{ij}(\eta_1)$ follows exactly the linear response $R_L$.
In addition, the response $R$ exhibits periodic oscillations over the time 
variable $a_1$ with a frequency which increases at high wavenumbers as 
$\omega\propto k$. Therefore, at one-loop order within the direct 
steepest-descent expansion the memory of initial conditions is not really 
erased at small scales or late times since the amplitude of the response 
$R$ does not decay but only exhibits strong oscillations.

\begin{figure}[htb]
\begin{center}
\epsfxsize=8 cm \epsfysize=7 cm {\epsfbox{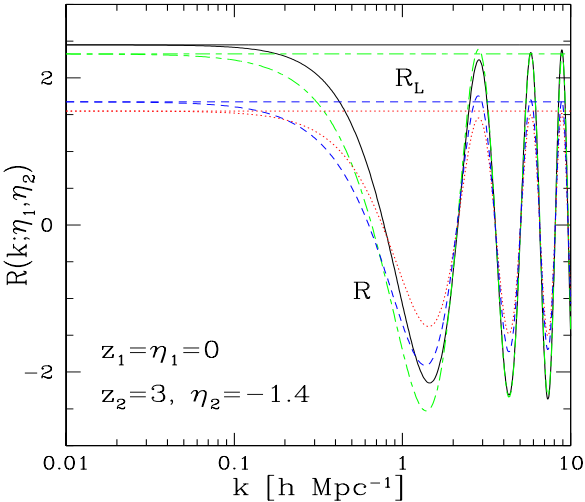}}
\end{center}
\caption{The non-linear response function $R(k;\eta_1,\eta_2)$ as a function 
of wavenumber $k$, at times $\eta_1=0,\eta_2=-1.4$, i.e. $z_1=0,z_2=3$. 
The horizontal lines are the linear response $R_L$ of eq.(\ref{RL}) which 
is independent of $k$}
\label{figR0k}
\end{figure}

We first display in Fig.~\ref{figR0t1} the evolution forward over time 
$\eta_1$ of the non-linear response $R(k;\eta_1,\eta_2)$ for wavenumbers $k=1$ 
(left panel) and $10\times h$ Mpc$^{-1}$ (right panel). We do not show the case
$k=0.1 h$ Mpc$^{-1}$ where the response $R$ closely follows the linear response
$R_L$ given in eq.(\ref{RL}). In agreement with the analysis performed
through eqs.(\ref{Rasymp})-(\ref{rho0}) we find that the non-linear response
$R$ exhibits oscillations $\sim \cos(\omega a_1)$ with a frequency $\omega$
which grows at higher $k$ and an amplitude which follows the linear response
$R_L$. Moreover, all components $R_{ij}$ oscillate in phase.

Next, we show in Fig.~\ref{figR0t2} the evolution backward over time $\eta_2$
of the non-linear response $R$. It shows a behavior similar to the
forward evolution displayed in Fig.~\ref{figR0t1} with respect to $\eta_1$. 
At large scales we recover the smooth increase (in absolute values) at early 
times $\eta_2$ of linear theory (\ref{RL}) whereas at small scales we obtain 
fast oscillations $\sim \cos(\omega a_2)$ related to 
eqs.(\ref{Rasymp})-(\ref{rho0}) with an amplitude which follows again 
the linear response $R_L$.

Finally, since at equal times the response function obeys 
eq.(\ref{Requaltimes}) we only display in Fig.~\ref{figR0k} the response at 
unequal times $(\eta_1=0,\eta_2=-1.4)$. At low $k$ we recover the linear 
response $R_L$.
At higher $k$ above $0.2 h$ Mpc$^{-1}$ the non-linear response obtained
at one-loop order within this steepest-descent approach departs from the
linear prediction and shows oscillations with increasingly high frequencies
while their amplitude follows the linear response $R_L$ at high $k$.
This behavior is a result of the oscillatory behavior with time $\eta_1$
displayed in Fig.~\ref{figR0t1} above and analyzed in 
eqs.(\ref{Rasymp})-(\ref{rho0}).

\subsection{Two-point correlation}
\label{Two-point-correlation0}

\begin{figure}[htb]
\begin{center}
\epsfxsize=9 cm \epsfysize=8 cm {\epsfbox{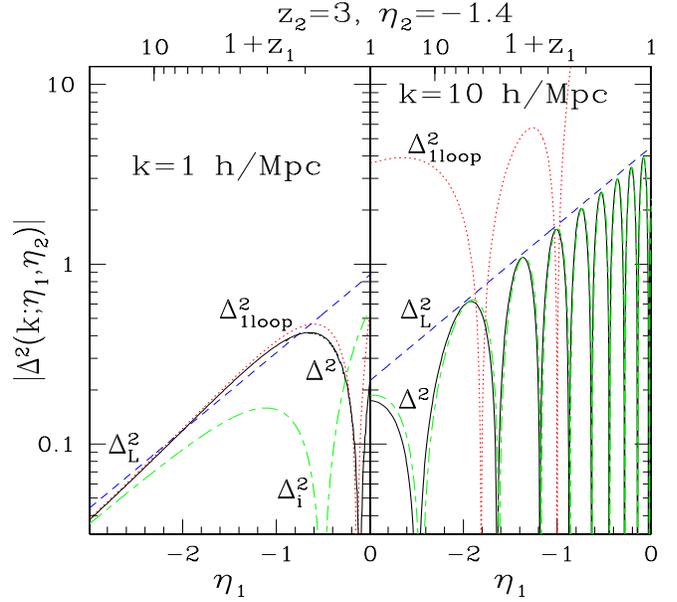}}
\end{center}
\caption{The density two-point correlation $G_{11}(k;\eta_1,\eta_2)$ as a 
function of time $\eta_1$, for $\eta_2=-1.4$ (i.e. $z_2=3$) and wavenumbers 
$k=1$ (left panel) and $10\times h$ Mpc$^{-1}$ (right panel). 
We plot the logarithmic power $\Delta^2=4\pi k^3G_{11}$. 
We display the linear power $\Delta^2_L$ of 
eq.(\ref{GL}) (dashed line), the usual one-loop perturbative result 
$\Delta^2_{\rm 1 loop}$ of eq.(\ref{P1loop}) (dotted line), the full non-linear
power $\Delta^2$ from eq.(\ref{GPi}) (solid line) and the contribution 
$\Delta^2_i$ from eq.(\ref{GRPiR})(dot-dashed line). Note that $G_{11}$ whence 
$\Delta^2$ can be negative, as shown by the oscillations displayed in the 
figure.}
\label{figG0t1}
\end{figure}

\begin{figure}[htb]
\begin{center}
\epsfxsize=8 cm \epsfysize=7 cm {\epsfbox{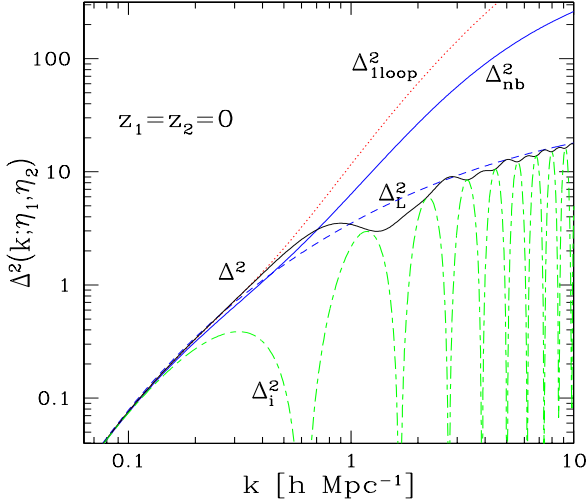}}
\end{center}
\caption{The logarithmic power $\Delta^2(k)$ of eq.(\ref{Delta2def}) at 
redshift $z=0$, that is for equal times $\eta_1=\eta_2=0$ ($z_1=z_2=0$). 
We display the linear power $\Delta^2_L$ (dashed line), the usual one-loop 
perturbative result $\Delta^2_{\rm 1 loop}$ (dotted line), the full non-linear
power $\Delta^2$ from eq.(\ref{GPi}) (solid line) and the contribution 
$\Delta^2_i$ from eq.(\ref{GRPiR}) (dot-dashed line). We also show the fit 
$\Delta^2_{\rm nb}$ (upper solid line) from numerical simulations 
(Smith et al. 2003). All quantities (including $\Delta^2_i$ which is the
square of an oscillating function) are positive.}
\label{figG0kz0z0}
\end{figure}

\begin{figure}[htb]
\begin{center}
\epsfxsize=8 cm \epsfysize=7 cm {\epsfbox{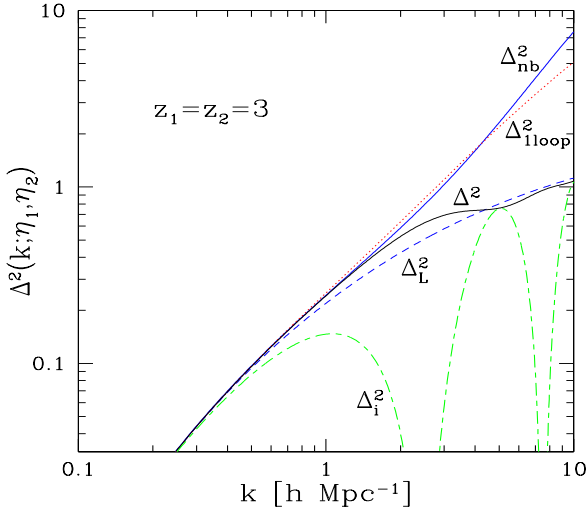}}
\end{center}
\caption{The logarithmic power $\Delta^2(k)$ as in Fig.~\ref{figG0kz0z0} but
at redshift $z=3$.}
\label{figG0kz3z3}
\end{figure}

\begin{figure}[htb]
\begin{center}
\epsfxsize=8 cm \epsfysize=7 cm {\epsfbox{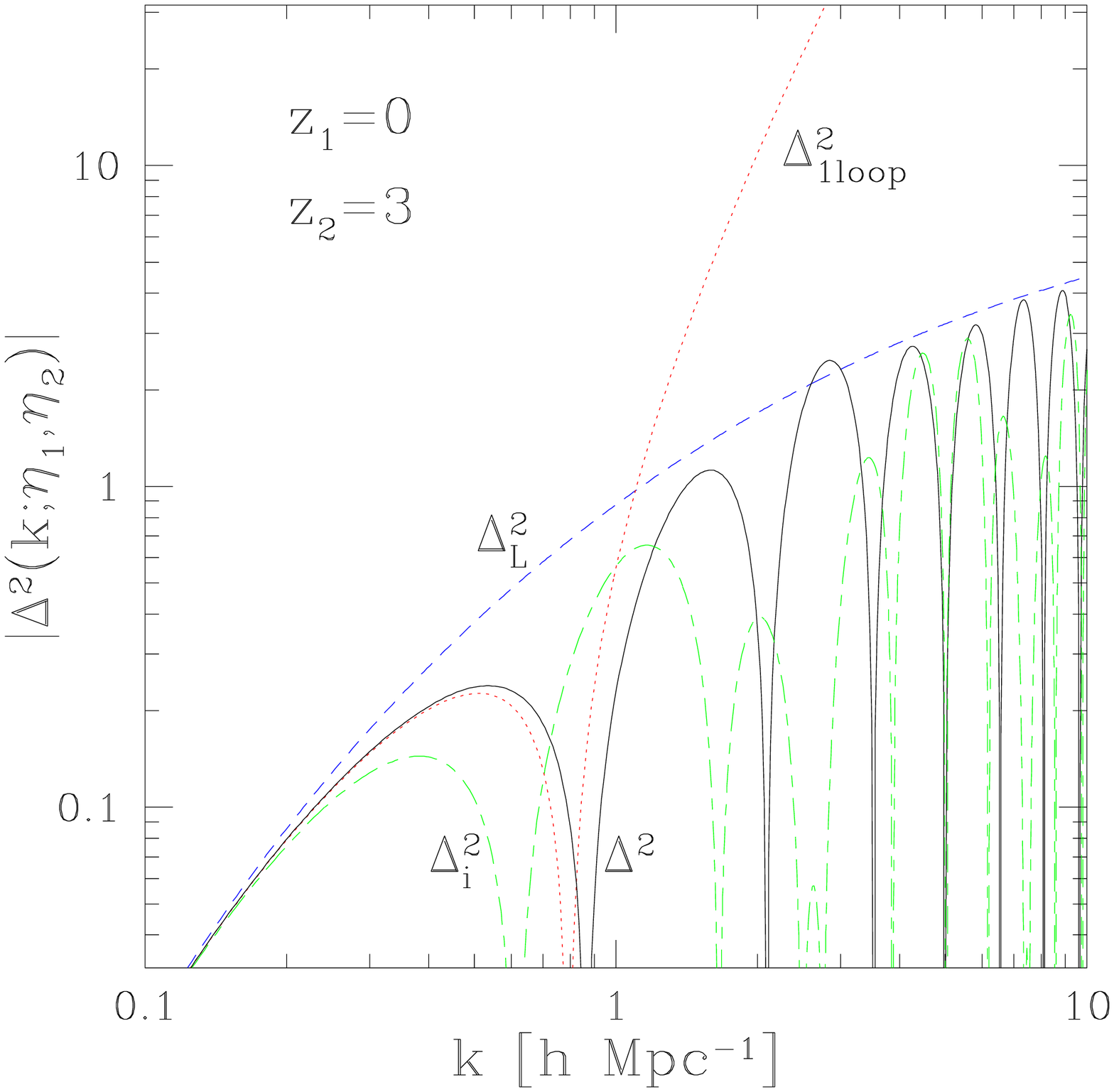}}
\end{center}
\caption{The logarithmic power $\Delta^2(k)$ as in Fig.~\ref{figG0kz0z0} 
but at unequal times $(\eta_1=0,\eta_2=-1.4)$ (i.e. $z_1=0,z_2=3$).}
\label{figG0kz0z3}
\end{figure}

Finally, from eq.(\ref{GPi}) we compute the non-linear two-point correlation
function $G$ obtained through the steepest-descent method at one-loop order.
We compare our results in Figs.~\ref{figG0t1}-\ref{figG0kz0z3} at redshifts
$z=0,3$ with a fit $G_{\rm nb}$ from numerical simulations (Smith et al. 2003),
the linear prediction $G_L$ (eq.(\ref{GL})) and the usual one-loop result 
$G_{\rm 1 loop}$ obtained from the standard perturbative expansion over powers
of the linear density field.
The standard one-loop power-spectrum can be written 
(e.g., Jain \& Bertschinger 1994, Scoccimarro \& Frieman 1996a,b) as:
\beq
P_{\rm 1 loop} = P_{L} + P_{22} + P_{13} ,
\label{P1loop}
\eeq
where $P_L$ is the linear contribution and $P_{22}$ and $P_{13}$ are of order
$P_L^2$ with:
\beqa
P_{22}(k;\eta_1,\eta_2) & = & e^{2\eta_1+2\eta_2} \int\d\bq P_{L0}(q) 
P_{L0}(\bk-\bq) \nonumber \\
&& \times 2 F_2^{(s)}(\bq,\bk-\bq)^2 ,
\label{P22}
\eeqa
and:
\beqa
P_{13}(k;\eta_1,\eta_2) & = & \frac{1}{2} \left( e^{3\eta_1+\eta_2} +
e^{\eta_1+3\eta_2} \right) P_{L0}(k)  \nonumber \\
&& \times \int\d\bq P_{L0}(q) 6 F_3^{(s)}(\bq,-\bq,\bk) .
\label{P13}
\eeqa
The angular kernels $F_2^{(s)}$ and $F_3^{(s)}$ are obtained from the
couplings $\alpha$ and $\beta$ of eq.(\ref{alphabeta}). In particular
we have (Goroff et al. 1986):
\beq
F_2^{(s)}(\bk_1,\bk_2) = \frac{5}{7}+\frac{\bk_1.\bk_2}{2k_1k_2} 
\left(\frac{k_1}{k_2} + \frac{k_2}{k_1} \right) 
+ \frac{2}{7} \left( \frac{\bk_1.\bk_2}{k_1k_2}\right)^2 .
\label{F2s}
\eeq
On the other hand, after integration over angles 
in eq.(\ref{P13}) one obtains (Makino et al. 1992):
\beqa
\lefteqn{P_{13}(k;\eta_1,\eta_2) = \frac{1}{2} \left( e^{3\eta_1+\eta_2} +
e^{\eta_1+3\eta_2} \right) P_{L0}(k) } \nonumber \\
&& \times \pi \int\d q P_{L0}(q) q^2 \left[ 
\frac{6k^6-79k^4q^2+50k^2q^4-21q^6}{63k^2q^4} \right. \nonumber \\
&& + \left. \frac{(q^2-k^2)^3(7q^2+2k^2)}{42k^3q^5} 
\ln\left|\frac{k+q}{k-q}\right| \right] .
\label{P13q}
\eeqa
From eqs.(\ref{P22})-(\ref{P13q}) we can see that at large wavenumbers
$k\rightarrow\infty$ the integrals over $q$ in eqs.(\ref{P22}), (\ref{P13q})
are dominated by the region where the slope $n$ of the linear power-spectrum
is of order $n \simeq -1$. This yields:
\beqa
k\rightarrow\infty : \;\; P_{\rm 1 loop} & \simeq & - e^{2\eta_1+2\eta_2} 
\left(\cosh(\eta_1-\eta_2)-1\right) \nonumber \\
&&  \times k^2 P_{L0}(k) \frac{4\pi}{3} \int \d q P_{L0}(q) .
\label{P1loophighk}
\eeqa
Therefore, at unequal times $\eta_1\neq\eta_2$ the standard one-loop 
prediction becomes negative at high $k$ and the logarithmic power (as defined
below in eq.(\ref{Delta2})) grows as 
$\Delta_{\rm 1 loop}^2 \propto - e^{3\eta_1} k^2$ with $k$ and $\eta_1$,
in agreement with Figs.~\ref{figG0t1}, \ref{figG0kz0z3}. Note that the
leading term (\ref{P1loophighk}) vanishes at equal times $\eta_1=\eta_2$.
This is related to the cancellation of the infrared divergences which appear
for linear power-spectra $P_{L}\propto k^n$ with $n \leq -1$ when one 
considers equal-times statistics (Vishniac 1983; Jain \& Bertschinger 1996).

The dependence $e^{3\eta_1} k^2$ merely reflects the order $P_L^2$ of the
perturbative expansion. In order to go deeper into the non-linear regime one 
may consider higher order terms but the latter grow increasingly fast and 
one would need at least partial resummations to obtain a well-behaved series 
(e.g. Crocce \& Scoccimarro 2006a). This is precisely what the large-$N$ 
expansions discussed in this article attempt to perform.

We first show in Fig.~\ref{figG0t1} the evolution over time $\eta_1$ of
the density two-point correlation $G_{11}(k;\eta_1,\eta_2)$, which is
symmetric with respect to $\eta_1\leftrightarrow\eta_2$. We actually
display the ``logarithmic power'' $\Delta^2(k;\eta_1,\eta_2)$ defined as
in eq.(\ref{Delta2def}) by:
\beq
\Delta^2(k;\eta_1,\eta_2) = 4\pi k^3 G_{11}(k;\eta_1,\eta_2) .
\label{Delta2}
\eeq
At equal times $\eta_1=\eta_2$ it is equal to the usual logarithmic power
and we have $\Delta^2>0$. However, at different times $\eta_1\neq\eta_2$
we can have $G_{11}<0$ whence $\Delta^2<0$. We present the cases $k=1$ 
(left panel) and $k=10\times h$ Mpc$^{-1}$ (right panel) in Fig.~\ref{figG0t1} 
for $\eta_2=-1.4$. For clarity we only display the density-density
correlation $G_{11}$. Other components $G_{ij}$ show a similar behavior.

The oscillations show that $G_{11}$ can indeed be negative at unequal times.
At these small scales the non-linear power $\Delta^2$ is already significantly
different from the linear power $\Delta^2_L$ but we can check in the left panel
that $\Delta^2$ matches the usual one-loop power $\Delta^2_{\rm 1 loop}$
of eq.(\ref{P1loop}) at small $\Delta^2$. As recalled in 
sect.~\ref{General-properties} the one-loop large-$N$ expansions indeed match
the usual perturbative expansion up to one-loop order (i.e. $P_L^2$). At later
times while $\Delta^2_{\rm 1 loop}$ grows very fast as $e^{3\eta_1}$ from
eq.(\ref{P13}) the prediction $\Delta^2$ of the one-loop steepest-descent 
approach remains well-behaved of order $\Delta^2_L$. 
Thus, we can see in the right panel that at very small scales where 
$\Delta^2_{\rm 1 loop}$ has become exceedingly large $\Delta^2$ remains 
well-controlled. In fact, following the behavior of the response function
analyzed in sect.~\ref{Response-function0} it exhibits fast oscillations with
an amplitude close to $\Delta^2_L$. We also show in Fig.~\ref{figG0t1}
the power $\Delta^2_i$ obtained from the first term alone of eq.(\ref{GPi})
which is detailed in eq.(\ref{GRPiR}). This corresponds to the ``non-linear
transport'' of the initial linear fluctuations without taking into
account the fluctuations $\Pi$ generated at all times by the non-linear 
dynamics. We can see that this term is sub-dominant in the mildly non-linear
regime (left panel) but happens to be again of same order as $\Delta^2$
in the highly non-linear regime. 

Next, we show in Fig.~\ref{figG0kz0z0} the logarithmic power $\Delta^2(k)$
today at equal redshifts $z_1=z_2=0$ (i.e. equal times $\eta_1=\eta_2=0$).
In agreement with eqs.(\ref{GPi})-(\ref{GRPiR}) and the subsequent discussion
both $\Delta^2$ and $\Delta^2_i$ are positive ($\Delta^2_i$ is the square of
an oscillating function).
At large scales $\Delta^2$ converges to the linear regime $\Delta^2_L$.
At smaller scales we can check again that $\Delta^2$ first follows the usual
one-loop power $\Delta^2_{\rm 1 loop}$ and next goes back to $\Delta^2_L$ 
whereas $\Delta^2_{\rm 1 loop}$ keeps increasing very fast in magnitude. 
Unfortunately, $\Delta^2$ does not agree with the fit $\Delta^2_{\rm nb}$ from
numerical simulations (Smith et al. 2003) on a larger range of $k$ as compared
with $\Delta^2_{\rm 1 loop}$ (although the rough agreement between 
$\Delta^2_{\rm nb}$ and $\Delta^2_{\rm 1 loop}$ at small scale is purely 
coincidental).
On the other hand, whereas adding higher-order terms in the standard 
perturbative expansion may not improve much the agreement with N-body 
simulations (especially since it would lead to increasingly steep terms at 
high $k$) the series obtained in the large-$N$ approach is likely to be 
well-behaved as various terms do not ``explode'' at small scales. 
However, this would require intricate calculations.
We can note that although $\Delta^2_i$ is of
order $\Delta^2$ one cannot neglect the fluctuations $\Pi$ generated by
the non-linear dynamics and the sum yields a smooth power $\Delta^2$.

We also display in Fig.~\ref{figG0kz3z3} the logarithmic power $\Delta^2(k)$
at equal redshifts $z_1=z_2=3$. We recover the same behaviors as those 
obtained at redshift $z=0$.

Finally, we show in Fig.~\ref{figG0kz0z3} the logarithmic power 
$\Delta^2(k;\eta_1,\eta_2)$ at unequal times $\eta_1=0,\eta_2=-1.4$, that is
at redshifts $z_1=0,z_2=3$. We find again that $\Delta^2$ matches the usual
one-loop power $\Delta^2_{\rm 1 loop}$ at large scales and follows its change
of sign at $k\sim 0.8 h$ Mpc$^{-1}$. Note indeed that at unequal times 
$\Delta^2$ need not be positive. At smaller scales, the steepest-descent
large-$N$ result
$\Delta^2$ departs from the fast growing $\Delta^2_{\rm 1 loop}$ and shows
a series of oscillations of moderate amplitude, which again follow 
$\Delta^2_L$. These features suggest that the first change of sign in 
Fig.~\ref{figG0kz0z3}, which is common to both $\Delta^2_L$ and
$\Delta^2_{\rm 1 loop}$, may be real and not a mere artefact of the
one-loop expansion. Note that in spite of the oscillations with time
seen in Figs.~\ref{figG0t1},\ref{figG0kz0z3} the large-$N$ approach 
automatically ensures that $\Delta^2>0$ at equal times thanks to the 
structure of eq.(\ref{GPi}). This property holds at all orders over $1/N$ 
since eq.(\ref{GPi}) is independent of the expressions of the self-energies 
$\Sigma,\Pi$.

\section{2PI effective action approach}
\label{2PI-effective-action-approach}

We now present the results obtained at one-loop order from the 2PI 
effective action approach recalled in sect.~\ref{2PI-effective-action}.
Thus, we need to solve the system of coupled equations (\ref{Geq})-(\ref{Peq})
where in the expression (\ref{Seq})-(\ref{Peq}) of the self-energy the
auxiliary two-point functions $G_0$ and $R_0$ are replaced by $G$ and $R$.
Thanks to causality, which leads to the Heaviside factor
$\theta(\eta_1-\eta_2)$  of eq.(\ref{Rk}) within both $R$ and $\Sigma$,
we solve the system (\ref{Geq})-(\ref{Peq}) by moving forward over time. 

Thus, we use a grid $\eta^{(n)}$ with $n=1,..,N_{\eta}$
for the time variables. At the earliest time-step $\eta^{(1)}$ we initialize
all matrices by their linear value at $\eta_1=\eta_2=\eta^{(1)}$.
Next, once we have obtained all two-point functions
up to time $\eta^{(n)}$ (i.e. over $\eta_1\leq\eta^{(n)}$ and 
$\eta_2\leq\eta^{(n)}$, initially $n=1$) we advance to the next time-step 
$\eta^{(n+1)}$ as follows. The response $R$ at equal times 
$\eta_1=\eta_2=\eta^{(n+1)}$ is first obtained from eq.(\ref{Requaltimes}).
Next, we move backward over time $\eta_2=\eta^{(n)},..,\eta^{(1)}$ at
fixed $\eta_1=\eta^{(n+1)}$ by using eq.(\ref{Rforward}) and eq.(\ref{Seq}).
That is, to compute $R(\eta^{(n+1)},\eta^{(i)})$ with $i=n,..,1$ we use for 
each value of $i$ the integro-differential eq.(\ref{Rforward}) to move over 
$\eta_1$ from $\eta_1=\eta^{(n)}$ up to $\eta_1=\eta^{(n+1)}$ at fixed 
$\eta_2=\eta^{(i)}$. Thanks to the Heaviside factors the r.h.s. of 
eq.(\ref{Rforward}) only involves $\Sigma(\eta_1,\eta)$ and $R(\eta,\eta_2)$ 
with $\eta_1\geq\eta\geq\eta_2$ which are already known. Besides, once 
$R(\eta^{(n+1)},\eta^{(i)})$ has been obtained we compute 
$\Sigma(\eta^{(n+1)},\eta^{(i)})$ from eq.(\ref{Seq}) before moving downward
to step $i-1$. Since $R$ and $\Sigma$ vanish for $\eta_2>\eta_1$ we have
actually obtained in this fashion $R$ and $\Sigma$ for all 
$\eta_1,\eta_2\leq\eta^{(n+1)}$ 
(indeed $R(\eta^{(i)},\eta^{(n+1)})=0$ for $i\leq n$).

Next, we compute $\Pi$ and $G$ from eq.(\ref{Peq}) and eq.(\ref{GPi}).
At fixed $\eta_1=\eta^{(n+1)}$ we now move forward over time 
$\eta_2=\eta^{(1)},..,\eta^{(n)}$. Indeed, the r.h.s of eq.(\ref{GPi})
only involves $\Pi(\eta_1',\eta_2')$ with $\eta_1'\leq\eta_1$ and
$\eta_2'\leq\eta_2$ which is already known. Besides, once 
$G(\eta^{(n+1)},\eta^{(i)})$ has been obtained we compute 
$\Pi(\eta^{(n+1)},\eta^{(i)})$ from eq.(\ref{Peq}) before moving forward
to step $i+1$ and we use the symmetry (\ref{Gsym}) to derive 
$G(\eta^{(i)},\eta^{(n+1)})$ as well as $\Pi(\eta^{(i)},\eta^{(n+1)})$.

In fact, at each step $i$ for $(\eta^{(n+1)},\eta^{(i)})$ the integrals
in the r.h.s. of eqs.(\ref{Rforward}), (\ref{GPi}) also involve the values
$R(\eta^{(n+1)},\eta^{(i)})$ and $G(\eta^{(n+1)},\eta^{(i)})$ which are
being computed if we use the boundary points in the numerical computation
of the time-integrals. Rather than using open formulae for the integrals we
perform a few iterations at each time-step $\eta^{(n)}$. This procedure
converges over a few loops.

Finally, to speed-up the numerical computation we use finite elements over
Fourier space $\bk$ to store the structure of the self-energy terms $\Sigma$
and $\Pi$. Thus, since all matrices only depend on the wavenumber modulus 
$k=|\bk|$ we use a grid $k^{(n)}$ with $n=1,..,N_k$. Next, in order to 
interpolate for all values of $k$ we can write for instance any matrix such 
as $G(k)$ as:
\beq
G(k) = \sum_{n=1}^{N_k} G^{(n)} \chi^{(n)}(k) , 
\label{Gkm}
\eeq
with:
\beq
G^{(n)}=G(k^{(n)}) \hspace{0.3cm} \mbox{and} \hspace{0.3cm} 
\chi^{(i)}(k^{(j)})= \delta_{ij} ,
\label{chi_ij}
\eeq
where $\delta_{ij}$ is the Kronecker symbol. The basis functions 
$\chi^{(n)}(k)$ define our interpolation scheme. We could use basis functions
which all extend over the full range $0<k<\infty$ (i.e. a spectral method
built from cardinal functions) but we prefer here to use simple finite 
elements with $\chi^{(n)}(k)=0$ if $k<k^{(n-1)}$ or $k>k^{(n+1)}$. 
For convenience we actually use triangular functions over the variable
$\kappa=(k-1)/(k+1)$ which maps $]0,\infty[$ to $]-1,1[$:
\beq
\chi^{(n)}(\kappa)= \left\{ \bea{ccc} 
{\displaystyle \frac{\kappa-\kappa^{(n-1)}}{\kappa^{(n)}-\kappa^{(n-1)}}} & \mbox{if} &  
\kappa^{(n-1)} \leq \kappa \leq \kappa^{(n)} \\
{\displaystyle \frac{\kappa^{(n+1)}-\kappa}{\kappa^{(n+1)}-\kappa^{(n)}}} & \mbox{if} &  
\kappa^{(n)} \leq \kappa \leq \kappa^{(n+1)}
\ea \right.
\eeq
which is equivalent to use linear interpolation over $\kappa$ for $G(k)$. 
Then, we can take advantage of the fact that the kernels 
$\gamma^s_{i;i_1,i_2}(\bk_1,\bk_2)$ in eqs.(\ref{Sgg})-(\ref{Pgg}) do not 
depend on time to compute once for all the integrals over wavenumber needed 
for the self-energy $\Sigma$ and $\Pi$. Thus, we define the quantities:
\beqa
\Sigma_{\nu_1 \nu_2}^{(n;n_1,n_2)} & = & 4 \int\d\bq \;
\gamma^s_{\nu_1}(\bq,\bk^{(n)}-\bq)
\gamma^s_{\nu_2}(\bk^{(n)},\bq-\bk^{(n)}) \nonumber \\
&& \times \chi^{(n_1)}(q) \chi^{(n_2)}(|\bk^{(n)}-\bq|)
\label{Schi1}
\eeqa
and:
\beqa
\Pi_{\nu_1 \nu_2}^{(n;n_1,n_2)} & = & 2 \int\d\bq \;
\gamma^s_{\nu_1}(\bq,\bk^{(n)}-\bq)
\gamma^s_{\nu_2}(-\bq,-\bk^{(n)}+\bq) \nonumber \\
&& \times \chi^{(n_1)}(q) \chi^{(n_2)}(|\bk^{(n)}-\bq|)
\label{Pchi1}
\eeqa
where we noted $\nu=(i;i_1,i_2)$ the indices of the vertices 
$\gamma^s_{i;i_1,i_2}$. These quantities $\Sigma_{\nu_1 \nu_2}^{(n;n_1,n_2)}$
and $\Pi_{\nu_1 \nu_2}^{(n;n_1,n_2)}$ only depend on the vertices $\gamma^s$
given in eqs.(\ref{gamma1})-(\ref{gamma2}) and on the grid which we use over
wavenumbers hence they are computed at the beginning of the code within
initialization subroutines. Next, the self-energy at a given time 
$\eta_1,\eta_2$ is obtained as:
\beqa
\Sigma_{i_1 i_2}(k^{(n)};\eta_1,\eta_2) & = & \sum_{j_1 j_2 l_1 l_2} 
\sum_{n_1,n_2} \Sigma_{\nu_1 \nu_2}^{(n;n_1,n_2)} 
R_{j_1 l_1}^{(n_1)}(\eta_1,\eta_2) \nonumber \\
&& \times G_{j_2 l_2}^{(n_2)}(\eta_1,\eta_2)
\label{Schi2}
\eeqa
with $\nu_1=(i_1;j_1,j_2)$ and $\nu_2=(l_1;i_2,l_2)$, and:
\beqa
\Pi_{i_1 i_2}(k^{(n)};\eta_1,\eta_2) & = & \sum_{j_1 j_2 l_1 l_2} 
\sum_{n_1,n_2} \Pi_{\nu_1 \nu_2}^{(n;n_1,n_2)} 
G_{j_1 l_1}^{(n_1)}(\eta_1,\eta_2) \nonumber \\
&& \times G_{j_2 l_2}^{(n_2)}(\eta_1,\eta_2)
\label{Pchi2}
\eeqa
with $\nu_1=(i_1;j_1,j_2)$ and $\nu_2=(i_2;l_1,l_2)$. Therefore, using
eqs.(\ref{Schi2})-(\ref{Pchi2}) we obtain the self-energy by advancing
over times $\eta_1,\eta_2$ as described above without performing new
integrations over wavenumbers at each step (they are replaced by the discrete
sums over $n_1,n_2$ with weights $\Sigma_{\nu_1 \nu_2}^{(n;n_1,n_2)}$
or $\Pi_{\nu_1 \nu_2}^{(n;n_1,n_2)}$).

\subsection{Response $R$ and self-energy $\Sigma$}
\label{Response-R-and-self-energy-Sigma}

\begin{figure}[htb]
\begin{center}
\epsfxsize=9 cm \epsfysize=8 cm {\epsfbox{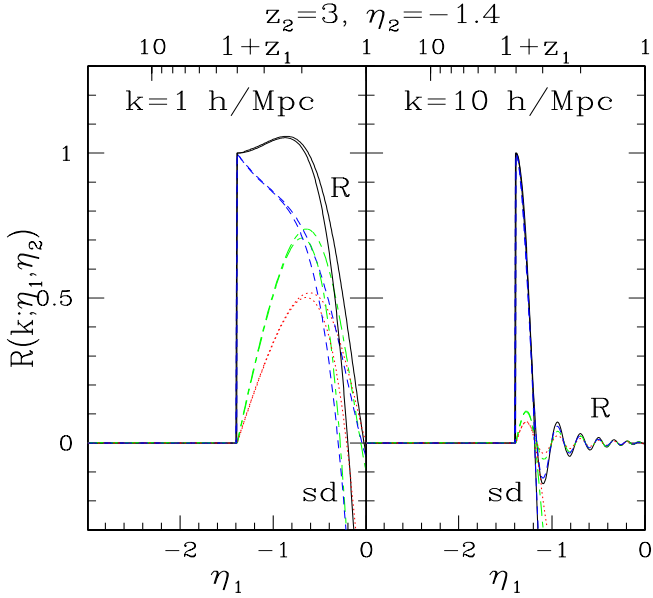}}
\end{center}
\caption{The non-linear response $R(k;\eta_1,\eta_2)$ as a function of forward
time $\eta_1$, for $\eta_2=-1.4$ (i.e. $z_2=3$) and wavenumbers $k=1$ 
(left panel) and $10 \times h$ Mpc$^{-1}$ (right panel), as in 
Fig.~\ref{figR0t1} but for the 2PI effective action method. 
For comparison we also plot the first half-oscillation of the response 
obtained in sect.~\ref{Direct-steepest-descent} from the direct 
steepest-descent method (curves labeled ``sd'') which was shown in 
Fig.~\ref{figR0t1}.}
\label{figRt1}
\end{figure}

We first display in Fig.~\ref{figRt1} the evolution forward over time $\eta_1$
of the response $R(k;\eta_1,\eta_2)$. We can see that the non-linear response
exhibits oscillations as for the steepest-descent result of Fig.~\ref{figR0t1}
but its amplitude now decays at large times $\eta_1$ instead of following the
linear envelope. This can be understood analytically from the 
following simple model. Let us consider the equation:
\beq
\frac{\pl R}{\pl\eta_1} = \sigma \int_{\eta_2}^{\eta_1} \d\eta 
\; e^{\eta_1+\eta} R(\eta_1,\eta) R(\eta,\eta_2)
\label{Rtoydef}
\eeq
for a one-component response $R(\eta_1,\eta_2)$. This is a simplified form
of the system (\ref{Rforward}), (\ref{Seq}) where we used the linear 
time-dependence (\ref{GL}) for $G$ to obtain a closed equation for $R$.
The parameter $\sigma\leq 0$ represents the amplitude of the self-energy 
$\Sigma$ at the wavenumber of interest. Then, the linear regime $R_L$ 
corresponds to $\sigma=0$ and yields the solution:
\beq
\eta_1 \geq \eta_2 : \;\;\; R_L(\eta_1,\eta_2) = 1 ,
\label{RLtoy}
\eeq
where we used the initial condition $R(\eta,\eta)=1$ at equal times.
Next, the steepest-descent method presented in 
sect.~\ref{Steepest-descent} and sect.~\ref{Direct-steepest-descent} 
amounts to replace $R(\eta_1,\eta)$ by $R_L(\eta_1,\eta)$ into the integral
in the r.h.s. of eq.(\ref{Rtoydef}). This yields a linear equation for $R$
which can be transformed into a differential equation as in 
sect.~\ref{Response-function0} by taking the derivative of eq.(\ref{Rtoydef})
with respect to $\eta_1$. Since we only have one component one differentiation
is sufficient to remove the integral and we obtain:
\beq
\frac{\pl^2 R}{\pl\eta_1^2} - \frac{\pl R}{\pl\eta_1}
= \sigma e^{2\eta_1} R  \;\;\; \mbox{whence} \;\;\;
\frac{\pl^2 R}{\pl a_1^2} = \sigma R ,
\label{R0toy1}
\eeq
where we introduced the scale-factor $a_1=e^{\eta_1}$ as the 
new time-variable. Therefore, we obtain the solution:
\beq
R(a_1,a_2) = \cos[\omega(a_1-a_2)] \;\;\; \mbox{with} 
\;\;\; \omega= \sqrt{-\sigma} .
\label{R0toy2}
\eeq
Thus, as in sect.~\ref{Response-function0} we obtain within the 
steepest-descent method an oscillating response function with an amplitude
given by the linear response $R_L$ and a frequency over the time-variable
$a_1$ given by $ \sqrt{-\sigma}$. Finally, in the 2PI effective action 
method we need to solve the non-linear equation (\ref{Rtoydef}) which also
reads:
\beq
\frac{\pl R}{\pl a_1} = \sigma \int_{a_2}^{a_1} \d a 
\; R(a_1,a) R(a,a_2) ,
\label{Rtoydefa1}
\eeq
where we changed variables from $\eta$ to $a=e^{\eta}$. We can look for
a solution of eq.(\ref{Rtoydefa1}) of the form $R(a_1,a_2)=R(a_1-a_2)$
so that the r.h.s. of eq.(\ref{Rtoydefa1}) becomes a simple convolution:
\beq
\frac{\d R}{\d a} = \sigma \int_0^a \d a' \; R(a-a') R(a') .
\label{Rtoydefa}
\eeq
Introducing the Laplace transform ${\tilde R}(s)$ defined by:
\beq
{\tilde R}(s)= \int_0^{\infty} \d a \; e^{-s a} R(a) ,
\label{Rsdef}
\eeq
and taking the Laplace transform of eq.(\ref{Rtoydefa}) we obtain:
\beq
s {\tilde R}(s) - R(0) = \sigma {\tilde R}(s)^2 .
\label{Rseq}
\eeq
Using $R(0)=1$ and requiring that ${\tilde R}(s)$ vanishes for 
$s\rightarrow +\infty$ yields:
\beq
{\tilde R}(s) = \frac{\sqrt{s^2+4\omega^2}-s}{2\omega^2} ,
\label{Rs}
\eeq
where we introduced $\omega$ as defined in eq.(\ref{R0toy2}).
This is a well-known Laplace transform and eq.(\ref{Rs}) directly gives:
\beq
R(a_1,a_2)= \frac{J_1[2\omega(a_1-a_2)]}{\omega(a_1-a_2)} ,
\label{Rtoy2PI}
\eeq
where $J_1$ is a Bessel function of the first kind.
Thus, we see that within the 2PI effective action method the response
function exhibits oscillations as in the direct steepest-descent method.
However, their amplitude is no longer given by the linear response and becomes 
much smaller at late times along with a higher frequency. For the simple 
model (\ref{Rtoydef}) the amplitude of the response function decays as 
$a_1^{-3/2}$ instead of being constant while the frequency becomes twice
larger. This explains the qualitative behavior seen in Fig.~\ref{figRt1}.
In particular, note that in agreement with the simple model 
(\ref{Rtoydef})-(\ref{Rtoy2PI}) the comparison of the right panel of 
Fig.~\ref{figRt1} with Fig.~\ref{figR0t1} shows that the frequency of the
oscillations is indeed about twice larger in the non-linear regime
for the 2PI effective action result than for the steepest-descent prediction.
Of course, the result shown in Fig.~\ref{figRt1} is more complicated than
the simple model (\ref{Rtoydef}) since the two-point correlation $G$
which enters $\Sigma$ whence $R$ also depends non-linearly on $R$ through 
eq.(\ref{Geq}) or the more explicit expression (\ref{GPi}).
However, as we shall check in sect.~\ref{Mixed-approaches} the dependence
of $\Sigma$ on $G$ does not change much the behavior of the response $R$
(see Fig.~\ref{figRG0k}).

\begin{figure}[htb]
\begin{center}
\epsfxsize=9 cm \epsfysize=8 cm {\epsfbox{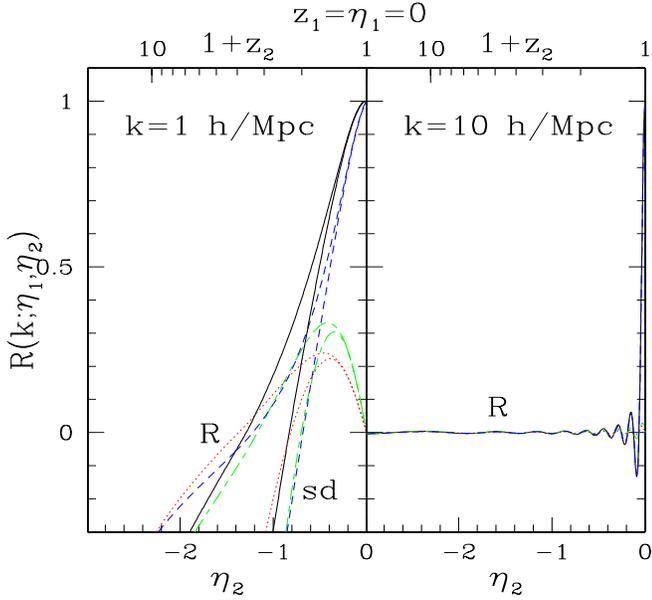}}
\end{center}
\caption{The non-linear response $R(k;\eta_1,\eta_2)$ as a function of backward
time $\eta_2$, for $\eta_1=0$ (i.e. $z_1=0$) and wavenumbers $k=1$ 
(left panel) and $10 \times h$ Mpc$^{-1}$ (right panel). 
We also plot in the left panel the first half-oscillation of the response 
obtained in sect.~\ref{Direct-steepest-descent} from the direct 
steepest-descent method (curves labeled ``sd'') which was shown in 
Fig.~\ref{figR0t2}.}
\label{figRt2}
\end{figure}

\begin{figure}[htb]
\begin{center}
\epsfxsize=8 cm \epsfysize=7 cm {\epsfbox{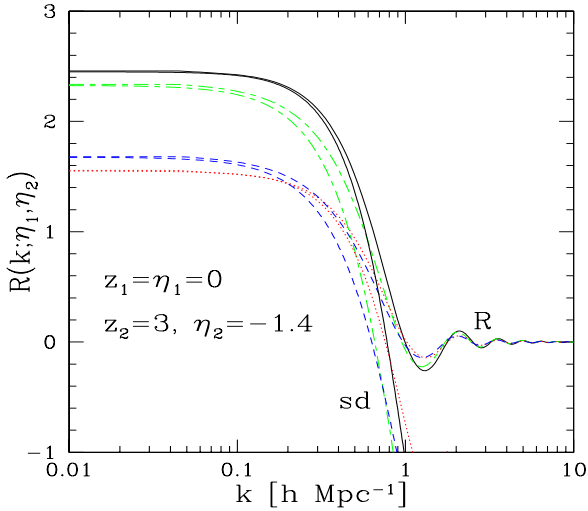}}
\end{center}
\caption{The non-linear response function $R(k;\eta_1,\eta_2)$ as a function 
of wavenumber $k$, at times $\eta_1=0,\eta_2=-1.4$. We also plot the first 
half-oscillation of the response obtained in 
sect.~\ref{Direct-steepest-descent} from the direct steepest-descent method
(curves labeled ``sd'') which was shown in Fig.~\ref{figR0k}.}
\label{figRk}
\end{figure}

On the other hand, we can see in Fig.~\ref{figRt1} that the first quarter
of oscillation, from 1 to 0, of the response $R$ follows rather closely
the curves labeled ``sd'' obtained from the steepest-descent method
presented in sect.~\ref{Direct-steepest-descent} which were also shown in 
Fig.~\ref{figR0t1}. Besides this agreement remains valid at quite large
$k$ (right panel) where two-point functions obtained from both methods
are generically very different as shown by the figures below.
This property can be understood from eq.(\ref{Sequaltimes}) which shows
that the dependence on wavenumber $\Sigma(k;\eta,\eta)\propto k^2$
of the self-energy $\Sigma$ at equal times is identical for both approaches
(at one-loop order). Moreover, eq.(\ref{Sequaltimes}) shows that the
normalization is governed by the value of the two-point correlation
$G_{22}(k';\eta,\eta)$ at the point where the logarithmic slope is $n=-1$
(this integral actually corresponds to the mean square velocity 
$\lag v^2\rag$). At redshift $z=0$ this wavenumber is still within the linear
regime (for the SCDM cosmology which we consider here) therefore the
self-energy $\Sigma$ at equal times $\eta_1=\eta_2$ obtained within the
steepest-descent method and the 2PI effective action approach are very close
until $z=0$. Then, the early time-evolutions at $\eta_1\simeq\eta_2$ of the
response $R$ obtained within both methods from eq.(\ref{Rforward}) are
very close. We can see from Fig.~\ref{figRt1} that this agreement holds 
until the response $R$ first vanishes. Beyond this point the steepest-descent 
method yields increasingly large oscillations (Fig.~\ref{figR0t1}) whereas
the 2PI effective action yields small oscillations which decay to 
zero (Fig.~\ref{figRt1}).

Next, we show in Fig.~\ref{figRt2} the evolution backward over time $\eta_2$
of the response $R$. In a manner consistent with Fig.~\ref{figRt1} it
exhibits oscillations which are strongly damped at large time separations
$|a_1-a_2|$, whence at early times $\eta_2$, while the frequency is also
larger than for the steepest-descent result shown in Fig.~\ref{figR0t2}.
Again the behavior at nearly
equal times $\eta_2\simeq\eta_1$ is close to the prediction of the 
steepest-descent method presented in sect.~\ref{Direct-steepest-descent}
until the response first vanishes. The response at unequal times
$(\eta_1=0,\eta_2=-1.4)$ is shown as a function of wavenumber $k$ in
Fig.~\ref{figRk}. At large scales we recover the linear regime whereas at small
scales we obtain oscillations which show a fast decay into the non-linear 
regime. This is consistent with the time-dependence displayed in 
Figs.~\ref{figRt1}, \ref{figRt2}. 

Therefore, in contrast with the 
steepest-descent approach we find that within the 2PI effective action method
the memory of initial conditions is in some sense erased as the response 
function decays for large time separation. This agrees with the intuitive
expectation that within the real collisionless gravitational dynamics
the details of the initial conditions are erased at small scales. Indeed,
after shell-crossing one can expect for instance that virialization processes
build halos which mainly depend on a few integrated quantities which 
characterize the collapsed region (such as the mass, initial overdensity,
angular momentum,..) and exhibit relaxed cores (e.g. isotropic velocity 
distributions), as suggested by numerical simulations. However, we must note
that the system studied in this paper is defined from the hydrodynamical
equations of motion (\ref{continuity1})-(\ref{Poisson1}) which break
down at shell-crossing. Therefore, there is no guarantee a priori that
such small-scale relaxation processes should come out from 
eqs.(\ref{continuity1})-(\ref{Poisson1}). It appears through the 
2PI effective action method presented in this paper and 
Figs.~\ref{figRt1}-\ref{figRk} that this is actually the case. In fact,
studying the large-$k$ behavior of the kernel $K_s$ Crocce \& Scoccimarro 
(2006b) managed to resum all dominant diagrams in this large-$k$ limit
and obtained a Gaussian decay $R(k) \sim e^{-k^2}$. In our case, the
simple model (\ref{Rtoydef}) suggests that the one-loop 2PI effective action 
method only yields a power-law decay as in eq.(\ref{Rtoy2PI}). Since at the
one-loop order considered in this article we do not resum all diagrams it is
not really surprising that we do not recover the ``exact'' Gaussian decay.
However, note that the large-$N$ expansion does not give a mere Taylor series
expansion of such a Gaussian decay (which would blow up at large $k$ or large
times) but already captures at a qualitative level the damping of the response 
$R$ at small scale or at large time separation thanks to the non-linearity of
the evolution equation obeyed by $R$.

\begin{figure}[htb]
\begin{center}
\epsfxsize=8 cm \epsfysize=8 cm {\epsfbox{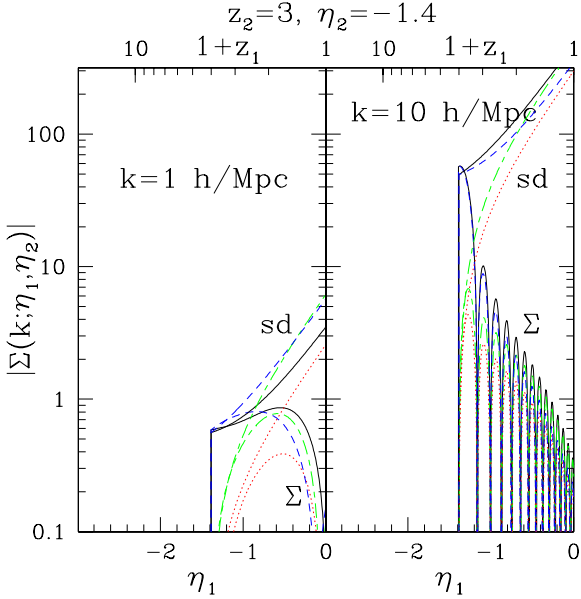}}
\end{center}
\caption{The self-energy $\Sigma(k;\eta_1,\eta_2)$ as a function of 
forward time $\eta_1$, for $\eta_2=-1.4$ (i.e. $z_2=3$) and wavenumbers 
$k=1$ (left panel) and $10 \times h$ Mpc$^{-1}$ (right panel). 
The curves labeled ``sd'' are the prediction of the steepest-descent method 
as in Fig.~\ref{figS0t1}.}
\label{figSt1}
\end{figure}

\begin{figure}[htb]
\begin{center}
\epsfxsize=8 cm \epsfysize=8 cm {\epsfbox{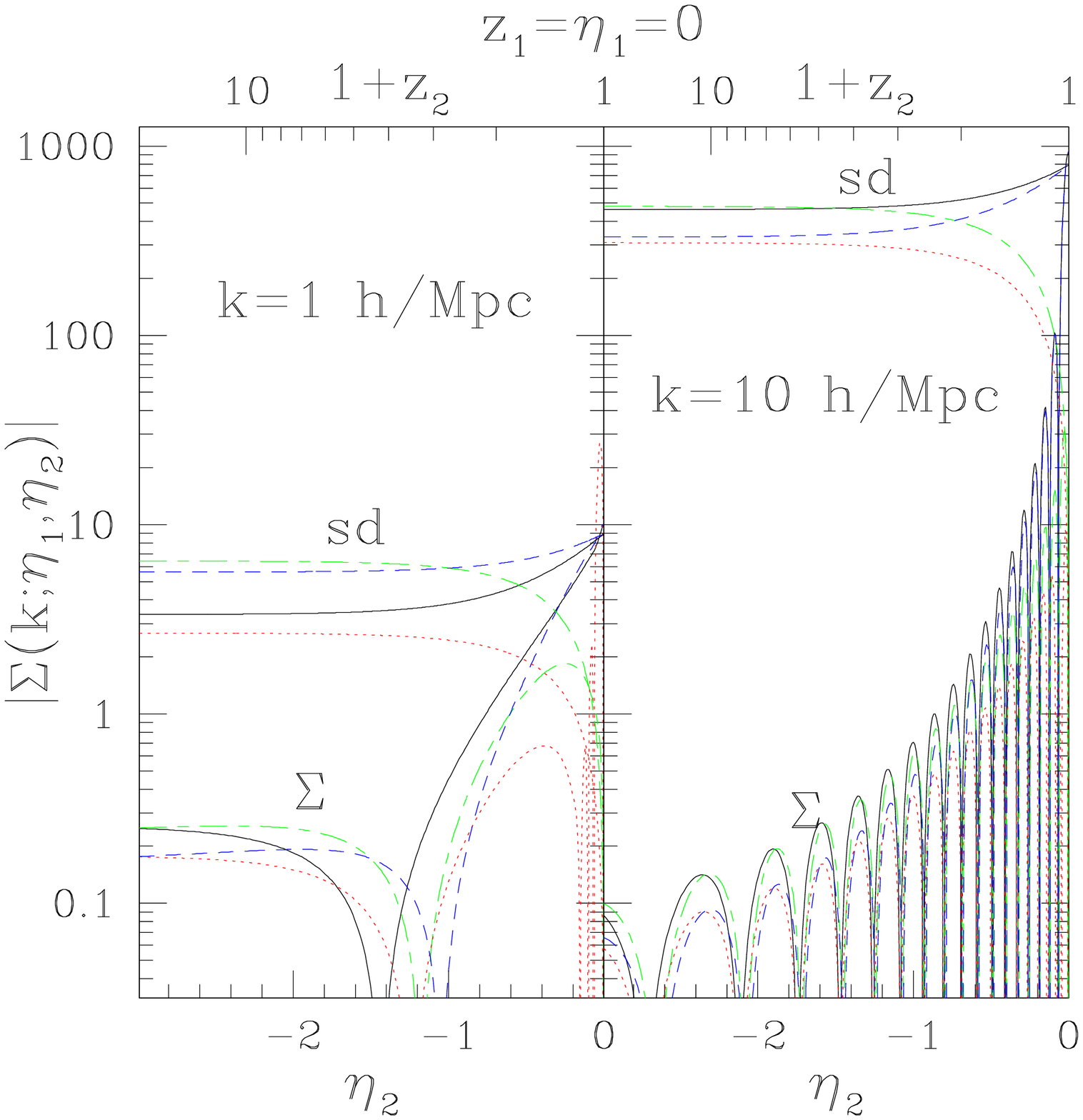}}
\end{center}
\caption{The self-energy $\Sigma(k;\eta_1,\eta_2)$ as a function of 
backward time $\eta_2$, for $\eta_1=0$ (i.e. $z_1=0$) and wavenumbers 
$k=1$ (left panel) and $10 \times h$ Mpc$^{-1}$ (right panel). 
The curves labeled ``sd'' are the prediction of the steepest-descent method 
as in Fig.~\ref{figS0t2}.}
\label{figSt2}
\end{figure}

\begin{figure}[htb]
\begin{center}
\epsfxsize=8 cm \epsfysize=7 cm {\epsfbox{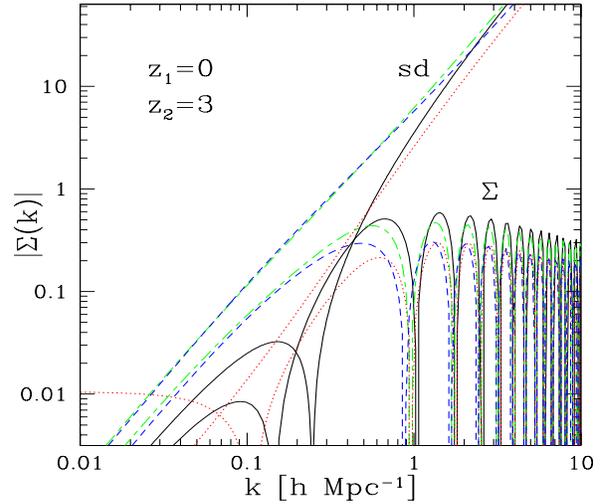}}
\end{center}
\caption{The self-energy $\Sigma(k)$ as a function of wavenumber $k$ at
unequal redshifts $z_1=0,z_2=3$. The curves labeled 
``sd'' are the prediction of the steepest-descent method.}
\label{figSk_z0_z3}
\end{figure}

We display in Fig.~\ref{figSt1} and Fig.~\ref{figSt2} the evolution with times
$\eta_1$ and $\eta_2$ of the self-energy $\Sigma$. At large scales 
($k<0.1 h$ Mpc$^{-1}$) and small time separations we recover the results 
obtained within the direct steepest-descent method but at smaller scales 
we obtain a series of oscillations with an amplitude which decreases at 
large time separations. 
This follows from the behavior of the non-linear response $R$ analyzed in 
Figs.~\ref{figRt1}-\ref{figRk} above.
Indeed, the self-energy $\Sigma$ depends linearly on the response $R$, see
eq.(\ref{Schi2}). Again the behavior at nearly equal times 
$\eta_1\simeq\eta_2$ is similar for both approaches (see in particular 
the left panel of Fig.~\ref{figSt1}).
The right panel of Fig.~\ref{figSt1} shows that at high wavenumbers the 
envelope of the oscillations of $\Sigma$ decays as a power-law of 
$a_1=e^{\eta_1}$ at late times. This agrees with the analysis of the simple
model (\ref{Rtoydef}).
In a similar fashion, Fig.~\ref{figSk_z0_z3} which displays the dependence 
on wavenumber $k$ of $\Sigma(k;\eta_1,\eta_2)$, at fixed times 
$\eta_1=0,\eta_2=-1.4$, matches the results obtained from the 
direct steepest-descent method at low $k$ (except for $\Sigma_{12}$) and 
exhibits decaying oscillations at high $k$.
We can note in Fig.~\ref{figSk_z0_z3} that the term $\Sigma_{12}$ (dotted line)
does not match the steepest-descent result in the limit $k \rightarrow 0$.
This is due to the behavior of the associated kernels 
$\Sigma_{\nu_1 \nu_2}^{(n;n_1,n_2)}$ of eq.(\ref{Schi1}). More precisely,
using eqs.(\ref{gamma1})-(\ref{gamma1}), eq.(\ref{Sgg}) reads in this case:
\beqa
\Sigma_{12}(k) & = & \int\d\bq \biggl \lbrace 
R_{11}(q) G_{21}(|\bk-\bq|) \frac{[\bk.(\bk-\bq)](\bk.\bq)}{k^2 |\bk-\bq|^2}
\nonumber \\
&& - R_{12}(q) G_{22}(|\bk-\bq|) \frac{[\bk.(\bk-\bq)]^2 q^2}{k^2 |\bk-\bq|^4}
\nonumber \\
&& + R_{21}(q) G_{11}(|\bk-\bq|) \frac{(\bk.\bq)^2}{k^2q^2} 
\nonumber \\
&& - R_{22}(q) G_{12}(|\bk-\bq|) 
\frac{(\bk.\bq)[\bk.(\bk-\bq)]}{k^2 |\bk-\bq|^2} \biggl \rbrace 
\eeqa
where we did not write the time-dependence. At large wavenumbers $q$ the
integrand $S_{12}(\bq)$ simplifies to:
\beq
S_{12}(\bq) \sim \frac{(\bk.\bq)^2}{k^2q^2} 
[ R_{21} G_{11} + R_{22} G_{12} - R_{11} G_{21} - R_{12} G_{22}] .
\label{S12q1}
\eeq
Therefore, in the linear regime (or for the steepest-descent self-energy 
$\Sigma_0$) where $G_{ij}(q;\eta_1,\eta_2) = e^{\eta_1+\eta_2} P_{L0}(q)$
the two-point correlations $G_{ij}$ factorize and we have:
\beq
q \gg \! k \! :  S_{0;12}(\bq) \sim P_{L0}(q) 
[R_{L;21}+R_{L;22}-R_{L;11}-R_{L;12}]
\label{S12q2}
\eeq
For a CDM-like power-spectrum $P_{L0}(q) \sim q^{-3}\ln q$ at large wavenumbers
while the linear response $R_L(q)$ is constant with respect to $q$. 
Hence the integration over $\bq$ yields at first sight a logarithmic divergence
for $q\rightarrow\infty$. In fact, we can check from eq.(\ref{RL}) that for
the linear response the terms in the bracket in eq.(\ref{S12q2}) actually
cancel so that the integral is well-defined for the linear two-point functions
(the integrand decreases as $1/q^4$ at large $q$) and it is dominated by
wavenumbers $q \sim k$, in agreement with the results obtained in 
sect.~\ref{Self-energy0}. However, for the 2PI effective action method where
we must use the non-linear two-point functions in eq.(\ref{S12q1}) this 
cancellation no longer holds at weakly non-linear scales $q_{NL}$ where they 
depart from the linear predictions. The integral still converges because of the
decay of the non-linear response at high $q$ but this implies that 
for $k \rightarrow 0$ the self-energy term $\Sigma_{12}$ is dominated by the 
contribution of wavenumbers $q \sim q_{NL}$ associated with the non-linear
transition and not by wavenumbers $q\sim k$. Therefore, $\Sigma_{12}$ does not
converge to the steepest-descent prediction $\Sigma_{0;12}$ at large scales
$k\rightarrow 0$. By contrast, for other terms the integrand behaves as:
\beq
q \! \gg \! k \! : \; S_{11}(\bq) \! \sim \! S_{22}(\bq) \! \sim 
\frac{\bk.\bq}{q^2} R G , \;\; S_{21}(\bq) \! \sim \! \frac{k^2}{q^2} R G .
\eeq
The additional power of $1/q$ ensures that the logarithmic divergence of
eq.(\ref{S12q2}) does not appear and at large scales the self-energy 
$\Sigma_{ij}$ is dominated by the contribution from wavenumbers $q\sim k$
hence it matches the results obtained within the steepest-descent approach.

\subsection{Correlation $G$ and self-energy $\Pi$}
\label{Correlation-G-and-self-energy-Pi}

\begin{figure}[htb]
\begin{center}
\epsfxsize=9 cm \epsfysize=8 cm {\epsfbox{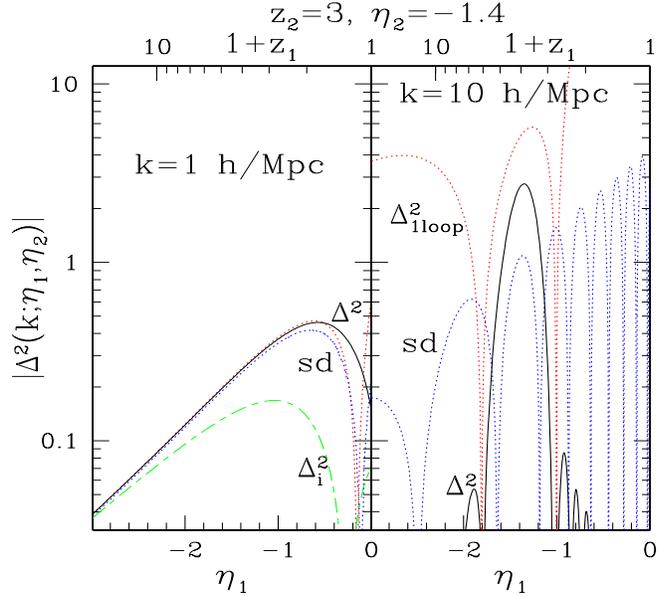}}
\end{center}
\caption{The density two-point correlation $G_{11}(k;\eta_1,\eta_2)$ as a 
function of time $\eta_1$, for $\eta_2=-1.4$ (i.e. $z_2=3$) and wavenumbers 
$k=1$ (left panel) and $10 \times h$ Mpc$^{-1}$ (right panel). We plot the 
logarithmic power $\Delta^2=4\pi k^3G_{11}$. We display the full non-linear 
power $\Delta^2$ from eq.(\ref{GPi}) (solid line) and the contribution 
$\Delta^2_i$ from eq.(\ref{GRPiR})(dot-dashed line), as well as the usual 
one-loop perturbative result $\Delta^2_{\rm 1 loop}$ of eq.(\ref{P1loop}) 
(dotted line) and the prediction of the steepest-descent method 
(``sd'', dotted line) also shown in Fig.~\ref{figG0t1}.}
\label{figGt1}
\end{figure}

\begin{figure}[htb]
\begin{center}
\epsfxsize=8 cm \epsfysize=7 cm {\epsfbox{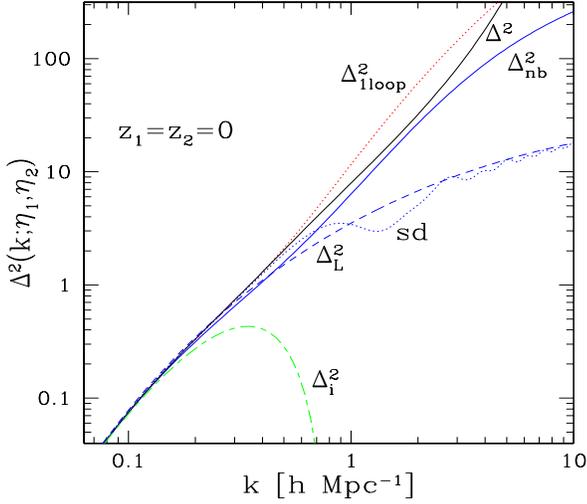}}
\end{center}
\caption{The logarithmic power $\Delta^2(k)$ at redshift $z=0$, that is
at equal times $z_1=z_2=0$. We display the full non-linear power $\Delta^2$ 
from eq.(\ref{GPi}) (solid line) and the contribution $\Delta^2_i$ from 
eq.(\ref{GRPiR})(dot-dashed line), as well as the linear power $\Delta^2_L$
(dashed line), the usual one-loop perturbative result $\Delta^2_{\rm 1 loop}$ 
of eq.(\ref{P1loop}) (dotted line), the prediction of the steepest-descent 
method (``sd'', dotted line) also shown in Fig.~\ref{figG0kz0z0} and the
fit $\Delta^2_{\rm nb}$ from numerical simulations (Smith et al. 2003). 
All quantities (including $\Delta^2_i$) are positive.}
\label{figGkz0z0}
\end{figure}

\begin{figure}[htb]
\begin{center}
\epsfxsize=8 cm \epsfysize=7 cm {\epsfbox{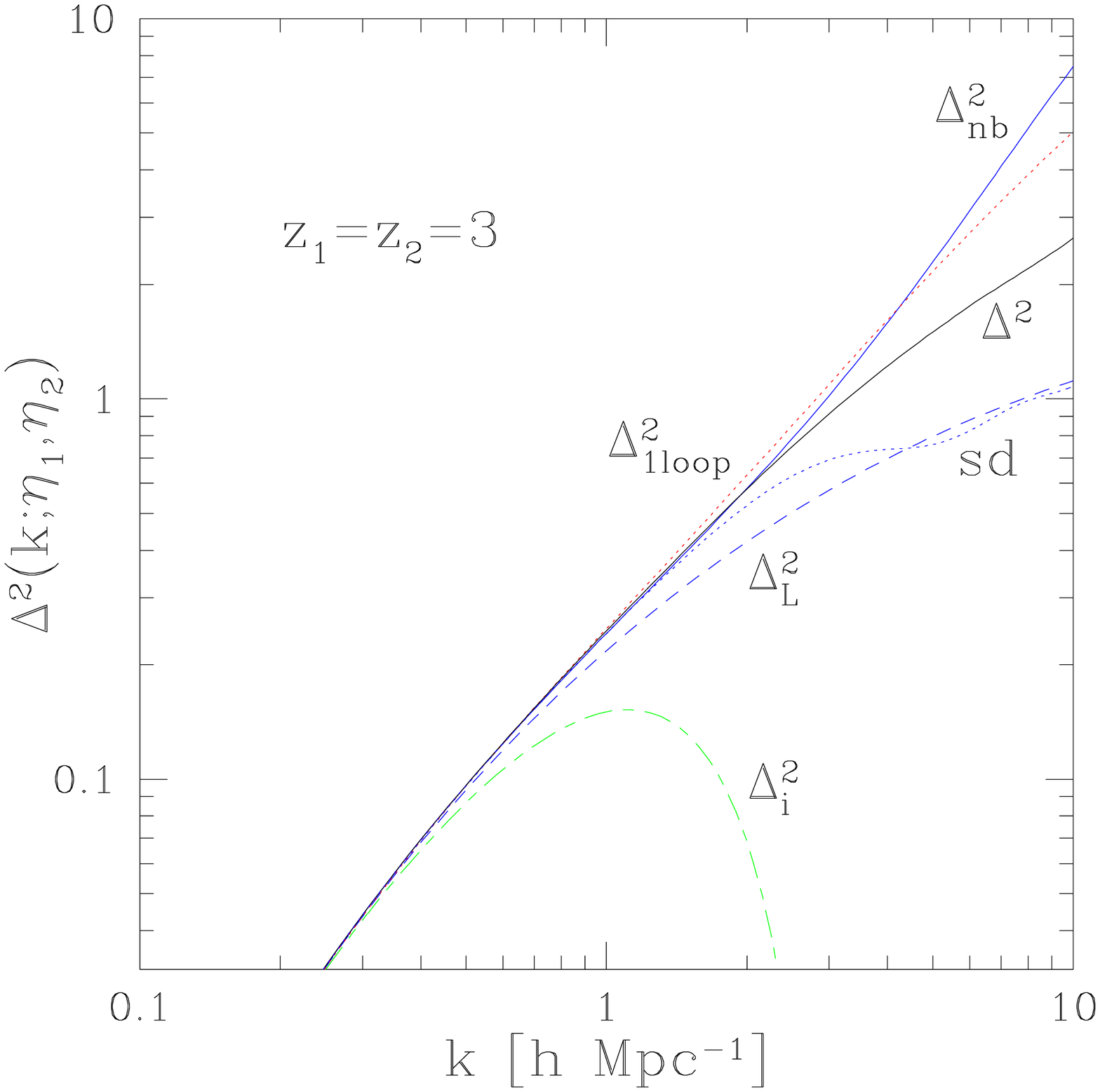}}
\end{center}
\caption{The logarithmic power $\Delta^2(k)$ at redshift $z=3$, that is
at equal times $z_1=z_2=3$.}
\label{figGkz3z3}
\end{figure}

\begin{figure}[htb]
\begin{center}
\epsfxsize=8 cm \epsfysize=7 cm {\epsfbox{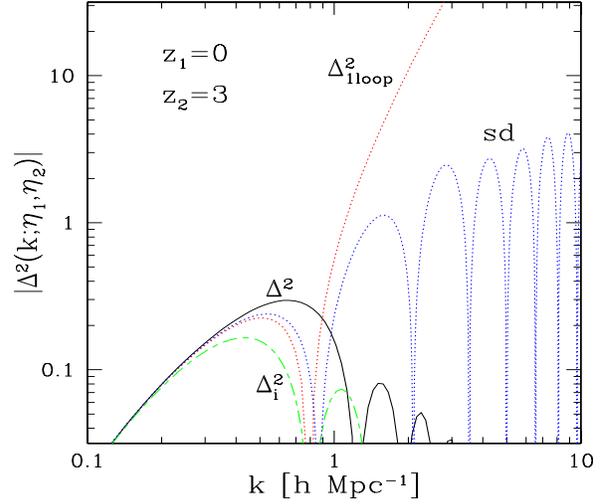}}
\end{center}
\caption{The logarithmic power $\Delta^2(k)$ at unequal times 
$(\eta_1=0,\eta_2=-1.4)$ (i.e. $z_1=0,z_2=3$).}
\label{figGkz0z3}
\end{figure}

Finally, we present in this section our results for the two-point correlation
$G$ and the self-energy $\Pi$. We first display in Fig.~\ref{figGt1} the
evolution with time $\eta_1$ of the density ``logarithmic power'' 
$\Delta^2(k;\eta_1,\eta_2)$ defined in eq.(\ref{Delta2}). As shown in the left
panel at early times and large scales we recover the steepest-descent result
since we must also match the usual one-loop expansion (\ref{P1loop}).
At later times or at smaller scales we obtained oscillations which are
strongly damped at large time separations in contrast with Fig.~\ref{figG0t1}. 
This difference of behaviors of the correlations $G$ between the direct 
steepest-descent method and the 2PI effective action scheme follows from the 
difference already seen in terms of the response $R$ analyzed in 
sect.~\ref{Response-R-and-self-energy-Sigma}.
The right panel of Fig.~\ref{figGt1} is particularly interesting as it shows
that in the highly non-linear regime the two-point correlation exhibits a peak
at equal times $\eta_1=\eta_2$ but whereas both the standard one-loop
expansion and the steepest-descent method yield large oscillations 
at larger time separations the 2PI effective action approach leads to a
strong damping. Thus, this is the only method (among those presented in this
paper) which yields (at one-loop order)
a damping of correlations at small scales and large time separations which
is expected to reflect the qualitative behavior of the exact gravitational 
dynamics.

Next, we show in Figs.~\ref{figGkz0z0}, \ref{figGkz3z3} the logarithmic power 
$\Delta^2(k;\eta_1,\eta_2)$ at equal redshifts $z_1=z_2=0$ and $z_1=z_2=3$
as a function of wavenumber $k$.
We find again that our results match the steepest-descent prediction as well
as the usual one-loop power (\ref{P1loop}) at large scales.
At small scales, despite the damping at unequal times shown in 
Fig.~\ref{figRk} for the response function and in Fig.~\ref{figGt1} for the
correlation function we obtain a steady growth of the power $\Delta^2(k)$
in between the linear prediction $\Delta_L^2$ and the usual one-loop 
prediction $\Delta^2_{\rm 1 loop}$. Besides, the agreement with the results
from numerical simulations is better than for the steepest-descent prediction.
Of course, because of the damping of the response $R$ at large time 
separations analyzed above we can check that the contribution $\Delta^2_i$ 
from eq.(\ref{GRPiR}) associated with the transport of the initial fluctuations
becomes negligible in the non-linear regime (contrary to what happens within
the steepest-descent approach, see Figs.~\ref{figG0kz0z0}, \ref{figG0kz3z3}).
However, the continuous generation of fluctuations, described by the
self-energy $\Pi$ in eq.(\ref{GPi}) and associated with the non-linearity
of the dynamics, sustains a high level of density fluctuations $\Delta^2(k)$
into the non-linear regime. The latter seen at equal times 
$\eta_1=\eta_2=\eta$ are related to the values of $\Pi(k;\eta_1',\eta_2')$
at nearly equal times $\eta_1'\simeq\eta_2'\simeq\eta$.

We show in Fig.~\ref{figGkz0z3} the logarithmic power 
$\Delta^2(k;\eta_1,\eta_2)$ at unequal times as a function of wavenumber $k$.
We find again that our result matches the steepest-descent prediction as well
as the usual one-loop power (\ref{P1loop}) at large scales.
However, the correlation now quickly decays at small scales in the non-linear
regime. This follows from the damping at large time separations analyzed above
for the response function which leads to a decorrelation at small scales and
large time separations.

\begin{figure}[htb]
\begin{center}
\epsfxsize=8 cm \epsfysize=7 cm {\epsfbox{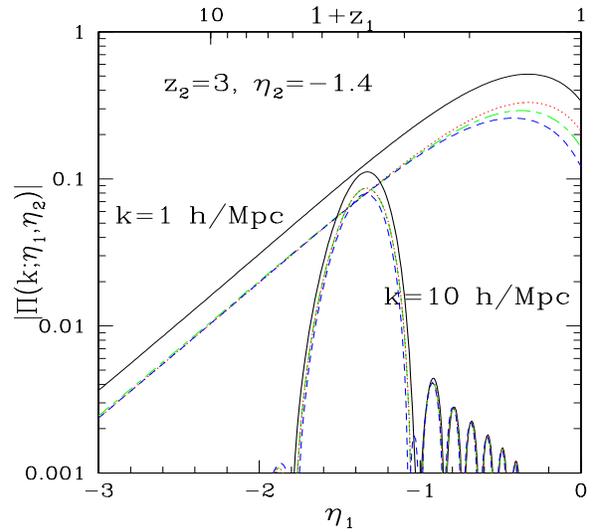}}
\end{center}
\caption{The self-energy $\Pi(k;\eta_1,\eta_2)$ as a function of 
forward time $\eta_1$, for $\eta_2=-1.4$ (i.e. $z_2=3$) and wavenumbers 
$k=1$ and $10 \times h$ Mpc$^{-1}$.}
\label{figPt1}
\end{figure}

\begin{figure}[htb]
\begin{center}
\epsfxsize=8 cm \epsfysize=7 cm {\epsfbox{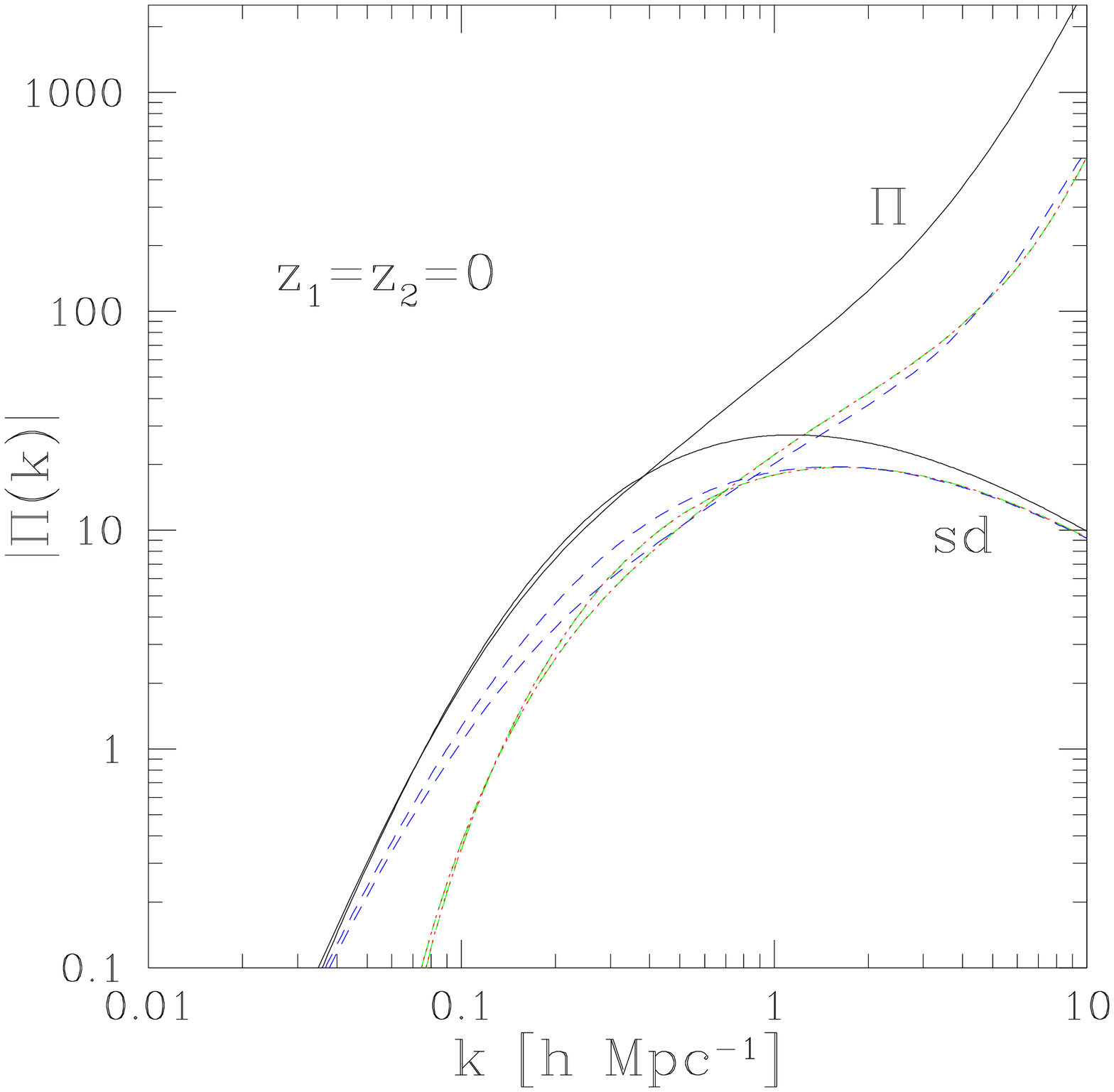}}
\end{center}
\caption{The self-energy $\Pi(k)$ as a function of wavenumber $k$ at equal
redshifts $z_1=z_2=0$. We also plot the prediction of the steepest-descent 
method (curves labeled ``sd'').}
\label{figPk_z0_z0}
\end{figure}

\begin{figure}[htb]
\begin{center}
\epsfxsize=8 cm \epsfysize=7 cm {\epsfbox{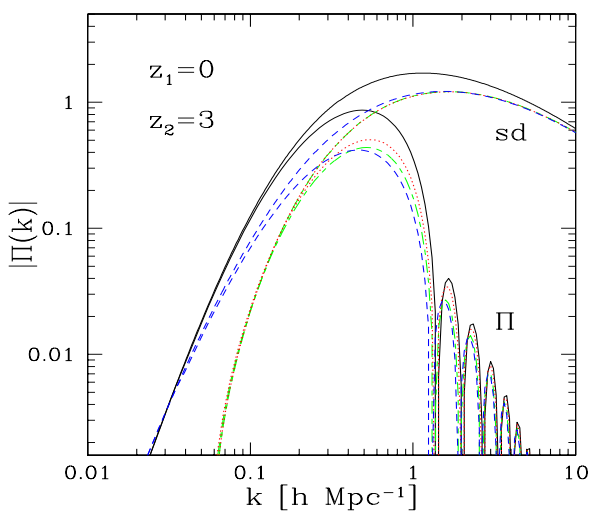}}
\end{center}
\caption{The self-energy $\Pi(k)$ as a function of wavenumber $k$ at unequal
redshifts $z_1=0,z_2=3$. We also plot the prediction of the steepest-descent 
method (curves labeled ``sd'').}
\label{figPk_z0_z3}
\end{figure}

We display in Fig.~\ref{figPt1} the self-energy $\Pi$ as a function 
of forward time $\eta_1$. We do not show the results of the steepest-descent
method which follow a mere exponential growth from eq.(\ref{Pi0})
At weakly non-linear scales ($k\sim 1 h$ Mpc$^{-1}$) we recover at early times
the exponential growth of the steepest-descent prediction whereas at late
times we obtain a fast decay. At highly non-linear scales ($k=10 h$ Mpc$^{-1}$)
we obtain a peak at equal times $\eta_1=\eta_2$ and strongly damped 
oscillations at large time separations. This is clearly consistent with the
results obtained for the correlation $G_{11}$ displayed in Fig.~\ref{figGt1}.

Next, we show in Fig.~\ref{figPk_z0_z0} the self-energy $\Pi$ at equal 
redshifts $z_1=z_2=0$ as a function of wavenumber $k$. We again recover the
steepest-descent prediction at large scales. However, at small scales we
now obtain a steady growth whereas the steepest-descent method yields a smooth
decline. Therefore, in agreement with Figs.~\ref{figGkz0z0}, \ref{figGkz3z3}
we find that the coupled system of equations (\ref{Peq}) (with $G_0$ replaced
by $G$) and (\ref{GPi}) entails at equal times a stronger growth of the 
correlation $G$ and of the self-energy $\Pi$ into the non-linear regime as 
compared with the non-coupled equations associated with the steepest-descent 
method.

Finally, we display in Fig.~\ref{figPk_z0_z3} the self-energy $\Pi$ at unequal 
redshifts $z_1=0,z_2=3$ as a function of wavenumber $k$. In agreement with 
our results for other two-point functions we recover the steepest-descent 
prediction at large scales and decaying oscillations at small scales.

\section{Simplified response approximations}
\label{Simplified-response-approximations}

The results obtained in the previous sections suggest that a convenient
approximation would be to use a simple analytical form for the response 
function $R$ and to compute the correlation $G$ and the self-energy $\Pi$
from eqs.(\ref{Peq}), (\ref{GPi}). This avoids the computation of the
self-energy term $\Sigma$ and the numerical integration of the
integro-differential eq.(\ref{Rforward}) which requires rather small 
time-steps. By contrast, eq.(\ref{GPi}) is a Volterra integral equation
of the second kind which is usually better behaved. We shall investigate
two such approximations associated with the direct steepest-descent approach
and the 2PI effective action scheme.

\subsection{Simplified response for the steepest-descent scheme}
\label{Simplified-response-for-the-steepest-descent-scheme}

Within the framework of the direct steepest-descent scheme studied in
sect.~\ref{Direct-steepest-descent} we found that the response function
exhibits in the non-linear regime a series of oscillations with an envelope
given by the linear theory prediction. Therefore, following eq.(\ref{rho0})
we consider the approximate response $R_{\rm app}$ defined as:
\beq
R_{\rm app}(x_1,x_2)= R_L(x_1,x_2) \cos[\omega(k)(a_1-a_2)] ,
\label{R0app}
\eeq
where $R_L(=R_0)$ is the linear response given in eq.(\ref{RL}), $a=e^{\eta}$
is the scale factor and $\omega(k)$ is given by eq.(\ref{omega}).

Here it is interesting to see how expression (\ref{R0app}) may be interpreted.
In the linear regime the density and velocity fluctuations $\psi_L$ are merely
amplified by a time-dependent factor (proportional to the scale-factor) which
implies in particular that the physics is exactly local: 
$\psi_L(\bx,\eta)$ only depends on the initial conditions (and a possible 
external noise) at the same location $\bx$. Thus in real space we have 
$R_L(\bx_1,\bx_2)\propto \delta_D(\bx_1-\bx_2)$ and in Fourier space,
using the factorization (\ref{Rk}) associated with statistical homogeneity,
$R_L(k)$ is constant with respect to $k$, in agreement with eq.(\ref{RL}). 
On the other hand, the cosine dependence of eq.(\ref{R0app}) yields in real
space:
\beq
R_{\rm app}(x_1,x_2) = R_L(a_1,a_2) \cP(\bx_1-\bx_2;a_1,a_2)
\label{R0appx}
\eeq
where $R_L(a_1,a_2)$ is the time-dependent factor of the linear response
(\ref{RL}) and we introduced the
distribution $\cP(\br;a_1,a_2)$ obtained from eq.(\ref{R0app}) as:
\beq
\cP(\br;a_1,a_2) = \frac{-1}{4\pi r} \delta_D'(r-d) \;\; 
\mbox{with} \;\; d= (a_1-a_2) r_0 
\label{P0d}
\eeq
where $r=|\br|$ and:
\beq
r_0= \left(\frac{4\pi}{3} \int \d q P_{L0}(q)\right)^{1/2} .
\label{r0}
\eeq
Thus, eqs.(\ref{R0app})-(\ref{r0}) may be interpreted as the displacement
of the linear fluctuations over a random distance $\br$ with a probability 
distribution $\cP(r)$ (whereas the linear 
regime obviously corresponds to the distribution $\delta_D(\br)$). 
This non-linear operation corresponds to some diffusion of the initial 
conditions (over angles of the shift $\br$) but
the singular distribution over the length $|\br|$ leads to the mere cosine
tail in Fourier space instead of a strong decay.
This can be contrasted to the Zeldovich approximation (Zeldovich 1970) where 
the trajectory of particles with initial comoving location $\bq$ is 
written as $\bx(\bq,a)= \bq+\br$ with $\nabla_{\bq}.\br=-\delta_L(\bq,a)$. 
Thus the 
displacement field is Gaussian which leads to a Gaussian decay at high $k$
(Crocce \& Scoccimarro 2006a).
However, we must emphasize that we only write eq.(\ref{R0appx}) for comparison
but this is not necessarily the best way to understand the result of the 
direct-steepest approach (at this stage expression (\ref{R0appx}) is only a 
mere re-writing of eq.(\ref{R0app})). In particular, there is no reason
why the non-linear response should be factorized as in eq.(\ref{R0appx})
as $R_L$ times a displacement term.

\begin{figure}[htb]
\begin{center}
\epsfxsize=8 cm \epsfysize=7 cm {\epsfbox{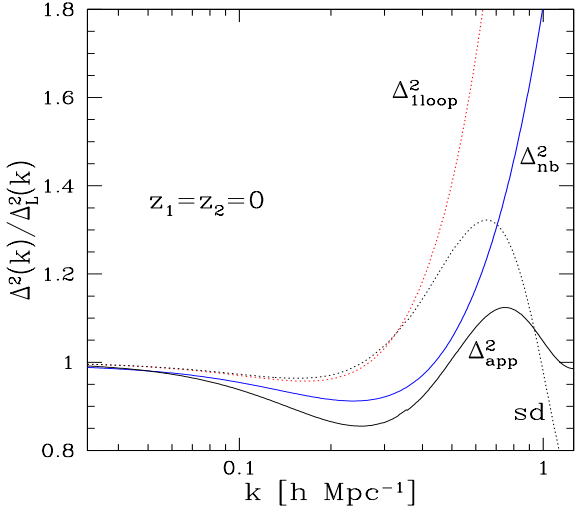}}
\end{center}
\caption{The logarithmic power $\Delta^2(k)$ at redshift $z=0$. 
We display the ratio $\Delta^2(k)/\Delta^2_L(k)$, where $\Delta^2_L$ is the 
linear power, for the usual one-loop perturbative result 
$\Delta^2_{\rm 1 loop}$ (dotted line), a fit $\Delta^2_{\rm nb}$ 
(upper solid line) from numerical simulations (Smith et al. 2003), the result 
of the steepest-descent method (``sd'', dotted line) shown in 
Fig.~\ref{figG0kz0z0} and the approximate power $\Delta^2_{\rm app}$ 
(solid line) obtained from the approximation (\ref{R0app}).}
\label{figG0sR_kz0z0}
\end{figure}

\begin{figure}[htb]
\begin{center}
\epsfxsize=8 cm \epsfysize=7 cm {\epsfbox{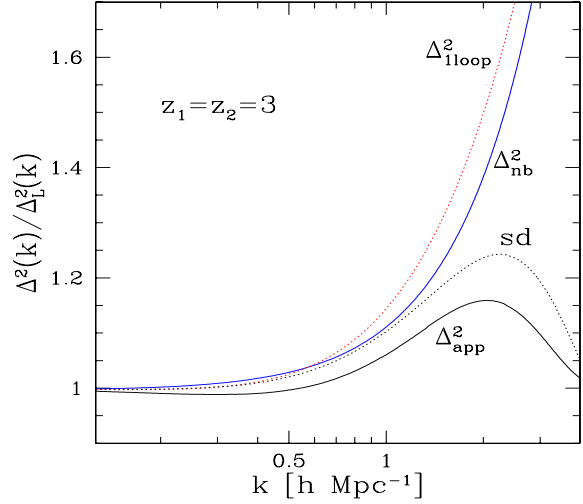}}
\end{center}
\caption{The ratio $\Delta^2(k)/\Delta^2_L(k)$ as in Fig.~\ref{figG0sR_kz0z0} 
but at redshift $z=3$.}
\label{figG0sR_kz3z3}
\end{figure}

Next, we use the steepest-descent prescription (\ref{Pi0})-(\ref{Pi0k})
for the self-energy $\Pi$ and we compute numerically the correlation $G$
from the explicit expression (\ref{GPi}). In fact, using eq.(\ref{Pi0}) the
integrals over time may be performed analytically which yields:
\beqa
G_{\rm app}(k;\eta_1,\eta_2) & = & G_L(k;\eta_1,\eta_2) \cos(\omega a_1)
\cos(\omega a_2) \nonumber \\
&& + a_1^2 a_2^2 R(\omega a_1).\Pi_0(k).R(\omega a_2)^T
\label{G0app}
\eeqa
with:
\beq
R(\omega a)= f_+(\omega a) R_0^+ + f_-(\omega a) R_0^- .
\label{Romega}
\eeq
The matrices $R_0^{\pm}$ are given in eqs.(\ref{R0pR0m}) and we
introduced the functions:
\beq
f_+(x)= \int_0^1 \d t \; \cos[x(1-t)] = \frac{\sin(x)}{x}
\label{fp}
\eeq
and:
\beqa
\lefteqn{f_-(x) = \int_0^1 \d t \; t^{5/2} \cos[x(1-t)] =  \frac{5}{2x^2} }
\nonumber \\
&& + \frac{15\sqrt{2\pi}}{8x^{7/2}} \left[ \cos(x) S(\sqrt{2x/\pi}) 
- \sin(x) C(\sqrt{2x/\pi}) \right] , 
\label{fm}
\eeqa
where $C(x)$ and $S(x)$ are the Fresnel integrals. At early times or large
scales the first term dominates in eq.(\ref{G0app}) and we recover the linear
regime $G\rightarrow G_L$. At late times and small scales we must take into
account both terms and for large $\omega a_i$ we obtain:
\beqa
\lefteqn{ \omega a_i \gg 1 : \;  G(k;a_1,a_2) \simeq 
a_1 a_2 \biggl \lbrace G_{L0}(k) \cos(\omega a_1)
\cos(\omega a_2) }\nonumber \\
&& + R_0^+.\Pi_0(k).R_0^{+T} \;
\frac{\sin(\omega a_1)\sin(\omega a_2)}{\omega^2} \biggl \rbrace .
\label{Gapp2}
\eeqa
Thus, we can see that at late times the non-linear correlation grows as
$G(k;a_1,a_2) \sim a_1 a_2$ with oscillations for $a_1\neq a_2$. This
agrees with the results presented in section~\ref{Two-point-correlation0}.

We show our results for the equal-time density correlation $G_{11}$ (i.e. the
power $\Delta^2(k)$) in Figs.~\ref{figG0sR_kz0z0}, \ref{figG0sR_kz3z3} at
redshifts $z=0$ and $z=3$. We display the ratio to the linear power 
$\Delta^2(k)/\Delta^2_L(k)$ to zoom over the weakly non-linear region and
we also plot the exact one-loop steepest-descent prediction (``sd'')
shown in Fig.~\ref{figG0kz0z0} for comparison. 
Figs.~\ref{figG0sR_kz0z0}, \ref{figG0sR_kz3z3} clearly show that the
exact one-loop steepest-descent prediction matches the standard one-loop
perturbative result at large scales. Moreover, Fig.~\ref{figG0sR_kz0z0}
suggests that at $z=0$ the fit from numerical simulations (Smith et al. 2003)
may be off by $10\%$ at $k \sim 0.3 h$ Mpc$^{-1}$. The power 
$\Delta^2_{\rm app}$ follows roughly the behavior of the ``exact'' one-loop 
steepest-descent prediction but it is smaller by up to $15\%$ and does
not match very well the one-loop perturbative result. This failure
at intermediate scales is not really surprising since eq.(\ref{rho0})
was derived in the large-$k$ limit using the asymptotic behaviors 
(\ref{S0k1121})-(\ref{S0k2212}). However, as can be seen in Fig.~\ref{figS0pk}
they are only reached for $k>10 h$ Mpc$^{-1}$. Therefore, the approximation
(\ref{R0app}) is not sufficient to obtain accurate results over weakly
non-linear scales. This shows that the results obtained for the two-point
correlation are rather sensitive to the approximations used for the response
$R$. In fact, since the direct steepest-descent method is quite 
simple and easy to compute the approximation (\ref{R0app}) is not very useful 
for precise quantitative predictions and it is best to use the rigorous method 
presented in sect.~\ref{Direct-steepest-descent}.

\subsection{Simplified response for the 2PI effective action scheme}
\label{Simplified-response-for-the-2PI-effective-action-scheme}

Within the 2PI effective action method we found in 
sect.~\ref{2PI-effective-action-approach}
that the response function exhibits oscillations as for the steepest-descent
prediction but their amplitude decays at large times. Following the analysis
of the simple model (\ref{Rtoydef}) and eq.(\ref{Rtoy2PI}) we consider the
approximate response $R_{\rm app}$ defined as:
\beq
R_{\rm app}(x_1,x_2)= R_L(x_1,x_2) \frac{J_1[2\omega(k)(a_1-a_2)]}
{\omega(k)(a_1-a_2)} .
\label{Rapp}
\eeq
Of course, this expression can again be written in the form (\ref{R0appx})
in real space, with a distribution $\cP(r)$ which is now given by:
\beq
\cP(|\br|;a_1,a_2) = \frac{\theta(r<2d)}{8\pi^2 d^3 \sqrt{1-(r/2d)^2}} ,
\label{Pd}
\eeq
where the top-hat factor $\theta(r<2d)$ obeys $\theta=1$ for $r<2d$ and 
$\theta=0$ for $r>2d$. The decay of the response function at high $k$ 
corresponds to the smoother distribution (\ref{Pd}) as compared with
eq.(\ref{P0d}), but the discontinuity at $r=2d$ leads to a slow power-law
damping instead of an exponential decay. Again, eq.(\ref{Pd}) is a mere
rewriting of eq.(\ref{Rapp}) and we do not claim here that the distribution
$\cP(r)$ is anything more than a phenomenological tool.

Then, we use eq.(\ref{Peq}) (with $G_0$ replaced by $G$) and eq.(\ref{GPi}) 
to compute the self-energy $\Pi$ and the correlation $G$, following the
procedure described in sect.~\ref{2PI-effective-action-approach}.

\begin{figure}[htb]
\begin{center}
\epsfxsize=8 cm \epsfysize=7 cm {\epsfbox{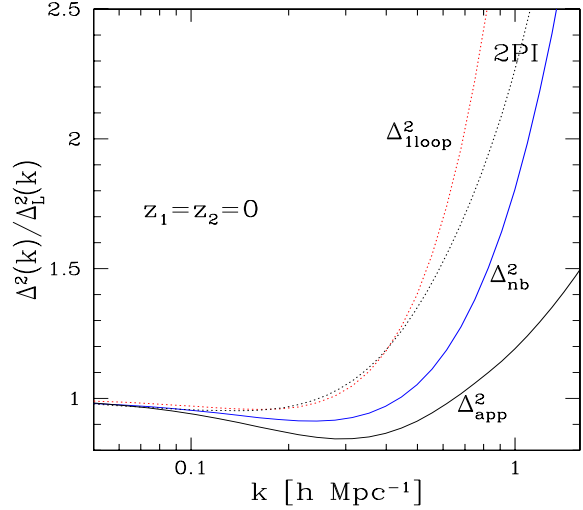}}
\end{center}
\caption{The logarithmic power $\Delta^2(k)$ at redshift $z=0$. 
We display the ratio $\Delta^2(k)/\Delta^2_L(k)$, where $\Delta^2_L$ is the 
linear power, for the usual one-loop perturbative result 
$\Delta^2_{\rm 1 loop}$ (dotted line), a fit $\Delta^2_{\rm nb}$ 
(upper solid line) from numerical simulations (Smith et al. 2003), the result 
of the 2PI effective action method (``2PI'', dotted line) shown in 
Fig.~\ref{figGkz0z0} and the approximate power $\Delta^2_{\rm app}$ 
(solid line) obtained from the approximation (\ref{Rapp}).}
\label{figGsR_kz0z0}
\end{figure}

\begin{figure}[htb]
\begin{center}
\epsfxsize=8 cm \epsfysize=7 cm {\epsfbox{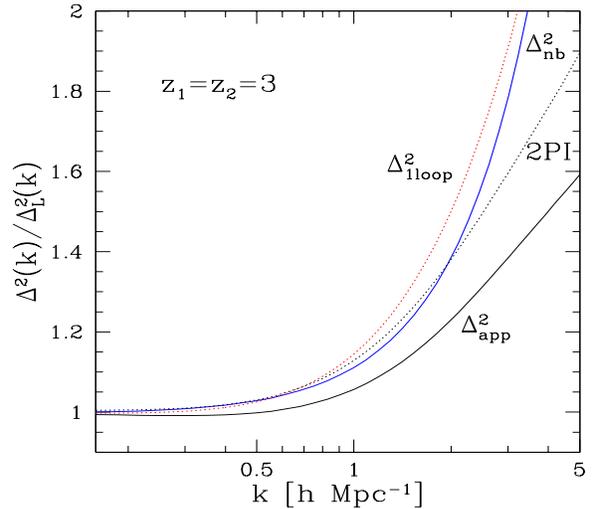}}
\end{center}
\caption{The ratio $\Delta^2(k)/\Delta^2_L(k)$ as in Fig.~\ref{figGsR_kz0z0} 
but at redshift $z=3$.}
\label{figGsR_kz3z3}
\end{figure}

We show our results for the power $\Delta^2(k)$) in 
Figs.~\ref{figGsR_kz0z0}, \ref{figGsR_kz3z3} at redshifts $z=0$ and $z=3$. 
We again display the ratio to the linear power $\Delta^2(k)/\Delta^2_L(k)$ and
we also plot the exact 2PI effective action prediction (``2PI'')
shown in Fig.~\ref{figGkz0z0} for comparison. 
Figs.~\ref{figGsR_kz0z0}, \ref{figGsR_kz3z3} again clearly show that the
exact 2PI effective action prediction matches the standard one-loop
perturbative result at large scales. In a fashion similar to the 
approximation (\ref{R0app}) we find that the power $\Delta^2_{\rm app}$ 
follows roughly the behavior of the exact one-loop 2PI effective action
prediction but it is smaller by up to $15\%$ and does not match very well 
the one-loop perturbative result. This shows again that the two-point 
correlation is rather sensitive to the value of the response function.

\subsection{Mixed approaches}
\label{Mixed-approaches}

\begin{figure}[htb]
\begin{center}
\epsfxsize=8 cm \epsfysize=7 cm {\epsfbox{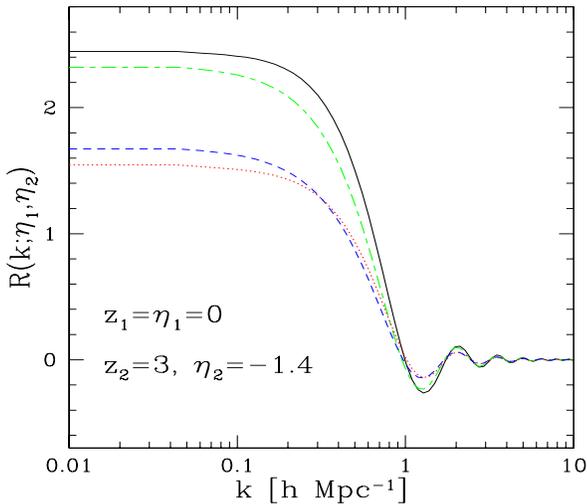}}
\end{center}
\caption{The non-linear response function $R(k;\eta_1,\eta_2)$ as a function 
of wavenumber $k$, at times $\eta_1=0,\eta_2=-1.4$, for the mixed approaches
where $G_0$ is used in the computation of $\Sigma$.}
\label{figRG0k}
\end{figure}

Finally, we consider in this section mixed approaches which use elements from
both the steepest-descent and 2PI effective action schemes.
The idea is to simplify the computation involved in the
2PI effective action method by separating the computations of the pairs
$(\Sigma,R)$ and $(\Pi,G)$. That is, we first compute the self-energy $\Sigma$
and the response $R$ from the coupled eqs.(\ref{Rforward}), (\ref{Seq}),
where we use $G_0$ and $R$ in the expression (\ref{Seq}) for $\Sigma$
(the steepest-descent scheme uses $G_0$ and $R_0$ whereas the 2PI effective 
action scheme uses $G$ and $R$). Thus, the pair $(\Sigma,R)$ becomes
independent of $(\Pi,G)$ and can be computed in a first step.
Since we keep the explicit dependence on $R$ in the self-energy $\Sigma$
the evolution equation (\ref{Rforward}) for the response function is no longer
linear but quadratic, hence we expect to recover the damping discussed in
sect.~\ref{Response-R-and-self-energy-Sigma} (see the analysis of 
eq.(\ref{Rtoydef})).
We display our results in Fig.~\ref{figRG0k} which shows that the non-linear
response exhibits indeed a damping close to the one obtained in 
Fig.~\ref{figRk} for the full 2PI effective action scheme.

Next, we compute the pair $(\Pi,G)$ from the equations (\ref{Peq}) and 
(\ref{GPi}). Here we can consider two procedures as we may use either 
$G_0$ or $G$ in the computation of $\Pi$.
Clearly, these mixed schemes still agree with the usual one-loop
perturbative results. These two-step approaches are slightly faster than the
full 2PI effective action method for numerical computations. Indeed, as 
explained in sect.~\ref{2PI-effective-action-approach}, within the 2PI 
effective action approach the equations obtained for two-point functions and
self-energies form a coupled system of non-linear equations. This leads to
an iterative scheme for their numerical computation (which converges within
$\sim 7$ steps at worst for the scales we consider here). On the other hand,
at each step the computation of the correlation $G$ takes most of the computer 
time (especially at late times) because of the double integral over times
in eq.(\ref{GPi}) but it converges faster (at fixed $R$) than the response 
$R$ (at fixed $G$). Therefore, it saves time to separate the computations
of $R$ (many fast iteration steps) and $G$ (few long iteration steps).

\begin{figure}[htb]
\begin{center}
\epsfxsize=8 cm \epsfysize=7 cm {\epsfbox{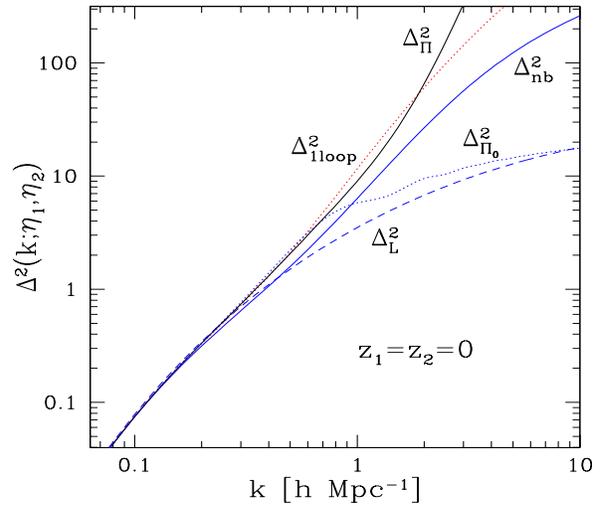}}
\end{center}
\caption{The logarithmic power $\Delta^2(k)$ at redshift $z=0$, that is
at equal times $z_1=z_2=0$. We display the linear power $\Delta^2_L$ 
(dashed line), the usual one-loop perturbative result $\Delta^2_{\rm 1 loop}$ 
(dotted line), a fit $\Delta^2_{\rm nb}$ (lower solid line) from numerical 
simulations (Smith et al. 2003), the mixed approach $\Delta^2_{\Pi_0}$ based 
on $\Pi_0$ (dashed line) and the mixed approach $\Delta^2_{\Pi}$ where 
$\Pi$ is coupled to $G$ (upper solid line).}
\label{figlGG0kz0z0}
\end{figure}

\begin{figure}[htb]
\begin{center}
\epsfxsize=8 cm \epsfysize=7 cm {\epsfbox{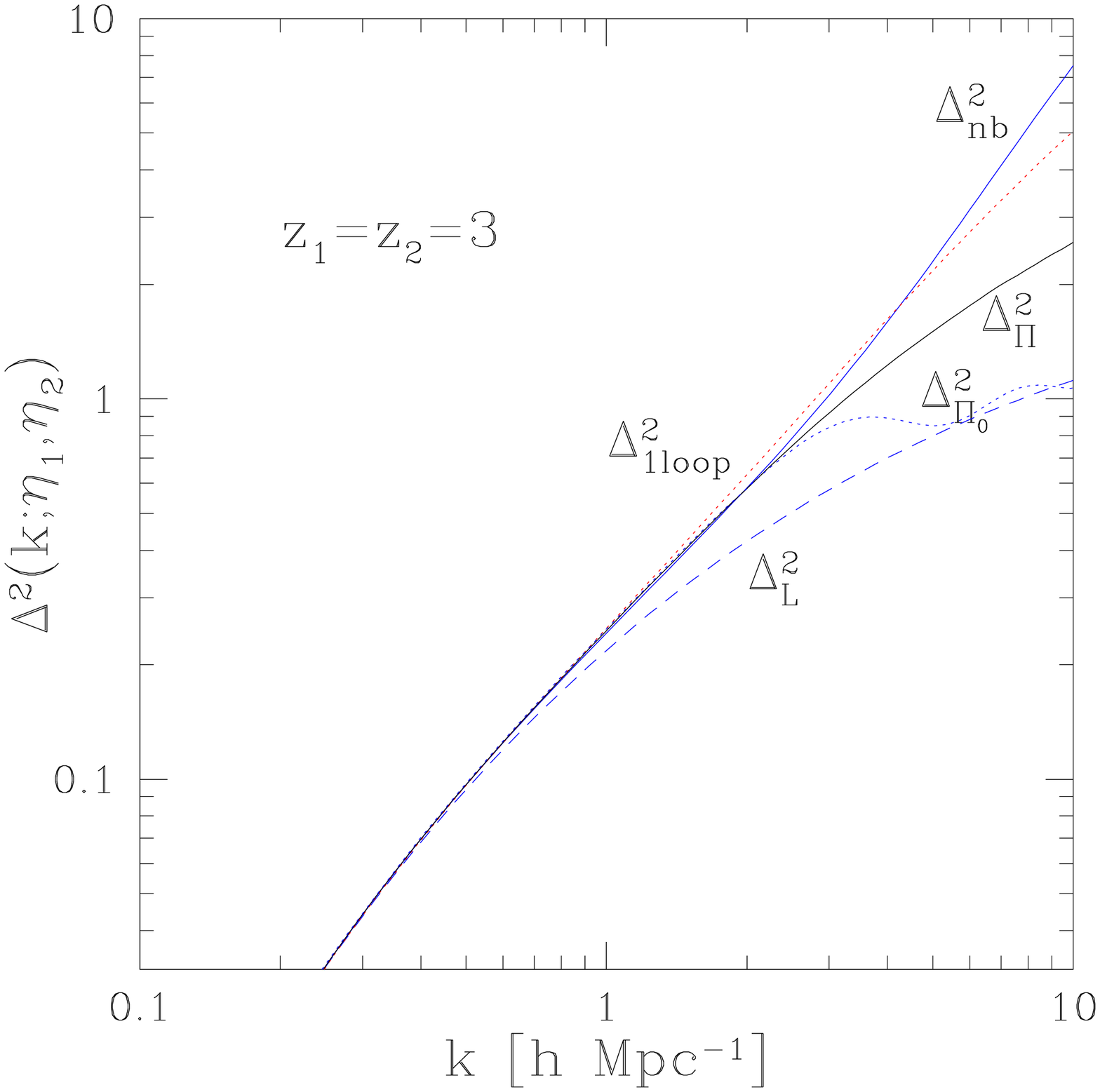}}
\end{center}
\caption{The logarithmic power $\Delta^2(k)$ at redshift $z=3$, that is
at equal times $z_1=z_2=3$.}
\label{figlGG0kz3z3}
\end{figure}

\begin{figure}[htb]
\begin{center}
\epsfxsize=8 cm \epsfysize=7 cm {\epsfbox{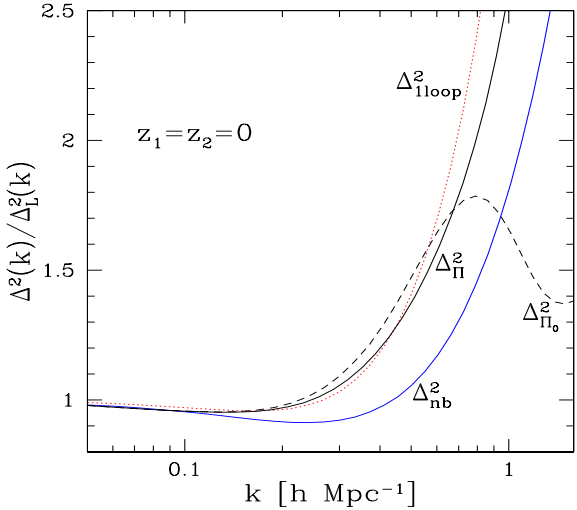}}
\end{center}
\caption{The ratio $\Delta^2(k)/\Delta^2_L(k)$ at redshift $z=0$ of the 
logarithmic power $\Delta^2$ to the linear power $\Delta^2_L$.
We display the ratio $\Delta^2/\Delta^2_L$ for the usual one-loop 
perturbative result $\Delta^2_{\rm 1 loop}$ (dotted line), a fit 
$\Delta^2_{\rm nb}$ (lower solid line) from numerical simulations 
(Smith et al. 2003), the mixed approach $\Delta^2_{\Pi_0}$ based on $\Pi_0$ 
(dashed line) and the mixed approach $\Delta^2_{\Pi}$ where $\Pi$ is coupled
to $G$ (upper solid line).}
\label{figGG0_kz0z0}
\end{figure}

\begin{figure}[htb]
\begin{center}
\epsfxsize=8 cm \epsfysize=7 cm {\epsfbox{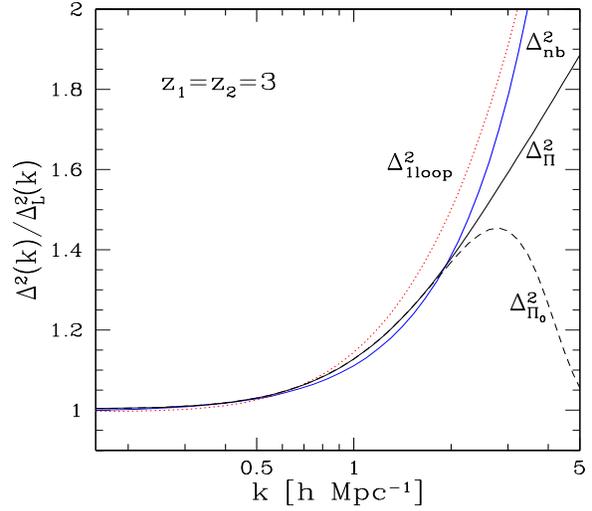}}
\end{center}
\caption{The ratio $\Delta^2(k)/\Delta^2_L(k)$ as in Fig.~\ref{figGG0_kz0z0} 
but at redshift $z=3$.}
\label{figGG0_kz3z3}
\end{figure}

We display in Figs.~\ref{figlGG0kz0z0}-\ref{figGG0_kz3z3}
our results for the power $\Delta^(k)$ at redshifts $z=0$ and $z=3$ for
both schemes. The dashed curve $\Delta^2_{\Pi_0}$ shows the power obtained
when the self-energy term $\Pi_0$ is used (as in the steepest-descent
approach) so that the correlation $G$ is directly given by the explicit
expression (\ref{GPi}). The solid curve $\Delta^2_{\Pi}$ corresponds to
the case where we use the non-linear correlation $G$ in the expression
(\ref{Peq}) of the self-energy, so that eqs.~(\ref{Peq}) and (\ref{GPi})
are a coupled non-linear system which we solve iteratively (as in the
2PI effective action scheme).
We can see that both methods exhibit a similar behavior until 
$\Delta^2(k)/\Delta^2_L(k)\sim 1.4$. Farther into the non-linear regime
the ``$\Pi_0$-scheme'' breaks from the non-linear growth of the power-spectrum
and shows oscillations which converge close to the linear power,
in a manner reminiscent of the steepest-descent
approximation, whereas the coupled ``$\Pi$-scheme'' follows this non-linear 
growth. On the other hand, in the weakly non-linear regime displayed in  
Figs.~\ref{figGG0_kz0z0}-\ref{figGG0_kz3z3} the predicted power $\Delta^2(k)$
is suppressed as compared with the usual one-loop result. Depending on the
redshift this can improve or worsen the global agreement with numerical 
simulations. 
However, Fig.~\ref{figGG0_kz3z3} suggests that when the usual one-loop 
result shows a better global agreement with simulations it is a mere 
coincidence as the simulation result separates last from the large-$N$
predictions (and the closer agreement with the usual one-loop result
at smaller scales has no theoretical basis). Nevertheless, the limited
accuracy of the numerical simulations prevents one from drawing definite
conclusions.
We can note in Fig.~\ref{figlGG0kz0z0} that in the highly non-linear regime 
the approximation $\Delta^2_{\Pi}$ grows larger than for the full 2PI 
effective action prediction displayed in Fig.~\ref{figGkz0z0}. Indeed,
in the latter case the non-linear response is further damped by the
strong growth of the two-point correlation $G$ (this would correspond to
a larger $\sigma$ whence $\omega$ in the simple model (\ref{Rtoy2PI})) and
this ``negative feedback'' within the coupled system $R,G$ leads to a smaller
$G$ as compared with the approximation $\Delta^2_{\Pi}$ where this feedback
is neglected. In fact, at very high $k$ (beyond the range shown in 
Fig.~\ref{figlGG0kz0z0}) the power $\Delta^2_{\Pi}$ appears to diverge. It is
not clear whether this is due to the finite resolution of the numerical 
computation or to a true divergence. However, for practical purposes this is
not a serious shortcoming since it occurs at small scales beyond the range of
validity of this approach (one would need to go beyond one-loop order and
probably beyond the hydrodynamical approach as shell-crossing can no longer
be neglected).

The good agreement in the weakly non-linear regime of these various schemes
enables one to evaluate their range of validity without the need to compare
their prediction with $N-$body simulations, which could be of great practical
interest. In fact, Fig.~\ref{figGG0_kz0z0} suggests that this procedure
can provide a better accuracy than numerical simulations in this regime,
where the latter may still be off by $10\%$.

\section{Application to a $\Lambda$CDM universe}
\label{LCDM}

We finally describe in this section how the results obtained in the previous
sections can be applied to more general cosmologies, such as a 
$\Lambda$CDM universe. As seen in sect.~\ref{Fluid} our formalism applies
equally well to any cosmology. We only need to use the relevant matrix $\cO$,
that is to take into account the time dependence of the ratio $\Om/f^2$ in
eq.(\ref{Odef}), and to recall that $\eta$ is the logarithm of the linear
growth factor, see eq.(\ref{eta}). In practice, a widely used approximation
is to take $\Om/f^2 \simeq 1$ so that the results obtained for the 
critical-density universe directly apply to the case $\Lambda\neq 0$ and the
only change is the time-redshift relation $\eta \leftrightarrow z$. 
We shall use this simple approximation here and compare our results to
numerical simulations. Thus, we consider the cosmological parameters
$\Om=0.3, \OL=0.7, \sigma_8=0.9$ and $H_0=70$ km.s$^{-1}$.Mpc$^{-1}$
as in Smith et al.(2003).

\begin{figure}[htb]
\begin{center}
\epsfxsize=8 cm \epsfysize=7 cm {\epsfbox{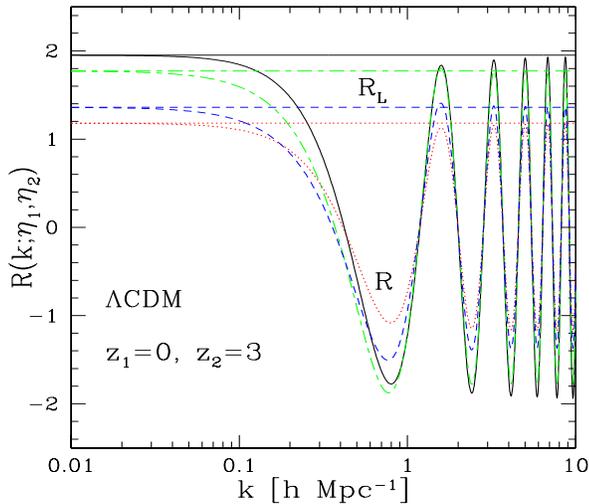}}
\end{center}
\caption{The non-linear response function $R(k;\eta_1,\eta_2)$ given by the
direct steepest-descent method as a function of wavenumber $k$ for a 
$\Lambda$CDM universe, at redshifts $z_1=0,z_2=3$. 
The horizontal lines are the linear response $R_L$ of eq.(\ref{RL}) which 
is independent of $k$.}
\label{figLCDM_R0k}
\end{figure}

\begin{figure}[htb]
\begin{center}
\epsfxsize=8 cm \epsfysize=7 cm {\epsfbox{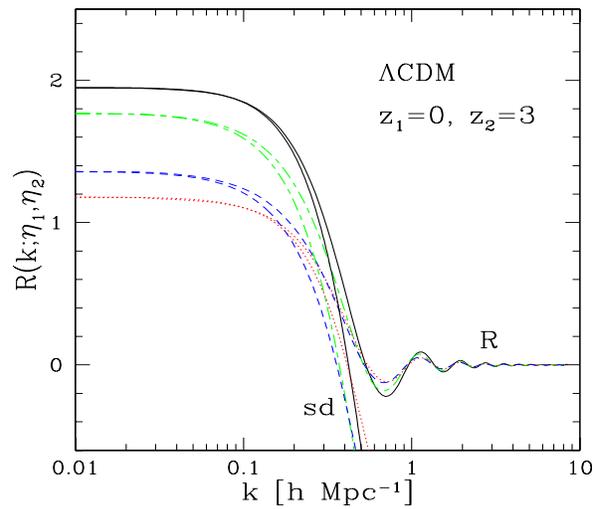}}
\end{center}
\caption{The non-linear response function $R(k;\eta_1,\eta_2)$ given by the
2PI effective action method as a function of wavenumber $k$ for a 
$\Lambda$CDM universe, at redshifts $z_1=0,z_2=3$.
We also plot the first half-oscillation of the response obtained
from the direct steepest-descent method (curves labeled ``sd'') which was 
shown in Fig.~\ref{figLCDM_R0k}.}
\label{figLCDM_Rk}
\end{figure}

We first plot in Figs.~\ref{figLCDM_R0k},\ref{figLCDM_Rk} the non-linear 
response functions we obtain within the one-loop steepest-descent
and 2PI effective action methods. We can check that we recover the behavior
obtained for the critical-density universe displayed in 
Figs.~\ref{figR0k},\ref{figRk}. The steepest-descent approach yields again
oscillations in the non-linear regime with an amplitude given by the
linear response at high $k$ whereas the 2PI effective action method
gives damped oscillations. The transition to the non-linear regime
occurs at slightly smaller wavenumber because of the larger normalization 
$\sigma_8$ of the linear power-spectrum.

\begin{figure}[htb]
\begin{center}
\epsfxsize=8 cm \epsfysize=7 cm {\epsfbox{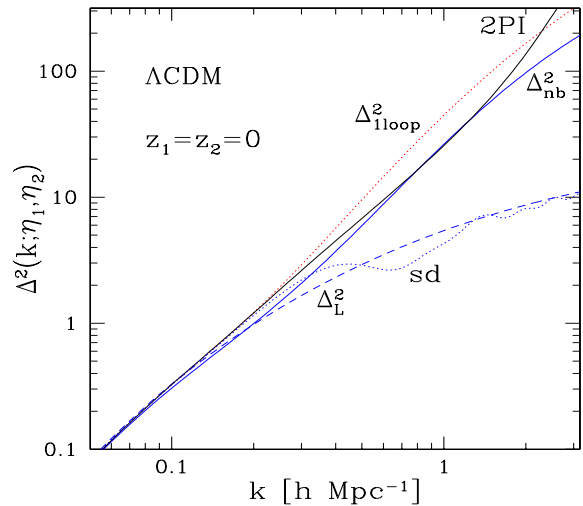}}
\end{center}
\caption{The logarithmic power $\Delta^2(k)$ at redshift $z=0$ for a
$\Lambda$CDM universe. We display the linear power $\Delta^2_L$ 
(dashed line), the usual one-loop perturbative result $\Delta^2_{\rm 1 loop}$ 
(dotted line), a fit $\Delta^2_{\rm nb}$ (lower solid line) from numerical 
simulations (Smith et al. 2003), the steepest-descent prediction (dotted line
``sd'') and the 2PI effective action result (solid line ``2PI'').}
\label{figLCDM_lGkz0z0}
\end{figure}

\begin{figure}[htb]
\begin{center}
\epsfxsize=8 cm \epsfysize=7 cm {\epsfbox{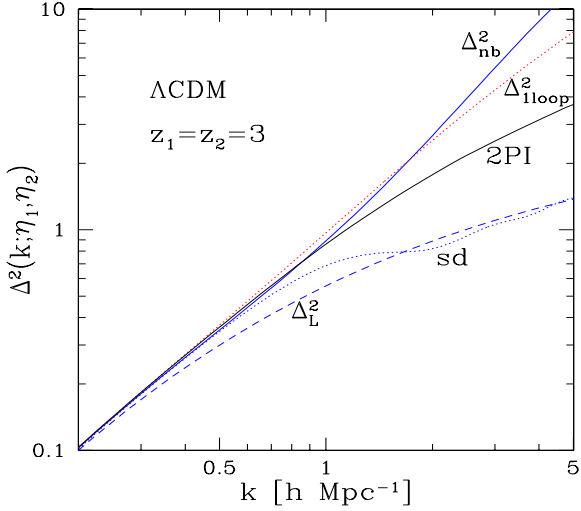}}
\end{center}
\caption{The logarithmic power $\Delta^2(k)$ at redshift $z=3$ for a
$\Lambda$CDM universe.}
\label{figLCDM_lGkz3z3}
\end{figure}

\begin{figure}[htb]
\begin{center}
\epsfxsize=8 cm \epsfysize=7 cm {\epsfbox{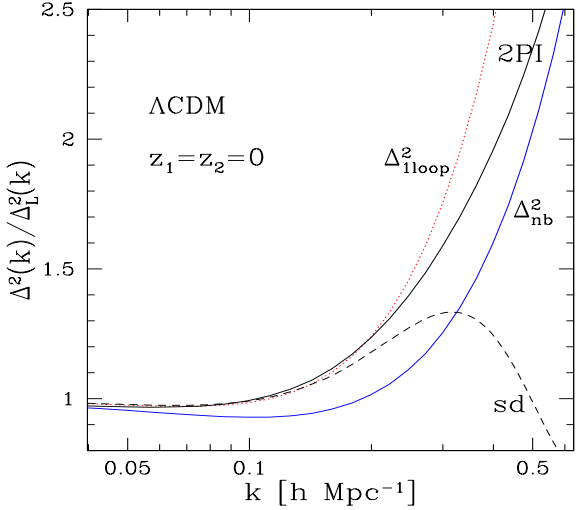}}
\end{center}
\caption{The ratio $\Delta^2(k)/\Delta^2_L(k)$ at redshift $z=0$ of the 
logarithmic power $\Delta^2$ to the linear power $\Delta^2_L$ for a
$\Lambda$CDM universe.
We display the ratio $\Delta^2/\Delta^2_L$ for the usual one-loop 
perturbative result $\Delta^2_{\rm 1 loop}$ (dotted line), a fit 
$\Delta^2_{\rm nb}$ (lower solid line) from numerical simulations 
(Smith et al. 2003), the steepest-descent prediction (dashed line ``sd'')
and the 2PI effective action result (solid line ``2PI'').}
\label{figLCDM_Gkz0z0}
\end{figure}

\begin{figure}[htb]
\begin{center}
\epsfxsize=8 cm \epsfysize=7 cm {\epsfbox{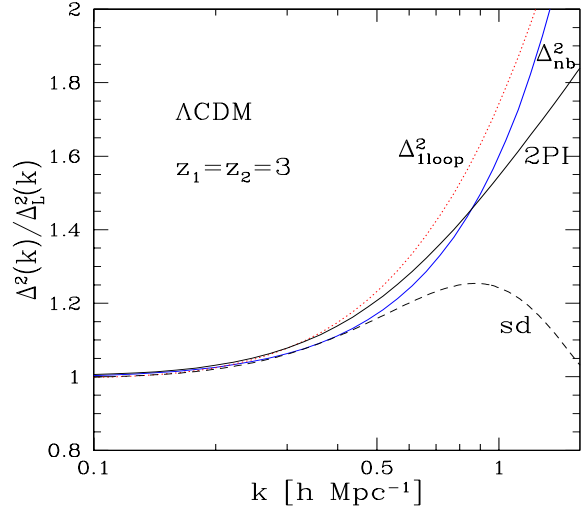}}
\end{center}
\caption{The ratio $\Delta^2(k)/\Delta^2_L(k)$ as in 
Fig.~\ref{figLCDM_Gkz0z0} but at redshift $z=3$.}
\label{figLCDM_Gkz3z3}
\end{figure}

We show in Figs.~\ref{figLCDM_lGkz0z0}-\ref{figLCDM_Gkz3z3} our results
for the two-point correlation $G_{11}$, that is the non-linear density
power spectrum. We again recover the behavior analyzed in previous sections
for the critical-density universe. Both large-$N$ expansion schemes agree
with the standard one-loop expansion at large scales. At small scales the 
steepest-descent result goes back to values close to the linear power
$\Delta^2_L$ whereas the 2PI effective action result keeps growing in a manner
similar to the usual perturbative result and numerical simulations.
The degree of agreement between the $N-$body simulations and either one
of the standard and 2PI results depends on redshift, that is on the slope of
the linear power-spectrum at the relevant scales. 

Thus, the large-$N$ expansions developed in this article apply in the same
manner for different cosmologies and exhibit the same properties.

\section{Conclusion}
\label{Conclusion}

In this article we have studied the predictions at one-loop order of large-$N$
expansions for the two-point correlations associated with the formation
of large-scale structures in the expanding universe. Focussing on weakly
non-linear scales we have used the hydrodynamical equations of motion
as a starting point. Then, we have recalled the path-integral formalism
which describes the statistical properties of the system, assuming Gaussian
initial conditions, and we have presented two possible large-$N$ expansions.
Next, we have described in details the numerical predictions obtained for
a SCDM cosmology.

First, we have presented a direct steepest-descent approach where the 
self-energy is expressed in terms of the linear response and correlation. 
This allows simple computations as the time-dependence of the 
self-energy can be factorized (for a critical density universe). Moreover, 
the non-linear response $R$ can be computed directly from this self-energy as 
it is not coupled to the non-linear correlation $G$ and it obeys a simple
linear equation. This also allows a detailed analysis of its behavior. 
We have found that at one-loop order this 
approach yields a non-linear response which exhibits strong oscillations
in the non-linear regime (small scales or late times) with an amplitude
which is given by the linear response. Therefore, the response does not
decay but only exhibits increasingly fast oscillations of the form
$R_L \cos[(a_1-a_2)k/k_*]$. However, once the response is integrated over
time such oscillations lead to some damping as compared with the linear 
response. Then, we have computed the non-linear correlation which can be
explicitly written in terms of this non-linear response which transports
forward over time the initial density and velocity fluctuations
as well as the fluctuations generated at all previous times by non-linear
couplings. We have checked that in the quasi-linear regime the non-linear 
correlation agrees with the usual one-loop prediction (obtained from the
standard perturbative expansion over powers of the initial fluctuations).
In the highly non-linear regime the correlation separates from the
steep growth displayed by the standard one-loop result and converges
back to the linear amplitude. Unfortunately, this behavior does not improve
the agreement with $N-$body numerical simulations. Nevertheless, it suggests
that this large-$N$ expansion is well-behaved as the two-point functions
do not ``explode'' in the highly non-linear regime. Contrary to the
standard perturbative expansion, where higher-order terms grow as increasingly
large powers of $k$ and $a(t)$, the higher-order terms obtained within this
method may remain of finite amplitude.  

Secondly, we have described a 2PI effective action approach where all
two-point functions are coupled. The numerical computation requires an 
iterative procedure but thanks to causality the equations can be solved
by integrating forward over time with the help of a few iterations at each
time-step (keeping an implicit scheme in the non-linear integro-differential 
equations). Then, we have found that the one-loop non-linear response again 
exhibits fast oscillations in the non-linear regime but its amplitude now 
shows a fast decay. We have explained how this damping is produced by the
non-linearity of the evolution equation obeyed by the response function. 
This yields
a power-law decay such as $R_L J_1[2(a_1-a_2)k/k_*]/[(a_1-a_2)k/k_*]$.
In the quasi-linear regime the non-linear correlation agrees again with the 
usual one-loop prediction but in the highly non-linear regime it no longer
converges back to the linear power. It rather shows a moderate growth
intermediate between the linear prediction and the result of numerical
simulations, with an amplitude which depends on the redshift (whence on the
slope of the linear power-spectrum). Therefore, this method appears to be
superior to the direct steepest-descent approach which could not recover
any non-linear damping (at one-loop order) but it requires more complex
numerical computations.

Thirdly, we have described other approximation schemes built from these
two approaches. Thus, we have found that using a simple explicit approximation
for the non-linear response within either large-$N$ scheme gives correct
qualitative results but does not manage to reproduce accurately the
quantitative results obtained from the full large-$N$ approaches. Next,
we have investigated the approximations obtained by separating the
computations of the response and the correlation (but keeping a non-linear
dynamics for the response). As expected we have found that this yields
a non-linear response which closely follows the damping obtained by the
full 2PI effective action approach and it gives good results (as compared
with the full large-$N$ schemes) for the correlation.

Finally, we have shown that these results also apply to more general 
cosmologies such as a $\Lambda$CDM universe, since our formalism extends
to any expansion history of the Universe provided we use the correct
time-redshift relation and linear growth factors.

Thus, we have shown in this article that large-$N$ expansions provide an
interesting systematic approach to the formation of large-scale structures
in the expanding universe. They already show at one-loop order some damping
in the non-linear regime for the response function and the amplitude of 
the two-point correlation remains well-behaved. Unfortunately, for practical
purposes the accuracy provided by these large-$N$ expansions at one-loop order
is not significantly better than the usual perturbative result for weakly 
non-linear scales. Nevertheless, the improvement may become greater at higher 
orders since the expansion is likely to be better behaved, but this requires
rather complex computations which are beyond the scope of this paper.

On the other hand, one of the goals of this article was to describe a different
approach to the problem of non-linear gravitational clustering than the standard
expansion schemes, namely a path-integral formalism. An important feature of 
this approach is that the statistical nature of the problem is already included
in the definition of the system, that is the action $S[\psi,\lambda]$ describes
both the equations of motion (here in the hydrodynamical limit) and the average
over Gaussian initial conditions. This also means that one directly works with
correlation functions (such as the power spectrum). By contrast, in usual
expansion schemes (or in $N$-body numerical simulations) one first works
with the density field associated with a specific initial condition, that is one
tries to express the non-linear density field in terms of the linear density
field (e.g. as a truncated power series) and next performs the average over the
initial condition. In principles, this formalism can present several advantages.
First, the statistical quantities such as the power-spectrum are precisely
the objects of interest for practical purposes. Secondly, they exhibit symmetries
(such as homogeneity and isotropy) which are not satisfied by peculiar realizations
of the initial conditions. Moreover, they are smooth functions (such as power laws)
rather than singular distributions. This could facilitate the numerical 
computations and the building of various approximations.

Of course, one can recover the standard perturbative results 
by expanding over the cubic part of the action, as recalled in this paper and
discussed in more details in Valageas (2004). However, the advantage of this
formalism is that it may serve as a starting point to other approximation schemes
by applying the methods of statistical mechanics and field theory. Thus, we have
described in details in this paper two large-$N$ expansion schemes. In addition
to the interesting properties of these two methods discussed above, we can note 
that it is useful to have several rigorous expansion schemes which only
differ by higher-order terms than the truncation order. Indeed, since one can expect
that they should be correct up to the scale where they start to depart from one 
another, this should allow one to estimate their range of validity without computing 
explicitly higher-order terms or performing $N$-body simulations.
In fact, in the weakly non-linear regime where all expansion schemes agree this 
procedure should provide a better accuracy than $N$-body simulations which may still be
inaccurate by up to $10\%$. Farther into the non-linear regime, it seems
that the usual one-loop result and the 2PI effective action scheme
give better results than the direct steepest-descent method.
On the other hand, one can hope that other methods than these two expansion schemes
could be applied to the action $S[\psi,\lambda]$. 

In regard to recent works we can
note that the Schwinger-Dyson equations obtained in Crocce \& Scoccimarro (2006a)
can also be obtained from the approach described in this paper, provided we use
the equations of motion in their integral form (that is we first integrate the
time-derivative), as also described in Valageas (2001). Whereas 
Crocce \& Scoccimarro (2006a) used a diagrammatic technique and the resummation
of all diagrams gives back the usual Schwinger-Dyson equations the path-integral
formalism directly gives these equations by expanding the action (or its Legendre 
transform) about a saddle-point. On the other hand, the diagrammatic approach
also provides a direct expression of the response function $R(\eta_1,\eta_2)$ (in the
limit $\eta_2\rightarrow-\infty$) as a series of diagrams (which could be recovered
here by writing the self-energy $\Sigma$ as a series of diagrams). Thus, 
Crocce \& Scoccimarro (2006a) managed to resum these diagrams in the high-$k$ limit
and to obtain the asymptotic behavior of $R(\eta_1,\eta_2\rightarrow-\infty)$.
It is not clear yet how this result could be directly obtained from the
path-integral formalism without expanding over diagrams.

Among the points which deserve further studies, 
it would be interesting to apply these approaches to 
higher-order correlations beyond the two-point functions investigated here. 
On the other hand, the analysis presented in this article
is based on the hydrodynamical equations of motion which break down at
small scales beyond shell-crossing. It is possible to apply these large-$N$ 
expansions to the Vlasov equation (Valageas 2004) but the numerical 
computations would be significantly more complex (since one needs to add
velocities to space coordinates). Nevertheless, the results obtained within
the hydrodynamical framework (damping and well-behaved non-linear 
asymptotics) can be expected to remain valid in the collisionless case and 
studies such as the present one may serve as a first step towards an 
application to the Vlasov dynamics. Such large-$N$ 
expansions could also be applied to simpler effective dynamics which attempt
to go beyond shell-crossing (based for instance on a Schroedinger equation,
Widrow \& Kaiser 1993). 
These works would be of great practical interest as several cosmological probes
(such as weak lensing surveys and measures of the baryon acoustic oscillations)
which aim to measure the recent expansion history of the Universe in order
to constrain the dark energy equation of state (as well as other cosmological
parameters) are very sensitive to the weakly non-linear scales where non-linear
effects can no longer be neglected.


\begin{thebibliography}{}

\bibitem[Bernardeau 1994]{Bernardeau1994}
Bernardeau F., 1994, A\&A, 291, 697

\bibitem[Bernardeau et al. 1997]{Bernardeau1997}
Bernardeau F., van Waerbeke L., Mellier Y., 1997, A\&A, 322, 1

\bibitem[Bernardeau et al. 2002]{Bernardeau2002}
Bernardeau F., Colombi S., Gaztanaga E., Scoccimarro R., 2002, Phys. Rept., 367, 1

\bibitem[Crocce \& Scoccimarro 2006a]{Crocce2006a}
Crocce M., Scoccimarro R., 2006a, Phys. Rev. D, 73, 063519

\bibitem[Crocce \& Scoccimarro 2006b]{Crocce2006b}
Crocce M., Scoccimarro R., 2006b, Phys. Rev. D, 73, 063520

\bibitem[Fry 1984]
Fry J.N., 1984, ApJ, 279, 499

\bibitem[Goroff et al. 1986]{Goroff1986}
Goroff M.H., Grinstein B., Rey S.-J., Wise M.B., 1986, ApJ, 311, 6

\bibitem[Jain \& Bertschinger (1994)]{Jain1994}
Jain B., Bertschinger E., 1994, ApJ, 431, 495

\bibitem[Jain \& Bertschinger (1996)]{Jain1996}
Jain B., Bertschinger E., 1996, ApJ, 456, 43

\bibitem[Koehler et al. 2007]{Koehler2007}
Koehler R.S., Schuecker P., Gebhardt K., 2007, astro-ph/0609354

\bibitem[Liddle \& Lyth (1993)]{Liddle93}
Liddle A.R., Lyth D.H., 1993, Phys. Rep., 231, 1

\bibitem[Makino et al. 1992]{Makino1992}
Makino N., Sasaki M., Suto Y., 1992, Phys. Rev. D, 46, 585

\bibitem[McDonald 2006]{McDonald2006}
McDonald P., 2006, astro-ph/0606028

\bibitem[Martin 1973]{Martin1973}
Martin P.C., Siggia E.D., Rose H.A., 1973, Phys. Rev. A, 8, 423 

\bibitem[Munshi 2007]{Munshi2007}
Munshi D., Valageas P., Van Waerbeke L., Heavens A., 2007, astro-ph/0612667

\bibitem[Peebles 1980]{Peebles1980}
Peebles P.J.E., 1980, The large scale structure of the universe,
Princeton University Press

\bibitem[Peebles 1982]{Peebles1982}
Peebles P.J.E., 1982, ApJ, 263, L1

\bibitem[Phythian 1977]{Phythian1977}
Phythian R., 1977, J. Phys. A, 10, 777

\bibitem[Scoccimarro \& Frieman 1996a]{Scoccimarro1996a}
Scoccimarro R., Frieman J., 1996a, ApJS, 105, 37

\bibitem[Scoccimarro \& Frieman 1996b]{Scoccimarro1996b}
Scoccimarro R., Frieman J., 1996b, ApJ, 473, 620

\bibitem[Seljak 2000]{Seljak2000}
Seljak U., 2000, MNRAS, 318, 203

\bibitem[Seljak \& Zaldarriaga 1996]{Seljak1996}
Seljak U., Zaldarriaga M., 1996, ApJ, 469, 437

\bibitem[Seo \& Eisenstein 2007]{Seo2007}
Seo H.-J., Eisenstein D.J., 2007, astro-ph/0701079

\bibitem[Smith et al. 2003]{Smith2003}
Smith R.E., Peacock J.A., Jenkins A., et al., 2003, MNRAS, 341, 1311

\bibitem[Valageas 2001]{Valageas2001}
Valageas P., 2001, A\&A, 379, 8 

\bibitem[Valageas 2002]{Valageas2002}
Valageas P., 2002, A\&A, 382, 477 

\bibitem[Valageas 2004]{Valageas2004}
Valageas P., 2004, A\&A, 421, 23 

\bibitem[Vishniac (1983)]{Vishniac1983}
Vishniac E.T., 1983, MNRAS, 203, 345

\bibitem[Widrow \& Kaiser 1993]{Widrow1993}
Widrow L.M., Kaiser N., 1993, ApJ, 416, L71

\bibitem[Zeldovich 1970]{Zeldovich1970}
Zeldovich Y.B., 1970, A\&A, 5, 84

\bibitem[Zinn-Justin (1989)]{Zinn1}
Zinn-Justin J., 1989, Quantum field theory and critical phenomena, Clarendon Press, Oxford 

\end{thebibliography}
\end{document}